\documentclass[final]{ustcstep}

\usepackage[dvipdfm]{graphicx}
\usepackage{verbatim}
\usepackage{amsmath}
\usepackage{multirow}
\usepackage{url}
\usepackage[square]{natbib}
\usepackage{color}
\usepackage{ulem}
\usepackage[figuresleft]{rotating}
\usepackage[displaymath]{lineno}

\DeclareGraphicsExtensions{.eps,.jpg,.png}
\newcommand\addr[2]{{\footnotesize \it $^{#1}$#2}\\}

\begin{document}

\title{Investigating plasma motion of magnetic clouds at 1 AU through a velocity-modified cylindrical force-free flux rope model}

\author{Yuming Wang,$^{1,2,*}$ Zhenjun Zhou,$^1$ Chenglong Shen,$^{1,2}$ Rui Liu,$^{1,3}$ and S. Wang$^{1}$\\[1pt]
\addr{1}{CAS Key Laboratory of Geospace Environment, Department
of Geophysics and Planetary Sciences, University of Science and}
\addr{ }{ Technology of China, Hefei, Anhui 230026, China}
\addr{2}{Synergetic Innovation Center of Quantum Information and
Quantum Physics, University of Science and Technology of China,}
\addr{ }{ Hefei, Anhui 230026, China}
\addr{3}{Mengcheng National Geophysical Observatory, School of
Earth and Space Sciences, University of Science and Technology of
China,}
\addr{ }{ Hefei, China}
\addr{*}{Corresponding Author, Contact: ymwang@ustc.edu.cn}}

\maketitle
\tableofcontents

\begin{abstract}
Magnetic clouds (MCs) are the interplanetary counterparts of coronal mass ejections
(CMEs), and usually modeled by a flux rope. By assuming the quasi-steady evolution and self-similar
expansion, we introduce three types of global motion into a
cylindrical force-free flux rope model, and developed a new velocity-modified
model for MCs. The three types of the global motion are
the linear propagating motion away from the
Sun, the expanding and the poloidal motion with respect to the axis
of the MC.
The model is applied to 72 MCs observed by Wind spacecraft
to investigate the properties of the plasma motion of MCs. First, we find that some
MCs had a significant propagation velocity perpendicular to the radial direction,
suggesting the direct evidence of the CME's deflected propagation and/or rotation
in interplanetary space. Second, we confirm the previous results that the expansion
speed is correlated with the radial propagation speed and most MCs did not expand
self-similarly at 1 AU. In our statistics, about 62\%/17\% of MCs underwent a
under/over-expansion at 1 AU and the expansion rate is about 0.6 on average. 
Third, most interestingly, we find that
a significant poloidal motion did exist in some MCs. Three speculations
about the cause of the poloidal motion are therefore proposed.
These findings advance our understanding of the MC's properties at 1 AU
as well as the dynamic evolution of CMEs from the Sun to interplanetary space.
\end{abstract}


\section{Introduction}
Since first identified by \citeauthor{Burlaga_etal_1981} in
1981, magnetic clouds (MCs) have been studied extensively in the
past decades. They are large-scale organized magnetic structures in interplanetary
space, developed from coronal mass ejections (CMEs), and play an
important role in understanding the evolution of CMEs from the Sun to
the heliosphere and the associated geoeffectiveness.

The current knowledge of MCs are mostly from
in-situ one-dimensional observations, and various MC fitting models
have been developed to reconstruct the global picture of MCs in two
or three dimensions. It is now believed that an MC is a loop-like magnetic flux rope with
two ends rooting on the Sun \citep[e.g.,][]{Burlaga_etal_1981,
Larson_etal_1997, Janvier_etal_2013}. The modeling efforts mainly focus on two
aspects. One is to reconstruct a realistic geometry and magnetic
field configuration. In past decades, MC fitting models have been developed
from cylindrically symmetrical force-free flux ropes
\citep[e.g.,][]{Goldstein_1983, Marubashi_1986, Burlaga_1988,
Lepping_etal_1990, Kumar_Rust_1996} gradually to asymmetrically
cylindrical (non-)force-free flux ropes
\citep[e.g.,][]{Mulligan_Russell_2001, Hu_Sonnerup_2002,
Hidalgo_etal_2002, Hidalgo_etal_2002a,
Cid_etal_2002, Vandas_Romashets_2003} and torus-shaped flux ropes
\citep[e.g.,][]{Romashets_Vandas_2003,
Marubashi_Lepping_2007, Hidalgo_Nieves-Chinchilla_2012}. Some
comparisons among various MC fitting models could be found in the
papers by, e.g., \citet{Riley_etal_2004},
\citet{Al-Haddad_etal_2011} and \citet{Al-Haddad_etal_2013}.

The other important aspect is to understand the expansion and distortion of the
cross-section of a MC \citep[e.g.,][]{Farrugia_etal_1993a,
Farrugia_etal_1995, Marubashi_1997, Shimazu_Vandas_2002,
Berdichevsky_etal_2003, Hidalgo_2003, Owens_etal_2006,
Dasso_etal_2007, Demoulin_Dasso_2009, Demoulin_Dasso_2009a, Demoulin_etal_2013}. This is generally an issue
about how the velocity is distributed in an MC. There are lots of
observational evidence that MCs expand when propagating away from
the Sun \citep[e.g.,][]{Klein_Burlaga_1982, Berdichevsky_etal_2003,
Wang_etal_2005a, Jian_etal_2006, Gulisano_etal_2010}. When the
expansion is anisotropic, the initially circular cross-section of a
MC will deform into a non-circular shape. The kinematic treatments
and MHD simulations suggested that MCs may develop into an ellipse or `pancake'
shape \citep{Riley_etal_2003, Riley_Crooker_2004, Owens_etal_2006}.
After exploring the parameter space of flux rope models, \citet{Demoulin_etal_2013}
concluded that the aspect ratio of the ellipse is around 2, but not too large.
On the other hand, the fitting results applying an ellipse
model to the observed MCs suggested that the cross-section of MCs
may be not far from a circle \citep[See Table 1 of
][]{Hidalgo_2003}. A similar result could be seen in the papers
by \citet{Hu_Sonnerup_2001} and \citet{Hu_Sonnerup_2002}, who used the Grad-Shafranov (GS)
technique to reconstruct MCs in a 2-dimensional plane. This technique does not
constrain the shape of the MC's cross-section, but we still can find a
nearly circular cross-section, particularly for the inner core.
Thus, it may be acceptable to assume
a circular cross-section of an MC at 1 AU.

Furthermore, expansion is only one type of the motion of the MC
plasma. Imagining a segment of an MC, which is approximately a
cylindrical flux rope (refer to Fig.\ref{fg_coord}b), one may assume that
there are at least three types of the global plasma motion: linear propagating motion, 
$\mathbf{v}_c=(v_X, v_Y, v_Z)$, expanding motion, $v_{e}$ and poloidal
motion, $v_{p}$. $\mathbf{v}_c$ is the velocity in a rest reference
frame, e.g., GSE coordinate system ($X$, $Y$, $Z$), in which spacecraft movement could be ignored
during the passage of an MC. $v_{e}$ and $v_{p}$ are both the
speeds in a plane perpendicular to the MC's axis. By applying a
cylindrical coordinate system, $(r,\varphi,z)$, sticking to the
axis of the MC, we have $v_{e}=v_r$ and $v_{p}=v_\varphi$.

Most of previous studies of fitting locally observed MCs simply
assumed that MCs propagate radially from the Sun, which means that $\mathbf{v}_c\approx v_X\mathbf{\hat{X}}$.
However, statistical and case studies
of the propagation and geoeffectiveness of CMEs suggested CME may
experience a deflected propagation in interplanetary space
\citep{Wang_etal_2002a, Wang_etal_2004b, Wang_etal_2006a,
Wang_etal_2014, kilpua_etal_2009, Rodriguez_etal_2011, Lugaz_2010,
Isavnin_etal_2013, Isavnin_etal_2014}. It will obviously provide a non-radial component
of the linear propagation motion. Besides, the orientation of the MC
axis may probably rotate when an MC propagates in interplanetary
space \citep[e.g.,][]{Rust_etal_2005, Wang_etal_2006c, Yurchyshyn_2008,
Yurchyshyn_etal_2009, Vourlidas_etal_2011, Nieves-Chinchilla_etal_2012,
Isavnin_etal_2014}. It is another source of
the non-radial motion. The pictures of both deflection and rotation
are mainly established on indirect evidence and models. Thus it is
interesting to seek any signatures of non-radial motion from in-situ
data.

As to the poloidal motion of plasma in MCs, there is so far no
particular study. But some phenomena hint at the possible existence of
such a motion. One of them is the strong field-aligned streams of
suprathermal electrons in MCs \citep[e.g.,][]{Larson_etal_1997}. Another
is the frequently observed plasma flows
in prominences and coronal loops in the solar corona.
Prominences and coronal loops may be a part of
a MC if they are involved in an eruption. If plasma poloidal motion
did exist in MCs, we will shed new light on the dynamic evolution
of CMEs.

The aim of the present work is to investigate the plasma motion of
MCs from in-situ data with the aid of a flux rope model. As the
first attempt, we utilize a relatively simple and ideal flux rope
model, which is cylindrically symmetrical and force-free, but with all the
three components of the plasma motion taken into account. One could
go to the web page \url{http://space.ustc.edu.cn/dreams/mc_fitting/} to run
our model.
The details of the model and method we employed in this work are
described in the next section.

\section{Model and Events}
\subsection{Velocity-modified cylindrical force-free flux rope model}\label{sec_model}
\subsubsection{Model description}\label{sec_model1}

\begin{figure*}[tbh]
  \centering
  \includegraphics[width=\hsize]{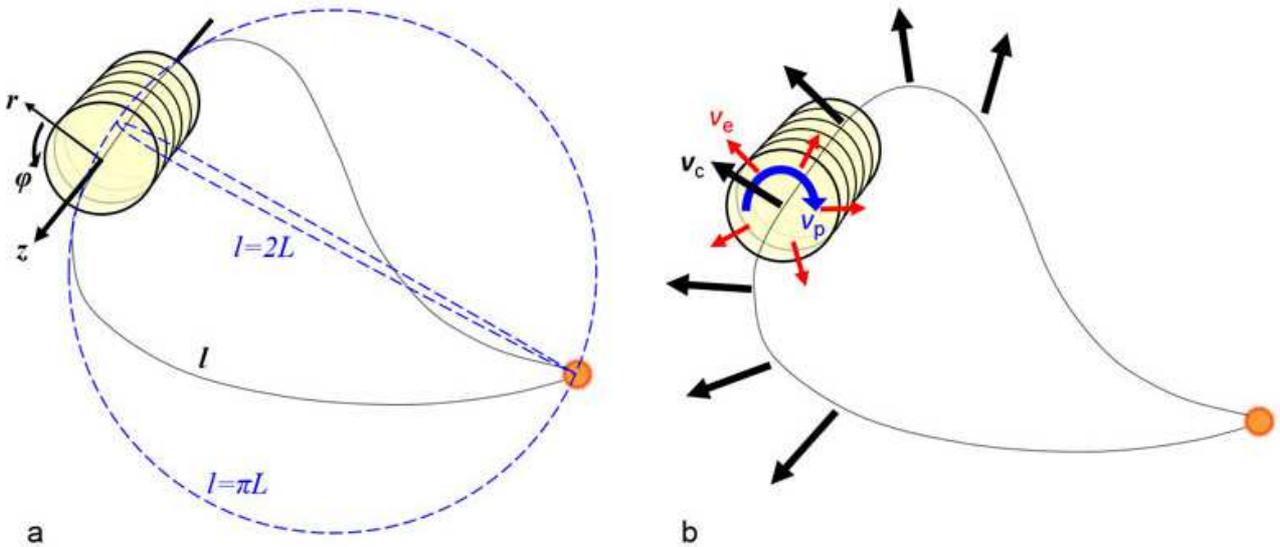}
  \caption{(a) Schematic picture of an MC at the heliocentric distance of $L$ (adapted from \citet{Wang_etal_2009}).
The black line indicates the looped axis of the MC with a length of $l$. The blue dashed lines suggest the upper
and lower limits of $l$. (b) Illustration of the three types of global motions of an MC. The black, red and blue
arrows denote the linear propagating motion, expanding motion and poloidal motion, respectively.}\label{fg_coord}
\end{figure*}

Our model is developed from the cylindrically symmetrical force-free
flux rope model that has been widely used in many previous studies
\citep[e.g.,][]{Lepping_etal_1990}. The following coordinate systems are used.
One is the GSE coordinate system, $(X, Y, Z)$, in which
$\mathbf{\hat{X}}$ is along the Sun-Earth (or Sun-spacecraft) line pointing toward the Sun,
$\mathbf{\hat{Z}}$ is a northward vector perpendicular to the ecliptic
plane and $\mathbf{\hat{Y}}$ completes the right-handed coordinate
system. Another is a cylindrical coordinate system in the MC
frame, $(r,\varphi,z)$, with $\mathbf{\hat{z}}$ along the axis of the
flux rope, as illustrated by Figure~\ref{fg_coord}a. Sometimes to
show the pattern of magnetic field and velocity, one
more Cartesian coordinates in the MC frame, $(x',y',z')$, is used,
in which $\mathbf{\hat{z}}'$ is identical with $\mathbf{\hat{z}}$,
$\mathbf{\hat{x}}'$-axis is the projection of the observational path on
the plane perpendicular to $\mathbf{\hat{z}}'$, approximately directed
toward the Earth and $\mathbf{\hat{y}}'$
completes the right-hand coordinate system (see
Fig.\ref{fg_vcircase1}e and $\ref{fg_vcircase2}e$).

The magnetic field of a cylindrical flux rope is described by
\citet{Lundquist_1950} solution in the coordinates $(r,\varphi,z)$
\begin{eqnarray}
\mathbf{B}_r&=&0 \label{eq_b1}\\
\mathbf{B}_\varphi&=&HB_0J_1(\alpha r)\boldsymbol{\hat{\varphi}} \label{eq_b2}\\
\mathbf{B}_z&=&B_0J_0(\alpha r)\mathbf{\hat{z}} \label{eq_b3}
\end{eqnarray}
in which $H=\pm1$ is the handedness or sign of the helicity, $B_0$
is the magnetic field strength at the axis, $\alpha$ is the constant
force-free factor, and $J_0$ and $J_1$ are the Bessel functions of
order 0 and 1, respectively. Like common treatment, we set the
boundary of a flux rope at the first zero point of $J_0$, which
leads to $\alpha=2.41/R$ and $R$ is therefore the radius of the flux
rope.

If any kinematic evolution of the flux rope could be ignored, the
magnetic field profile along an observational path is determined
once the following six parameters are known: (1)
the orientation of the flux rope's axis, i.e., the $\mathbf{\hat{z}}$-aixs, 
which is given by the
elevation and azimuthal angles, $\theta$
and $\phi$, in GSE coordinate system; $\theta=0$ is in the ecliptic
plane and $\theta=0$ and $\phi=0$ is toward the Sun, (2) the closest
approach (CA) of the observational path to the flux rope's axis,
given by $d$ in units of $R$; a positive/negative value of $d$ means
that the observational path is above/below the axis in $(x',y',z')$
coordinates, i.e., $y'>0$/$y'<0$ as shown by the examples of the positive $d$ 
in Fig.\ref{fg_vcircase1}e
and \ref{fg_vcircase2}e, and (3) the free parameters in
Eq.\ref{eq_b1}--\ref{eq_b3}, which are $H$, $B_0$ and $R$.

After the possible plasma motion is taken into account, five
additional parameters should be considered, which are $\mathbf{v}_c=(v_X,
v_Y, v_Z)$, $v_{e}$ and $v_{p}$. Here we assume that the flux
rope propagates and expands uniformly and experiences a quasi-steady,
self-similar expansion (or contraction) during the period of
interest. Thus, $\mathbf{v}_c$ is a constant vector, describing a global
linear motion, that does not change the internal magnetic field of
the flux rope.

The parameter $v_{e}$ is a constant expansion speed of the
boundary of the cross-section of the flux rope.
The self-similar assumption gives the expansion speed at any radial
distance, $r$, away from the flux rope's axis as
\begin{eqnarray}
v_r(x)=xv_{e} \label{eq_vr}
\end{eqnarray}
in which $x$ is the normalized radial distance,
equal to $r/R$. As a
consequence, the radius of the cross-section of the flux rope
evolves as follows
\begin{eqnarray}
R(t)=R(t_0)+v_{e}(t-t_0)\label{eq_rt}
\end{eqnarray}
in which $t_0$ is the initial time (or the time the observer first
encountering the MC flux rope), and the magnetic field as
\begin{eqnarray}
B_0(t)=B_0(t_0)\left[\frac{R(t_0)}{R(t)}\right]^2
\label{eq_flux_cons}
\end{eqnarray}
Eq.\ref{eq_flux_cons} follows the magnetic flux conservation.

It should be noted that, due to the curvature of the whole looped
flux rope, the propagation velocity of the different segment of the
flux rope is along the different direction as illustrated in Figure~\ref{fg_coord}b,
which may actually cause the expansion along the axis of the flux rope.
That is also why the length of the flux rope increases as indicated by Eq.\ref{eq_rl} below.
The axial expansion rate could be on the same order of the expansion rate
of the flux rope's cross-section \citep[e.g.,][]{Dasso_etal_2007,
Demoulin_etal_2008, Nakwacki_etal_2011}, which guarantee a roughly self-consistent 
evolution of a force-free flux rope \citep{Shimazu_Vandas_2002}.

Under the assumption of self-similar
expansion and the conservations in both mass and angular momentum,
the poloidal speed in the flux rope can be derived (see Appendix \ref{app_poloidal}) as
\begin{eqnarray}
v_\varphi(t, x)=k_1f_p(x)R(t)^{-1}\label{eq_vphi}
\end{eqnarray}
in which $k_1$ is a constant and $f_p(x)$ is a function dependent only on the relative position $x$.
However, the expression of $f_p(x)$ cannot be specified theoretically. As the first attempt,
here we tentatively assume $f_p(x)=1$. It should be noted that the point at $x=0$ is a singularity
under this assumption because $v_\varphi(0)$ has a non-zero value that is not physically meaningful. 
We just ignore this singularity, and as will be seen in Sec.\ref{sec_cir},
we prove that this assumption can be treated as an acceptable approximation. Then Eq.\ref{eq_vphi}
can be rewritten as \begin{eqnarray}
v_\varphi(t, x)=v_{p}(t)=v_{p}(t_0)\frac{R(t_0)}{R(t)}\label{eq_vp}
\end{eqnarray}
where the parameter $v_{p}(t)$ defines the poloidal speed of the plasma at
the boundary of the flux rope in the direction of
$\boldsymbol{\hat{\varphi}}$.

\subsubsection{Parameters}\label{sec_para}
In summary, Table~\ref{tb_para} lists all the 11 free parameters in
the velocity-modified cylindrical force-free flux rope model.
Besides, based on the cylindrical force-free flux rope assumption,
we may derive more parameters, which have been also summarized in
Table~\ref{tb_para}. It is straightforward to obtain the first two
derived parameters, $t_c$ and $\Theta$. The next four parameters are
derived as follows.

The axial magnetic flux, $\Phi_z$, is given by
\begin{eqnarray}
\Phi_z=\int_0^{2\pi}\int_0^{R} B_zrdrd\varphi= B_0R^2\kappa_1\label{eq_fluxz}
\end{eqnarray}
in which $\kappa_1=\frac{2\pi}{x_0^2}\int_0^{x_0}xJ_0(x)dx=1.35$ and $x_0=2.41$
is the first zero point of Bessel function $J_0$.
The poloidal magnetic flux, $\Phi_\varphi$, is given by
\begin{eqnarray}
\Phi_\varphi=\int_0^l\int_0^R B_\varphi drdz =B_0Rl\kappa_2\label{eq_fluxp}
\end{eqnarray}
in which $l$ is the length of the loop as denoted in Figure~\ref{fg_coord}a and
$\kappa_2=\frac{1}{x_0}\int_0^{x_0}J_1(x)dx=0.416$.
The helicity, $H_m$, is given by \citep[e.g.,][]{Berger_2003}
\begin{eqnarray}
H_m=\int_0^l\int_0^{2\pi}\int_0^R \mathbf{A}\cdot\mathbf{B} rdrd\varphi dz=B_0^2R^3l\kappa_3 \label{eq_helicity}
\end{eqnarray}
in which $\mathbf{A}$ is a vector potential of $\mathbf{B}$ and $\kappa_3=\frac{2\pi}{x_0^3}\int_0^{x_0}x[J_0^2(x)+J_1^2(x)]dx=0.701$.
The magnetic energy, $E_m$, is given by
\begin{eqnarray}
E_m=\frac{1}{2\mu}\int_0^l\int_0^{2\pi}\int_0^R B^2rdrd\varphi dz=B_0^2R^2l\kappa_4\label{eq_em}
\end{eqnarray}
in which $\kappa_4=\frac{x_0}{2\mu}\kappa_3=6.72\times10^5$ m H$^{-1}$.

We do not know the value of $l$, but we may assume that it is bounded between $\pi L$ and $2L$,
$L$ is the heliocentric distance of the flux rope, as shown in Figure~\ref{fg_coord}a.
In practice, we let 
\begin{eqnarray}
l=\left(\frac{\pi+2}{2}\pm\frac{\pi-2}{2}\right)L \label{eq_length}
\end{eqnarray}
The uncertainty, $\frac{\pi-2}{2}L$, in the length of the flux rope
will result in the uncertainty in $\Phi_\varphi$, $H_m$ and $E_m$. It should be noted that (1)
the uncertainty here is not from the fitting procedure but the length of the flux rope, and
(2) this treatment will underestimate the real length if the leg of an MC rather than its leading
part was observed.

Considering the conservation of magnetic flux, we may infer from Eq.\ref{eq_fluxz} and
\ref{eq_fluxp} that
\begin{eqnarray}
R\propto l \rm{\ \ \ or\ \ \ } R\propto L \label{eq_rl}
\end{eqnarray}
and consequently, the helicity is conserved too. 
Further, we may infer $E_m\propto L^{-1}$, suggesting that magnetic energy continuously decreases
when an MC propagates away from the Sun. In this work, we calculate its initial value, $E_{m0}$,
as listed in Table~\ref{tb_para}, which is the magnetic energy when the MC is only one solar radius
away from the Sun, and the associated uncertainty of about $\pm22\%$ comes from the uncertainty
in $l$. Be aware that the decay index of the magnetic energy of an MC may not be $-1$. The case study
by \citet{Nakwacki_etal_2011}, for example, has shown that the decay index for the MC observed
by two spacecraft at 1 and 5.4 AU, respectively, on 1998 March is about $-0.9$, a little bit slower
than the expectation from our model. If the decay index varies from $-0.9$ to $-1.1$, the extrapolated $E_{m0}$
suffers an additional uncertainty of about $^{+42\%}_{-71\%}$, which we do not taken into account in
our model.

\begin{table*}
\linespread{1.5} \caption{Parameters involved in the
velocity-modified cylindrical force-free flux rope model$^*$}
\begin{center}
\begin{tabular}{c|p{430px}}
\hline Parameter & Explanation \\
\hline
\multicolumn{2}{c}{Free parameters in the model} \\
\hline
$B_0(t)$ & Magnetic field strength at the axis of the flux rope.\\
$R(t)$ & Radius of the cross-section of the flux rope.\\
$H$ & Handedness or sign of helicity, must be $1$ (Right) or $-1$ (Left).\\
$\theta$ & Elevation angle of the axis of the flux rope in GSE.\\
$\phi$ & Azimuthal angle of the axis of the flux rope in GSE.\\
$d$ & The closest approach of the observational path to the axis of
the flux rope.\\
$v_X$ & Propagation speed of the flux rope in the direction of $\mathbf{\hat{X}}$.\\
$v_Y$ & Propagation speed of the flux rope in the direction of $\mathbf{\hat{Y}}$.\\
$v_Z$ & Propagation speed of the flux rope in the direction of $\mathbf{\hat{Z}}$.\\
$v_{e}$ & Expansion speed of the boundary of the flux rope in the direction of $\mathbf{\hat{r}}$.\\
$v_{p}(t)$ & Poloidal speed at the boundary of the flux rope in the direction of $\mathbf{\hat{\varphi}}$.\\
\hline
\multicolumn{2}{c}{Other derived parameters from the model} \\
\hline
$t_c$ & The time when the observer arrives at the closest approach.\\
$\Theta$ & Angle between the axis of the flux rope and
$\mathbf{\hat{X}}$-axis.\\
$\Phi_z$ & Axial magnetic flux of the flux rope.\\
$\Phi_\varphi$ & Poloidal magnetic flux of the flux rope.\\
$H_m$ & Magnetic helicity of the flux rope.\\
$E_{m0}$ & Initial magnetic energy, i.e., the magnetic energy of the flux rope when it was one solar radius away from the Sun.\\
$\chi_n$ & Normalized root mean square (rms) of the difference between the
modeled results and observations.\\
\hline
\end{tabular}
$^*$ See Sec.\ref{sec_model} for the definition of coordinate systems, derivations of some parameters and other details.
\end{center}
\label{tb_para}
\end{table*}

The last derived parameter, $\chi_n$, is used to evaluate the goodness-of-fit, which will be
introduced in the following section.

\subsubsection{Evaluation of the goodness-of-fit}\label{sec_good}
Our fitting procedure is designed to fit the observed magnetic field,
$\mathbf{B}^o$, and velocity, $\mathbf{v}^o$, together, which means that all
the three components of $\mathbf{B}^o$ and $\mathbf{v}^o$ are used to
constrain the model parameters.
There are 11 free parameters, among which three of them are time-dependent (see Table~\ref{tb_para})
and can be determined by Eq.\ref{eq_rt}, \ref{eq_flux_cons} and \ref{eq_vp}.
Here, we use the first contact of the MC, $t_0$, as the reference time.
The detailed fitting process is as follows. First, by given a set of 11 free parameters, 
the imaginarily observational path relative to the axis of the flux rope can be derived from 
the orientation of the axis, $(\theta, \phi)$, the closest approach, $d$, the propagation velocity, 
$(v_X, v_Y, v_Z)$, and the expansion speed, $v_e$. Second, the coordinates of the observational path are then
transformed from the GSE coordinate system to the MC frame, $(r,\varphi,z)$. In the MC frame, 
the three components of magnetic field along the path can be determined by Eq.\ref{eq_b1}--\ref{eq_b3}, 
in which the free parameters $B_0$, $R$ and $H$ are involved, and the velocity
can be calculated as $v_r\hat{\mathbf{r}}+v_\varphi\hat{\boldsymbol{\varphi}}$, in 
which $v_r$ and $v_\varphi$ are given by Eq.\ref{eq_vr} and \ref{eq_vp}. Third, we transform 
the derived magnetic field, $\mathbf{B}^m$, and velocity, $\mathbf{v}^m$, in the MC frame 
back to the GSE coordinate system, and evaluate the goodness-of-fit by calculating 
the normalized root mean square (rms) of the difference between the
modeled and observed values of magnetic field and velocity,
which is given as
\begin{eqnarray}
\chi_n&=&\sqrt{\frac{1}{2N}\sum_{i=1}^N\left[\left(\frac{\mathbf{B}_i^m-\mathbf{B}_i^o}{|\mathbf{B}_i^o|}\right)^2+\left(\frac{\mathbf{v}_i^m-\mathbf{v}_i^o}{|\mathbf{v}_i^o|-v_{\rm ref}}\right)^2\right]}\nonumber\\
&=&\sqrt{\frac{1}{2}(\chi_{Bn}^2+\chi_{vn}^2)} \label{eq_chi}
\end{eqnarray}
where $N$ is the number of measurements, and $v_{\rm ref}$ is a reference velocity.

To find the best-fit, we use the IDL package, MPFIT (refer to \url{http://purl.com/net/mpfit}), 
to perform least-squares fitting \citep{Markwardt_2009, More_1978}. The initial value of $B_0(t_0)$ is 
set to be the maximum value of the magnetic field strength during the interval of interest, 
the initial value of $R(t_0)$ is estimated as $\frac{1}{2}\overline{v_X}\Delta t$, in which $\Delta t$ is the duration
of the MC interval and $\overline{v_X}$ is the mean value of the observed $v_X$, the initial value
of the propagation velocity $(v_X,v_Y,v_Z)$ is the mean value of the observed velocity, 
the initial value of the expansion speed is estimated from the slope of the observed radial velocity, and
the initial value of the poloidal speed is set to be zero.
For the free parameter $H$, we just fix its value to $1$ or $-1$ by 
adding a loop in our fitting procedure. The elevation and azimuthal angles, $\theta$ and $\phi$, 
are two most important free parameters. In order to get the best fitting result, 
we test the initial value of $\theta$ every $15^\circ$ from $-90^\circ$ to $90^\circ$, 
and do the same thing for $\phi$ from $0^\circ$ to $360^\circ$. 
Besides, we assume that the
front and rear boundaries of an observed MC define the interval of the flux rope, and then the 
closest approach, $d$, could be uniquely determined based on the preset velocity and the axis orientation 
of the flux rope. In summary, we try 576 attempts of fitting (i.e., 576 sets of the initial values of the free parameters)
for an MC, and select the case with the smallest value of $\chi_n$ as 
the best-fit.

The reference velocity in Eq.\ref{eq_chi} is used to adjust the weight of the velocity in evaluating the goodness-of-fit, 
and is set as 
\begin{eqnarray}
\frac{\max(|\mathbf{v}^o|)-\min(|\mathbf{v}^o|)}{\max(|\mathbf{v}^o|)-v_{\rm ref}}=\frac{\max(|\mathbf{B}^o|)-\min(|\mathbf{B}^o|)}{\max(|\mathbf{B}^o|)}
\end{eqnarray}
If there was no reference velocity, $\chi_{Bn}$ and $\chi_{vn}$ may not have the same weight and cannot be 
inserted into one formula, because the dynamic range of the velocity is 
much different from that of the magnetic field. For example, assuming an MC interval during 
which the magnetic field varies from 5 to 35 nT and the velocity varies from 300 to 600 km s$^{-1}$, 
and at a given point $B^o=25$ nT, $B^m=20$ nT, $v^o=500$ km s$^{-1}$ and $v^m=450$ km s$^{-1}$,
we can get that the relative error between the modeled and observed values in magnetic field 
is $\frac{5}{25}=20\%$ and that in velocity is $\frac{50}{500}=10\%$. The values of the 
relative errors are different. But if considering the range of 30 nT in magnetic field and 
the range of 300 km s$^{-1}$ in velocity, one can find that the relative difference between the modeled and observed values
in magnetic field is the same as that in velocity.
Thus the value of relative error depends on the dynamic range. 
To remove this effect, we use the reference velocity, which is 250 km s$^{-1}$ in the above case.
The corrected relative error of the velocity becomes $\frac{50}{500-250}=20\%$, the same as that of the magnetic field.
After this treatment, $\chi_{Bn}$ and $\chi_{vn}$ can be fitted into one formula, Eq.\ref{eq_chi}, to assess the goodness-of-fit.

The normalized rms, $\chi_n$, has a definite meaning here. It measures
the average relative error of modeled vectors, $\mathbf{B}$ and $\mathbf{v}$, with both length and
direction taken into account. 
It should be noted that the normalized rms here is different from those used in some studies.
For example, in \citet{Lepping_etal_1990} and
\citet{Lepping_etal_2006}, $\mathbf{B}^o$ and $\mathbf{B}^m$ are normalized by
the strength of themselves, and that rms merely reflects the average
deviation of vector directions; while in
\citet{Marubashi_etal_2012}, $\mathbf{B}^o$ and $\mathbf{B}^m$ are normalized
by the maximum value of observed $B$ during the whole interval of
the flux rope, which might imply that a stronger MC has a smaller value of rms.

\subsection{Events and model testing}
The MC list compiled by~\citet{Lepping_etal_2006} is the basis of this study, and hereafter
we call it Lepping list (see \url{http://lepmfi.gsfc.nasa.gov/mfi/mag_cloud_S1.html}; the list including the fitted parameters are kept being
updated until 2011 December 13).
We use this list not only because it is well established but also because we can test
the procedure of our model by comparing our fitting results with \citet{Lepping_etal_2006}
results. There are a total of 121 MCs in the list. The start time, end time and model-derived
parameters are all given in the list. The goodness-of-fit is estimated by $Q_0$,
which is 1, 2 or 3 for good, fair and poor, respectively. Our study only
considers the MCs in the first two categories. It is noticed that
there are large data gaps in the published Wind data during 2000 July 15 -- 16.
Their event No.45 and 46 are in that period, and are therefore removed
from our study. Besides, the event No.85 in their list was studied by
\citet{Dasso_etal_2009}, and is believed to consist of two MCs. Thus we discard this
event too. Finally, a total of 72 events are available for our study as listed in
Table~\ref{tb_parbv1}--\ref{tb_parbv3} in Appendix \ref{app_tables}.

\begin{figure*}[p]
  \centering
  \includegraphics[width=0.9\hsize]{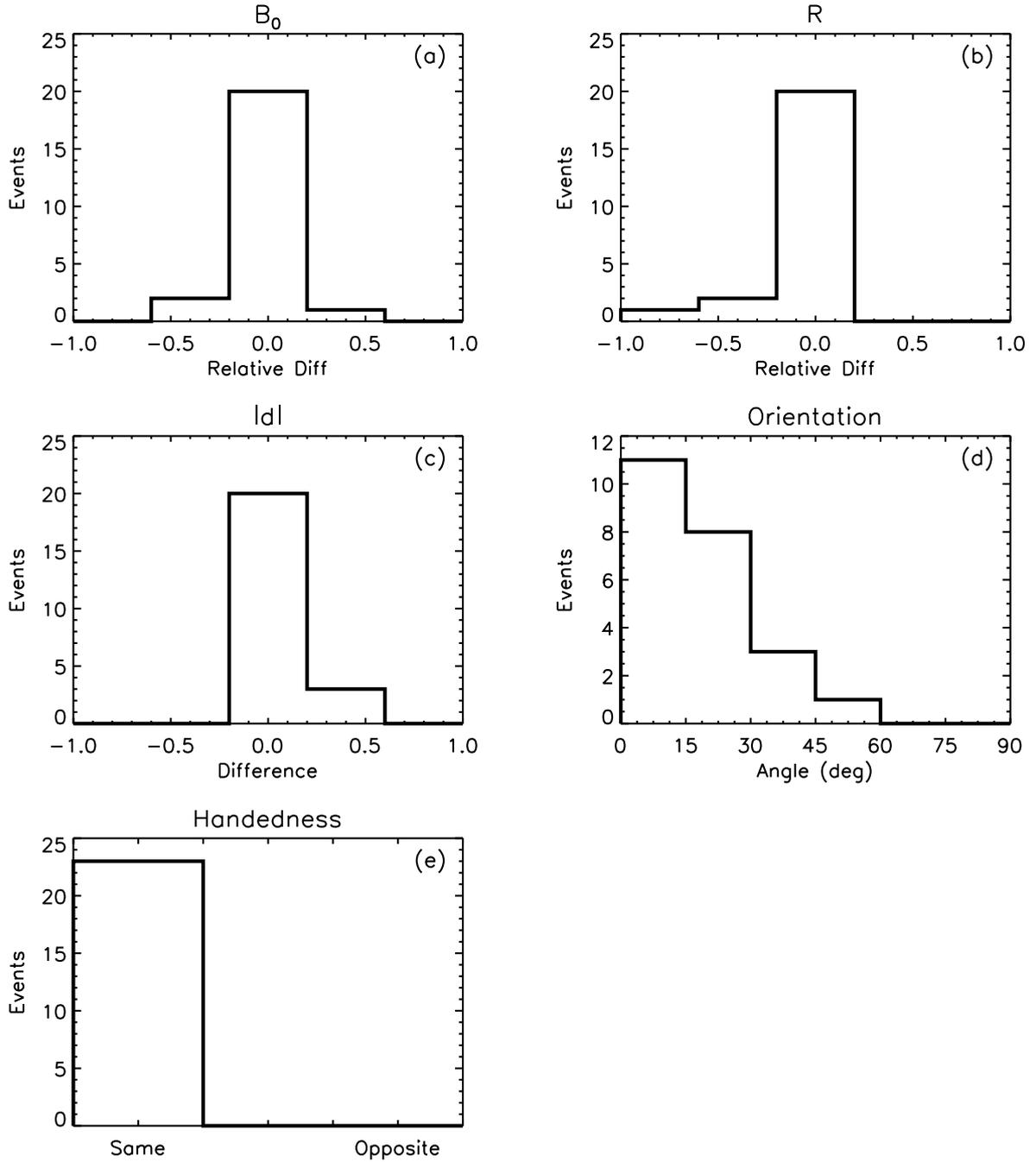}
  \caption{Histograms showing the differences of the values of fitting parameters between the
  Lepping's model and our model without velocity measurements taken into account. From top to
  bottom and left to right, there are the relative difference of the magnetic field
  strength ($B_0$) at the MC's axis between the two models, the relative difference of
  the radius ($R$), the difference of the closest approach $|d|$, the acute angle between the MC's
  axes derived from the two models, and the difference of the handedness.}\label{fg_errlepping}
\end{figure*}

To test our procedure, we switch off the velocity option in our model, and compare
the fitted parameters with those given in Lepping list.
When fitting an observed MC
with our model, we use the time-resolution of the Wind data as the same as that
indicated in the list, which is 15 minutes for some events and 30 minutes for others.
The front and rear boundaries of the MCs given in Lepping list are used and fixed in our model.
It is different from the procedure taken by \citet{Lepping_etal_2006}, in which
the preset boundaries may be modified by their model. The difference between their
fitted and observed boundaries is evaluated by the parameters, ck and asf,
given in their list (see Eq.7 and 9 in \citealt{Lepping_etal_2006} for their definitions).
Larger values of ck and asf mean larger difference. Thus to make a mostly reliable
comparison, we further tentatively exclude events whose ck or asf is larger than 10\%, that
results in a sample of 23 events. It should be noted that those excluded events will be
included again in the next section though they are not used for testing our procedure here.

The parameters characterizing the configuration and strength of an MC are compared,
which are $H$, $B_0$, $R$, $d$, and orientation ($\theta$ and $\phi$) of an MC.
Figure~\ref{fg_errlepping} shows the differences of the values between the parameters from
our and \citet{Lepping_etal_2006} procedures. It is found that both of them suggest the same
sign of helicity, $H$, for each MC (Fig.~\ref{fg_errlepping}e). For $B_0$ and $R$, we
calculate the relative difference, $\frac{f-f_L}{f_L}$, where $f$ is
from our results and $f_L$ from Lepping list. Our procedure gets very similar
results with Lepping list (Fig.\ref{fg_errlepping}a and \ref{fg_errlepping}b).
There are only 3 events, in which the relative difference in $B_0$ and $R$ is outside of
$\pm20$\%. For $d$, we do not use the above equation
to calculate the relative difference, because $d$ is a ratio of the closest approach
to the radius of the flux rope and its value could be zero which may cause the value of the
relative difference to be extremely large. Instead, we directly use the difference of absolute values of them, $|d|-|d_L|$.
A good agreement also can be obtained between the two procedures
(Fig.\ref{fg_errlepping}c). The two parameters $\theta$ and $\phi$, which define the orientation
of a flux rope's axis, are compared together. We calculate the angle between the orientation
derived by our procedure and that given in Lepping list. It is found that the angle is smaller
than $60^\circ$, and for most (19 out of 23) events, the angle is less than $30^\circ$.
Overall, the comparison shows that our procedure can almost reproduce the results derived
by \citet{Lepping_etal_2006}, confirming the validity of our procedure.

Now we switch on the velocity option, and apply the model to the 72 high-quality events
in Lepping list. The time-resolution of the data input into
our model is set to 10 minutes.
All the parameters derived from our fitting
procedure of these events have been listed in Table \ref{tb_parbv1}--\ref{tb_parbv3} in Appendix \ref{app_tables} for reference.
In the following two sections, we will evaluate
the effects of velocity on the goodness-of-fit, and show the statistical properties of plasma motion.

\section{Effects of velocity on the fitting results}
\begin{figure}[tb]
  \centering
  \includegraphics[width=0.8\hsize]{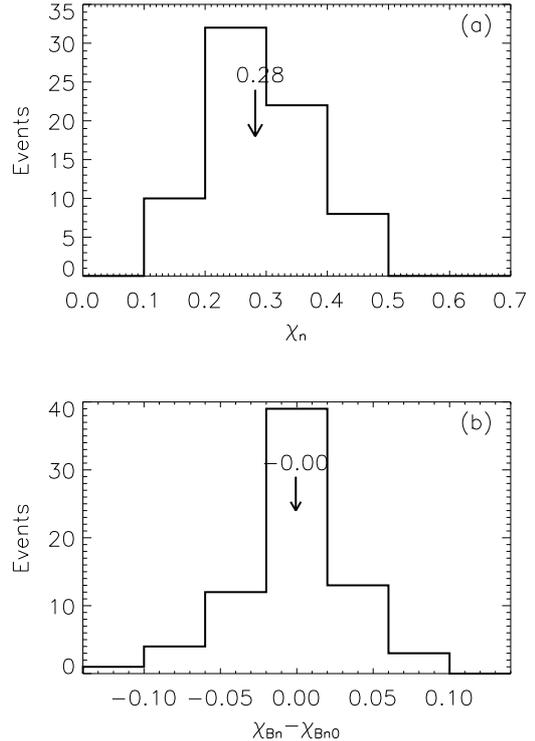}
  \caption{The distributions of normalized rms, which show the goodness-of-fit. 
(a) the normalized rms ($\chi_n$) calculated
  from both magnetic field and velocity, and (b) the difference of the normalized rms
  in magnetic field between the model results with velocity-on ($\chi_{Bn}$) and velocity-off ($\chi_{Bn0}$).}\label{fg_chi}
\end{figure}

\begin{figure*}[tb]
  \centering
  \includegraphics[width=\hsize]{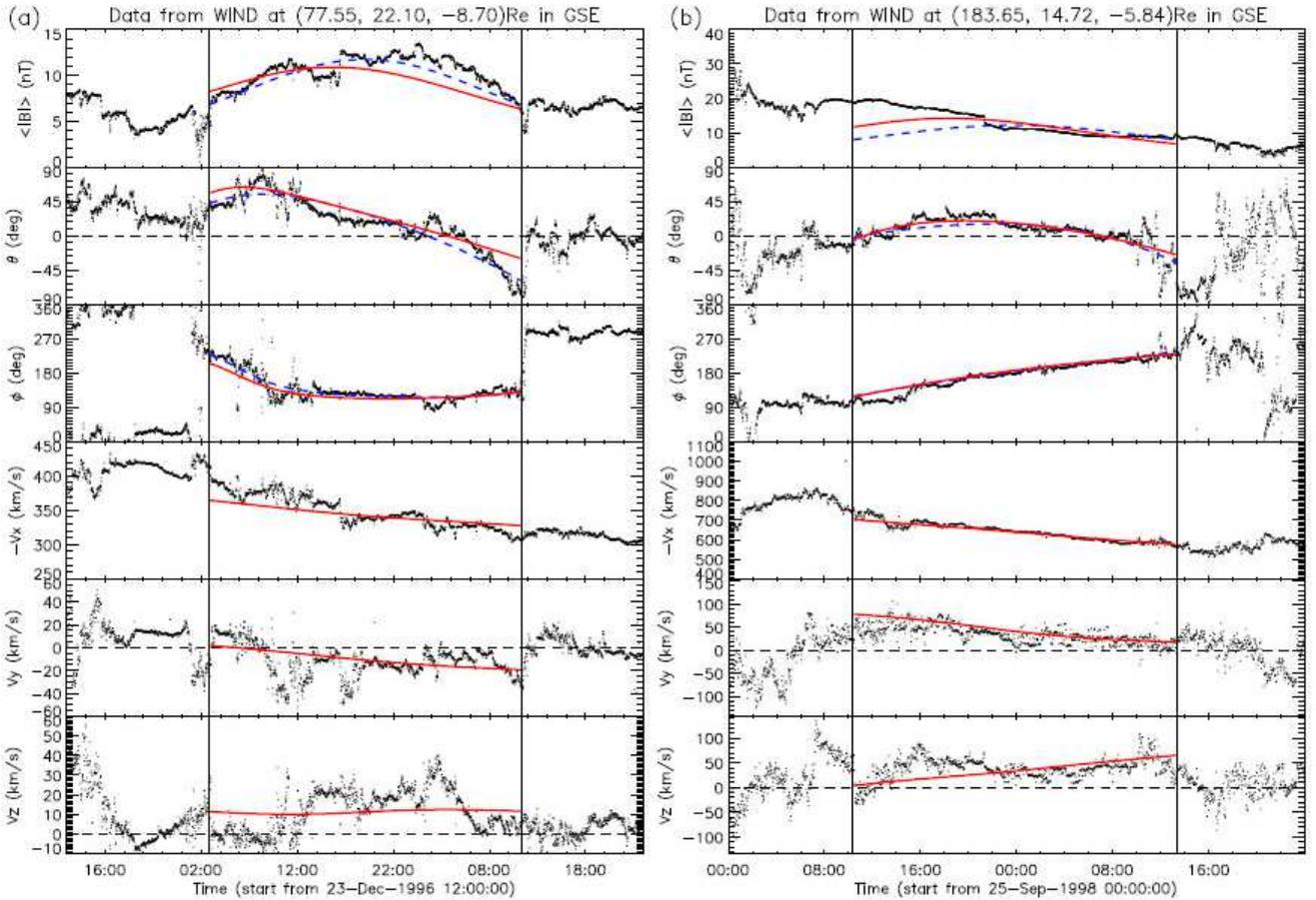}
  \caption{Two cases showing the comparison of the fitting results of our model with velocity-on (the red lines) and off (the blue dashed lines). Case (a) shows that the fitting results become worse when the velocity is taken into account, and Case (b) shows an opposite situation. In both cases, from the top to bottom, the panels are total magnetic field strength, elevation and azimuthal angles of magnetic field vector and three components of bulk velocity of solar wind plasma, respectively. The two vertical lines mark the front and rear boundaries of the MCs.}\label{fg_casebv}
\end{figure*}

The goodness-of-fit is evaluated by the average relative error, $\chi_n$, given by Eq.\ref{eq_chi}.
For all the 72 events, the distribution of $\chi_n$ has been displayed in Figure~\ref{fg_chi}a.
It is found that the values of $\chi_n$ are all less than 60\%,
and on average, $\chi_n$ is about 28\%. Further, the comparison between the
cases of velocity-on and velocity-off shows that considering velocity may 
get the value of $\chi_{Bn}$ larger or smaller (see Fig.\ref{fg_chi}b).
As examples, Figure~\ref{fg_casebv}a and \ref{fg_casebv}b show two cases; for one of them the fitting
to the magnetic field gets worse, and for the other, the fitting looks better.
In a typical MC measurements, a declining velocity means the expansion of the MC and as a consequence,
the magnetic field strength of the MC should decrease as well. The former case does not follow
the picture. The strongest magnetic field appeared in the rear part of the MC, but the velocity profile
indicates a clear expansion, suggesting that the strongest magnetic field should appeared in the front
part of the MC. The fitting without velocity being taken into account gives the blue dashed curves, which
match the observed magnetic field much better than red solid curves given by our velocity-modified model.
The latter case is a typically expanding MCs, and therefore the magnetic field is fitted better.

\begin{figure*}[p]
  \centering
  \includegraphics[width=0.9\hsize]{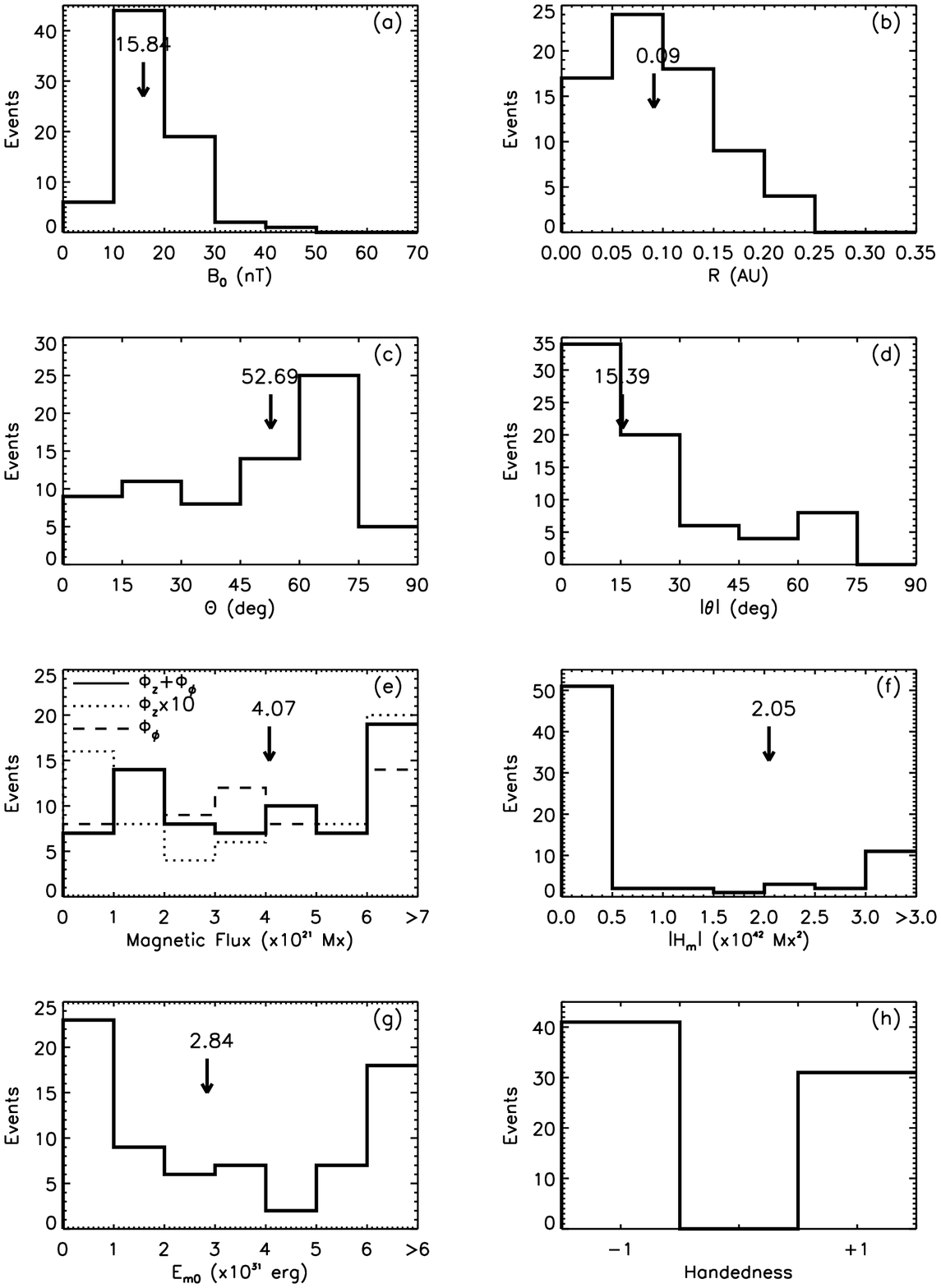}
  \caption{The distributions of the values of the fitting parameters from the velocity-modified model
  for 72 MC events. From the top to bottom and left to right, the panels show the magnetic field at 
the MC's axis ($B_0$), the radius the MC's cross-section ($R$), the acute angle between the axis 
and the Sun-spacecraft line ($\Theta$), the absolute value of the elevation angle of the axis ($|\theta|$), 
the total magnetic flux ($\Phi_z+\Phi_\varphi$), the unsigned helicity ($|H_m|$), the initial 
magnetic energy ($E_{m0}$) and the handedness ($H$). The median values of these parameters 
are indicated by the arrows.}\label{fg_mcprop}
\end{figure*}

Except those velocity-related parameters (which will be presented in the next section),
the parameters of MCs derived from our velocity-modified model are summarized in Figure~\ref{fg_mcprop}.
For the magnetic field strength at the MC axis,
most events fall into the range from 10 to 30 nT with an median value of about 16 nT.
For the radius, all the MCs are less than 0.25 AU, and its median value is about 0.09
AU. Note, the values of the above two parameters are all adopted at the time, $t_c$,
when the observer arrives at the closest approach to the MC axis. The angle between
the MC axis and the Sun-spacecraft line, $\Theta$, could be any value from $0^\circ$ to
$90^\circ$. The most probable angle is within $45^\circ-75^\circ$, and the median angle
is about $53^\circ$. It suggests that observed MCs are more likely to transversely cross
over the observer. The elevation angle of the MC axis tends to be small, with a median value
of about $15^\circ$, implying low tilt angles of flux ropes when they erupted from
the Sun.

Further by using the total length of the flux rope, $l$, given by Eq.\ref{eq_length}, 
we estimate that the total magnetic flux, which includes axial flux and poloidal
flux, is on the order of $10^{21}$ Mx with the median value of about $4.1\times10^{21}$ Mx.
The distributions of the axial flux and poloidal flux are also indicated by dotted and dashed lines (Fig.\ref{fg_mcprop}e).
Their median values are $0.4\times10^{21}$ and $3.6\times10^{21}$ Mx, respectively.
The estimation of the magnetic flux of MCs were made before by, e.g., \citet{Dasso_etal_2005},
\citet{Dasso_etal_2007} and \citet{Nakwacki_etal_2008}.
In \citet{Dasso_etal_2005}, for example, eight well-defined MCs were investigated and it was found
that the axial flux is around $0.4\times10^{21}$ Mx, highly consistent with the result obtained here.
Our results are also roughly in agreement with previous studies about the magnetic flux of MCs and reconnection flux
of solar eruptions by, e.g., \citet{Qiu_etal_2007}.

The helicity is shown in Figure~\ref{fg_mcprop}f and \ref{fg_mcprop}h.
The number of right-handed MCs is almost
equal to the number of left-handed MCs, and the absolute value
of the helicity is about $2.05\times10^{42}$ Mx$^2$ on average. For the 8 events
studied by \citet{Dasso_etal_2005}, the helicity per unit length is about $1\times10^{42}$ Mx$^2$/AU.
According to Eq.\ref{eq_length}, the helicity in their study is about $2.57\times10^{42}$ Mx$^2$,
consistent with the average value we obtained here.
Moreover, the initial magnetic energy is found to be about $2.84\times10^{31}$ erg, nearly one order higher than
the kinetic energy of a typical CME, which is about $10^{29}-10^{30}$ erg
\citep[e.g.,][]{Vourlidas_etal_2010}, even considering the uncertainty in the decay index when
we extrapolate the initial magnetic energy.

\begin{figure*}[p]
  \centering
  \includegraphics[width=0.9\hsize]{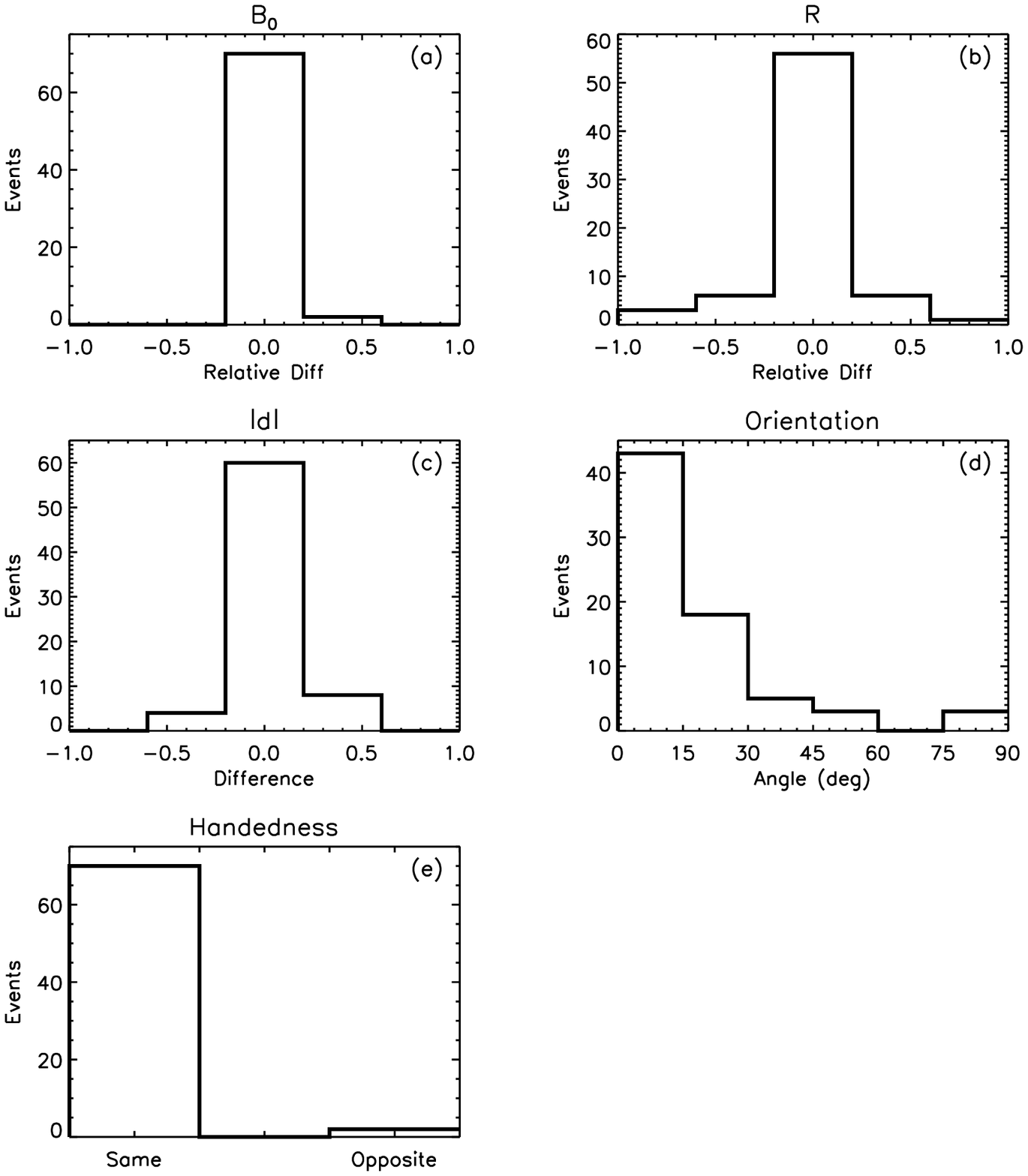}
  \caption{Histograms showing the differences of the values of fitting parameters between our
   model with and without velocity measurements taken into account. The arrangement is as the same as that in Fig.~\ref{fg_errlepping}.}\label{fg_errbv}
\end{figure*}

\citet{Nakwacki_etal_2008} compared the static and expansion models and found that the modeled
values of the MC parameters are not changed too much. Here we consider not only the expansion
but also other types of motion. Will the values of parameters change more significantly?
By comparing the values of the fitting parameters obtained from the velocity-modified and
non-velocity models as shown in Figure~\ref{fg_errbv}, we find that the values are more or less
changed, but not too significantly except the orientation.
The magnetic field strength is almost unaffected. For the radius, about 13\% (9 out of 72)
of the events get a smaller value after velocity is taken into account, and about 10\% of
the events get a larger value. For the closest approach, the velocity-modified model believes
that the observer should be slightly farther away from the MC axis for 8 events and closer for
only 4 events. The largest difference appears in the orientation. There are about 40\%
of the events, for which the orientation is changed by more than $15^\circ$, and
particularly, there are 6 events with the difference in orientation larger than $45^\circ$.
Besides, for two cases, the handedness is also changed.

\section{Statistical properties of plasma motion}\label{sec_vel}
\subsection{Linear propagating motion}\label{sec_prop}
One important implication to consider the velocity in fitting procedure is to reveal the properties
of the plasma motion of MCs. First, we investigate the axial component, $v_{axis}$, and perpendicular
component, $v_{perp}$, of the linear propagating motion  of an MC. The former is a velocity along the MC's axis,
i.e., in the $z$-direction of the MC frame, and the latter is a velocity perpendicular to 
the radial direction, i.e., the
Sun-spacecraft line, $v_{perp}=\sqrt{v_Y^2+v_Z^2}$. 

\begin{figure*}[tb]
  \centering
  \includegraphics[width=0.95\hsize]{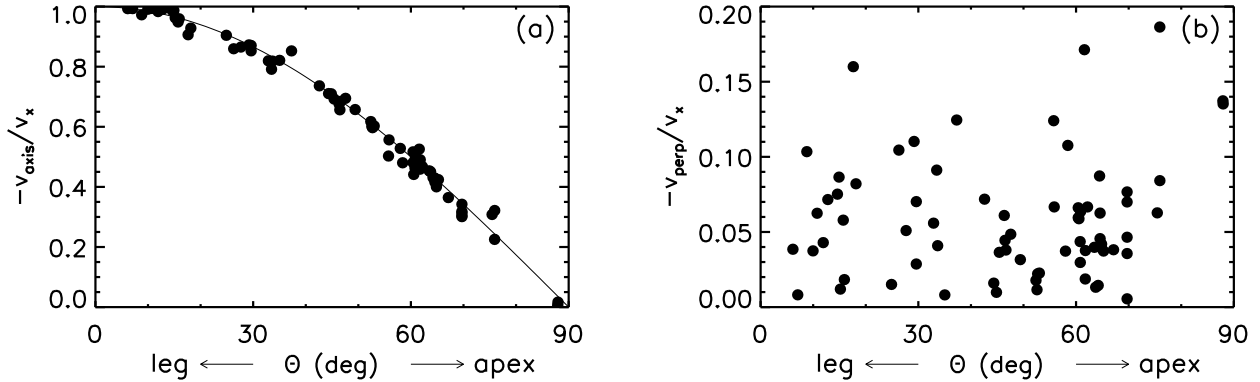}
  \caption{(a) The axial velocity and (b) the perpendicular velocity as a function of the angle between the MC's axis and the Sun-spacecraft line. Both the velocities are normalized by the value of radial propagation velocity. 
A small value of $\Theta$ means an encounter of the leg of an MC and a large value of $\Theta$ means an encounter of the apex of an MC.}\label{fg_vaxis}
\end{figure*}

The ratio of the axial velocity to the radial velocity, $-v_{axis}/v_X$,
as a function of the orientation of the MC's axis is shown in Figure~\ref{fg_vaxis}a.
A very nice correlation between them could be found.
When the axis is almost aligned with Sun-spacecraft line, the absolute value of
$v_{axis}$ is almost equal to that of $v_X$, and when the axis becomes more and more perpendicular to the Sun-spacecraft
line, $v_{axis}$ approaches zero. It well follows the cosine function as indicated
by the solid line. The small deviation away from the cosine function is no more than $0.06v_X$, which 
corresponds to a very small speed on the order of 10 km s$^{-1}$. This result suggests that the 
apparent axial velocity is mostly a consequence of the propagation of the MC though an insignificantly pure-axial 
flow might exist inside an MC.

\begin{figure*}[tb]
  \centering
  \includegraphics[width=0.95\hsize]{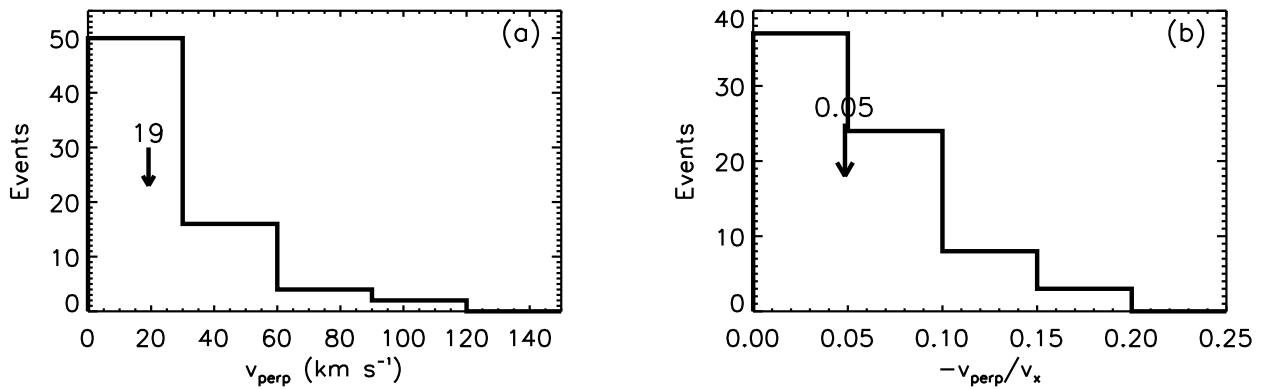}
  \caption{Distributions of (a) the propagation velocity perpendicular to the radial direction, i.e., the Sun-spacecraft line and (b) its value relative to the radial propagation velocity. Median value is indicated by the arrow in each panel.}\label{fg_vperp}
\end{figure*}

The distribution of $v_{perp}$ is given in Figure~\ref{fg_vperp}.
Panel (a) shows that except 6 events, all the other events have a perpendicular velocity less
than 60 km s$^{-1}$, and Panel (b) suggests that in 61 (about 85\% of) events the perpendicular
velocity is no
more than 10\% of radial velocity. A noteworthy thing is that some events have a significantly
perpendicular velocity with respect to the radial velocity. The MC shown in Figure~\ref{fg_casebv}b
is an example, which occurred on 1998 September 25. Our model infers that the orientation of the MC
axis is $\theta=60^\circ$ and $\phi=196^\circ$,
and the plasma motion of the MC is the combination of the linear motion at $-624$,
95 and 50 km s$^{-1}$ in $\mathbf{\hat{X}}$, $\mathbf{\hat{Y}}$ and $\mathbf{\hat{Z}}$ directions and
the expansion and poloidal motion at 90 and $-23$ km s$^{-1}$, respectively.

Considering the expansion of the whole looped structure as shown in Figure~\ref{fg_coord}b and 
that there is no significantly pure-axial flow inside an MC,
one may find that a significantly perpendicular velocity
may be present if not the front part (or the apex) but the flank (or the leg) of the MC was detected locally. Which part
of a looped MC is detected by spacecraft could be roughly inferred from the orientation
of the MC axis. We think that the apex of an MC is encountered if the inferred
axis of the MC is almost perpendicular to the Sun-spacecraft line, and the leg is encountered
if the axis is almost parallel to the Sun-spacecraft line.
Thus, if the expansion of the whole looped structure was the reason of the presence of the
perpendicular velocity, we may expect a very small perpendicular velocity for those 
apex-encountered events. Figure~\ref{fg_vaxis}b shows the scatter plot of
$-v_{perp}/v_X$ versus $\Theta$. Obviously, there is no dependence of the perpendicular velocity
on the axis orientation. For those events with $\Theta$ close to $90^\circ$, the perpendicular velocity
could be larger than 10\% of the radial velocity. Thus, there must be some other reasons for the
significantly perpendicular motion.

We think that the presence of a significantly perpendicular velocity might be an in-situ evidence of deflected
propagation of a CME in interplanetary space \citep[e.g.,][]{Wang_etal_2004b, Wang_etal_2014,
Lugaz_2010, Isavnin_etal_2014}. The positive value of $v_Y$ of the 1998 September 25 event
suggests an eastward deflection in the ecliptic plane. This is in agreement with the proposed
picture by \citet{Wang_etal_2004b} that a CME faster than background solar wind will be
deflected toward the east. 
If a CME propagated in interplanetary space with a perpendicular
velocity at a tenth of the radial velocity, i.e., $-\frac{v_{perp}}{v_X}=\frac{1}{10}$,
the deflection angle is then given by
\begin{eqnarray}
\int_{t_0}^{t_1} \frac{v_{perp}}{L} dt=\int_{L_0}^{L_1}-\frac{v_{perp}}{v_XL} dL=\int_{L_0}^{L_1}\frac{1}{10L}dL
\end{eqnarray}
Assuming $L_1=1$ AU and $L_0$ less than 5 solar radii, we can estimate that
the deflection angle is more than $20^\circ$.
This is quite consistent with our recent case study of a CME, which
was found to be deflected by more than $20^\circ$ on its way from the corona to 1 AU
\citep{Wang_etal_2014}. Alternatively, the perpendicular velocity might also be the result of
the rotation of the whole structure of a CME with respect to the radial direction 
in interplanetary space, just as the rotation in
the middle and outer corona \citep[e.g.,][]{Yurchyshyn_etal_2009, Vourlidas_etal_2011,
Isavnin_etal_2014}.

\subsection{Expanding motion}\label{sec_exp}

\begin{figure*}[tb]
  \centering
  \includegraphics[width=0.95\hsize]{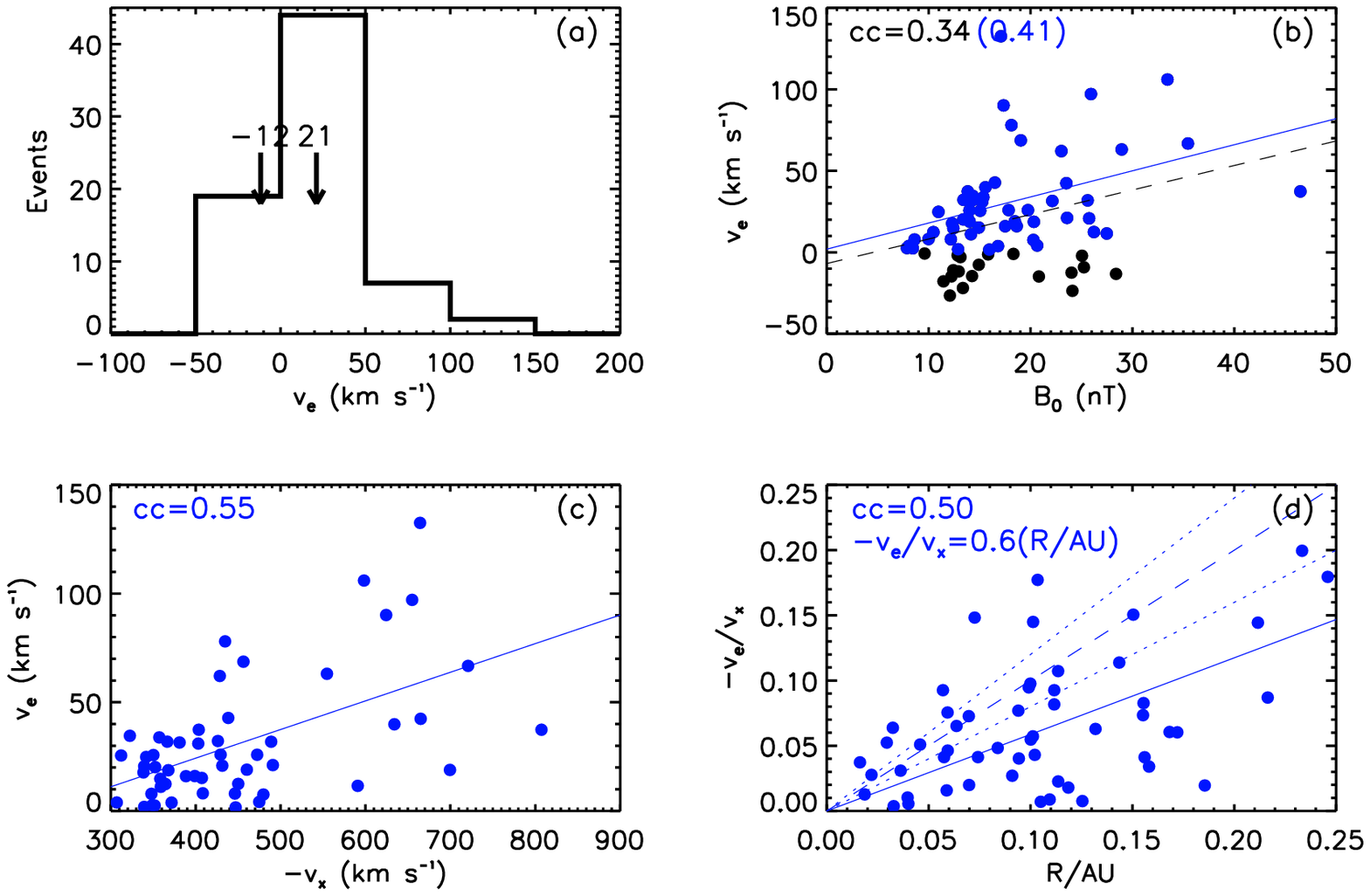}
  \caption{(a) Distribution of the expansion speed. The two arrows indicate the mean values for negative and positive 
expansion speeds, respectively. (b) Scatter plot showing the correlation between the expansion speed and magnetic field strength. Blue dots indicate the expansion events and black dots the contraction events. The black dashed and blue solid lines are the linear fitting to the all data points and the blue data points, respectively. (c) Scatter plot showing the correlation between the expansion speed and the radial propagation speed. The line gives the linear fitting result. (d) Scatter plot showing the correlation between the expansion speed in units of the radial propagation speed and the MC's radius in units of AU. The solid line gives the fitting result by a function of $y=ax$ as marked in the upper-left corner of the panel. The dashed line indicates the self-similar expansion, i.e., $R$ evolving proportionally to $L$, and the two dotted lines give the 20\% uncertainty. For the panel (c) and (d), only expansion events are included. }\label{fg_vexp}
\end{figure*}

\begin{figure}[tbh]
  \centering
  \includegraphics[width=0.95\hsize]{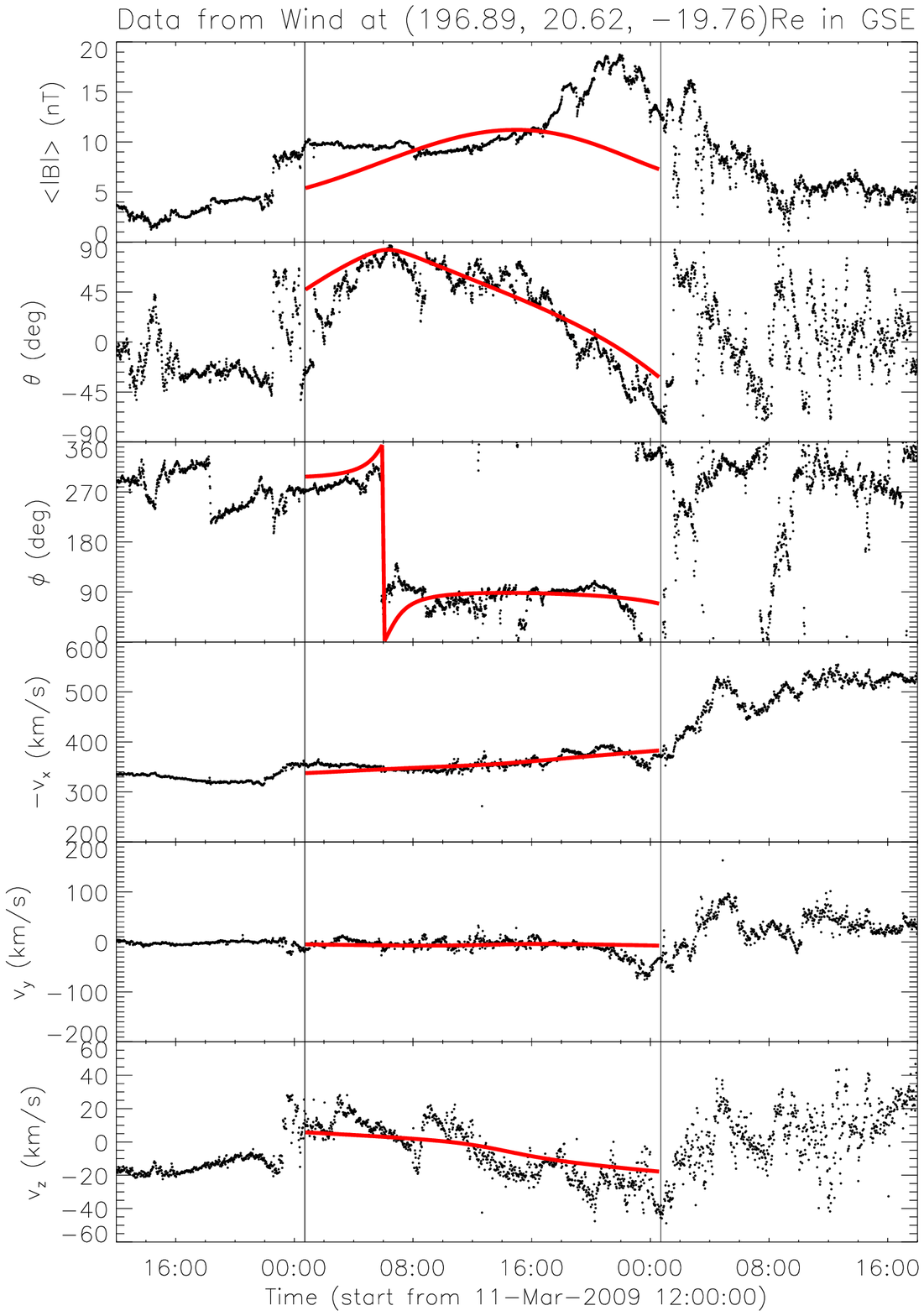}
  \caption{An example of contraction events. The arrangement is as the same as that in Fig.~\ref{fg_casebv}.}\label{fg_contraction}
\end{figure}

The distribution of the expansion speed is shown in Figure~\ref{fg_vexp}a. It
is noteworthy that a significant fraction (about 26\%) of the events experienced a contraction
process with a median value of about 12 km s$^{-1}$. We check the large contraction events having
$v_{e}<-20$ km s$^{-1}$, and find unsurprisingly that they were all caused by the overtaking 
of faster solar wind stream, as illustrated by the example
in Figure~\ref{fg_contraction}. In that event, the solar wind speed after the trailing edge
of the MC is much larger than that before the leading edge, which cause the magnetic field
strength increase with time, and reach the maximum near the rear boundary. Under this circumstance,
the MC cannot expand freely, but will be compressed by ambient solar wind. Our model suggests
that the contraction speed of the MC is about 27 km s$^{-1}$.

\begin{figure}[tb]
  \centering
  \includegraphics[width=0.95\hsize]{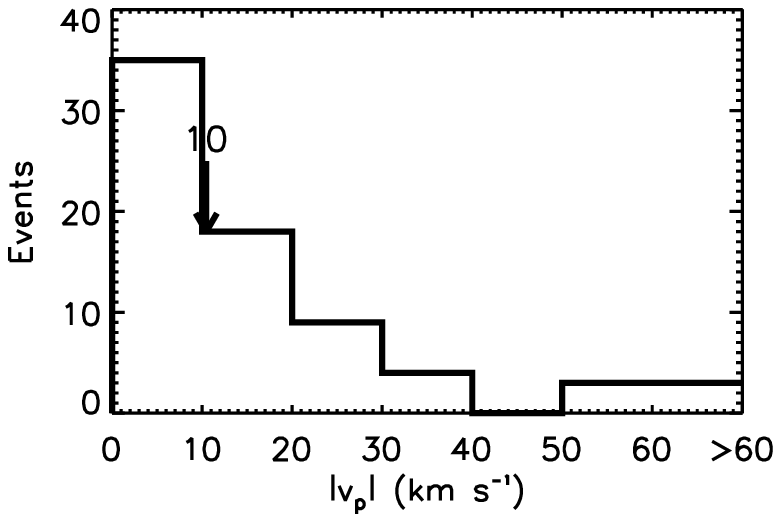}
  \caption{Distribution of the poloidal speed.}\label{fg_vcir}
\end{figure}

For the rest of the events suggested to be expanding at 1 AU, the median speed is
about 21 km s$^{-1}$. The
expansion substantially depends on the balance between the internal and external forces.
The internal force is partially characterized by magnetic field strength. Figure~\ref{fg_vexp}b shows
the correlation between $v_{e}$ and $B_0$. The black dashed line is the linear fitting to all the
data points and the blue solid line is the linear fitting to the data points of $v_{e}>0$ (blue dots).
A weak
correlation could be found, but it is not significantly different between all data points and data
points of $v_{e}>0$. The correlation coefficient is not so high, because the external condition is not considered.
With the increasing distance away from the Sun, both the magnetic and thermal pressures in the solar wind
usually decrease. It will consequently cause the unbalance between the internal and external forces, which
drives a CME/MC expanding. Thus we may expect that the expansion speed should be correlated with
propagation speed. As shown in Figure~\ref{fg_vexp}c, there does exist a stronger correlation between
$v_{e}$ and $v_X$ for expansion events. The correlation coefficient is about 0.55. It is consistent
with the previous result by, e.g., \citet{Demoulin_Dasso_2009} that rapid decrease of the total solar
wind pressure with heliocentric distance is the main driver of the MC expansion. Similar results could also
be found for the events in the inner and outer heliosphere \citep{Gulisano_etal_2010, Gulisano_etal_2012}.

On the other hand, the MC expansion could be classified as self-similar expansion, overexpansion and
underexpansion. Note, the term `self-similar' here only refers to that the radius, $R$, of an MC
evolves proportionally to the distance $L$, which is a subset of that defined in Sec.2.1 where
`self-similar' means that not only the size but also the internal plasma parameters of the MC evolve 
self-similarly.
Imaging data have suggested that most CMEs undergo a self-similar expansion in the outer
corona \citep[e.g.,][]{Schwenn_etal_2005}, but whether or not they maintain the self-similar expansion in
interplanetary space? To measure the expansion rate of MCs, \citet{Gulisano_etal_2010} used a dimensionless quantity
$\zeta=\frac{\Delta u_X}{\Delta t}\frac{L}{v_X^2}$, in which $\Delta u_X$ is the difference of the measured
solar wind speed along the Sun-spacecraft line between the front and rear boundaries of an MC.
They found that the value of $\zeta$ is about 0.7 after analyzing all the MCs observed by Helios spacecraft.
Considering $\Delta u_X$ is a proxy of the expansion speed of an MC
and $v_X\Delta t$ approximates the size of the MC, we may infer that $\zeta\approx\frac{v_{e}}{-v_X}\frac{L}{R}$,
and $\zeta<1$ means an underexpansion. Thus, the parameter, $\zeta$,
has the same physical meaning of the power index, $n$, appearing in the power law of the
heliocentric distance dependence of the CME/MC size \citep[e.g.,][]{Bothmer_Schwenn_1998, 
Demoulin_etal_2008, Savani_etal_2009}, namely $S=cL^n$, where $S$ is the MC size and $c$ is a constant.
\citet{Bothmer_Schwenn_1998} found $n\approx0.78$ and \citet{Demoulin_etal_2008} gave $n\approx0.8$.
Here we use model derived parameters to investigate the expansion rate again. Since all the MCs
in this study located at 1 AU, we have $L=1$ AU.
Figure~\ref{fg_vexp}d shows the plot of $-v_{e}/v_X$ versus $R/$AU.
The self-similar expansion is given by the dashed line. The data points above the line suggest an overexpansion,
and those below the line a underexpansion. By considering a 20\% uncertainty as indicated by the two
dotted lines, we find that only 21\% of the events underwent a nearly self-similar expansion, and 62\%/17\%
of the events have and expansion rate lower/larger than 0.8/1.2. By using a function of $\frac{v_e}{-v_X}=\zeta \frac{R}{1\rm AU}$
to fit the all these data points as indicated by the solid line, we get
$\zeta\approx0.6$ on average, generally consistent with that obtained by, e.g.,
\citet{Bothmer_Schwenn_1998}, \citet{Demoulin_etal_2008} and \citet{Gulisano_etal_2010}, if the uncertainty is considered.
A possible reason of why the expansion rate is significantly below unity is that the MCs are
perturbed by ambient solar wind and/or other transients as shown in observational statistics \citep{Gulisano_etal_2010},
which may cause the external pressure surrounding the MC decreasing with distance more slowly than usual. The numerical simulations 
by, e.g., \citet{Xiong_etal_2006b, Xiong_etal_2007}, \citet{Lugaz_etal_2013} 
also suggested that CME-CME interaction may affect the expansion rate of the preceding CME.

It should be noted that in our model the expansion speed is derived based on the measured
solar wind velocity along the Sun-spacecraft line and therefore the expansion speed here is more likely to
reflect the radial expansion rather than lateral expansion.
Lots of studies have shown that the cross-section of a CME will more or less distorted from the circular shape to a
pancake shape \citep[e.g.,][]{Riley_etal_2003, Riley_Crooker_2004, Owens_etal_2006}, suggesting that the lateral expansion is probably faster than the
radial expansion. Thus, the lateral expansion may probably still be self-similar as that in the outer corona
though the radial expansion is not.

One may notice that our model implies that $R\propto L$, meaning a self-similar expansion (see Sec.\ref{sec_para}).
It is inconsistent with the result here. This inconsistency may come from various sources, e.g., the 
assumption of the magnetic flux and/or helicity conservation, the assumption of $L\propto l$, 
and the assumption of the uniformly straight cylindrical geometry. It will consequently
affect the values of some derived parameters, e.g., $\Phi_\varphi$, $H_m$ and $E_m$. Although such an
inconsistency exists, we think that the values of these derived parameters still could be treated as a
first-order approximation.

\subsection{Poloidal motion}\label{sec_cir}

\begin{figure*}[tb]
  \centering
  \includegraphics[width=0.39\hsize]{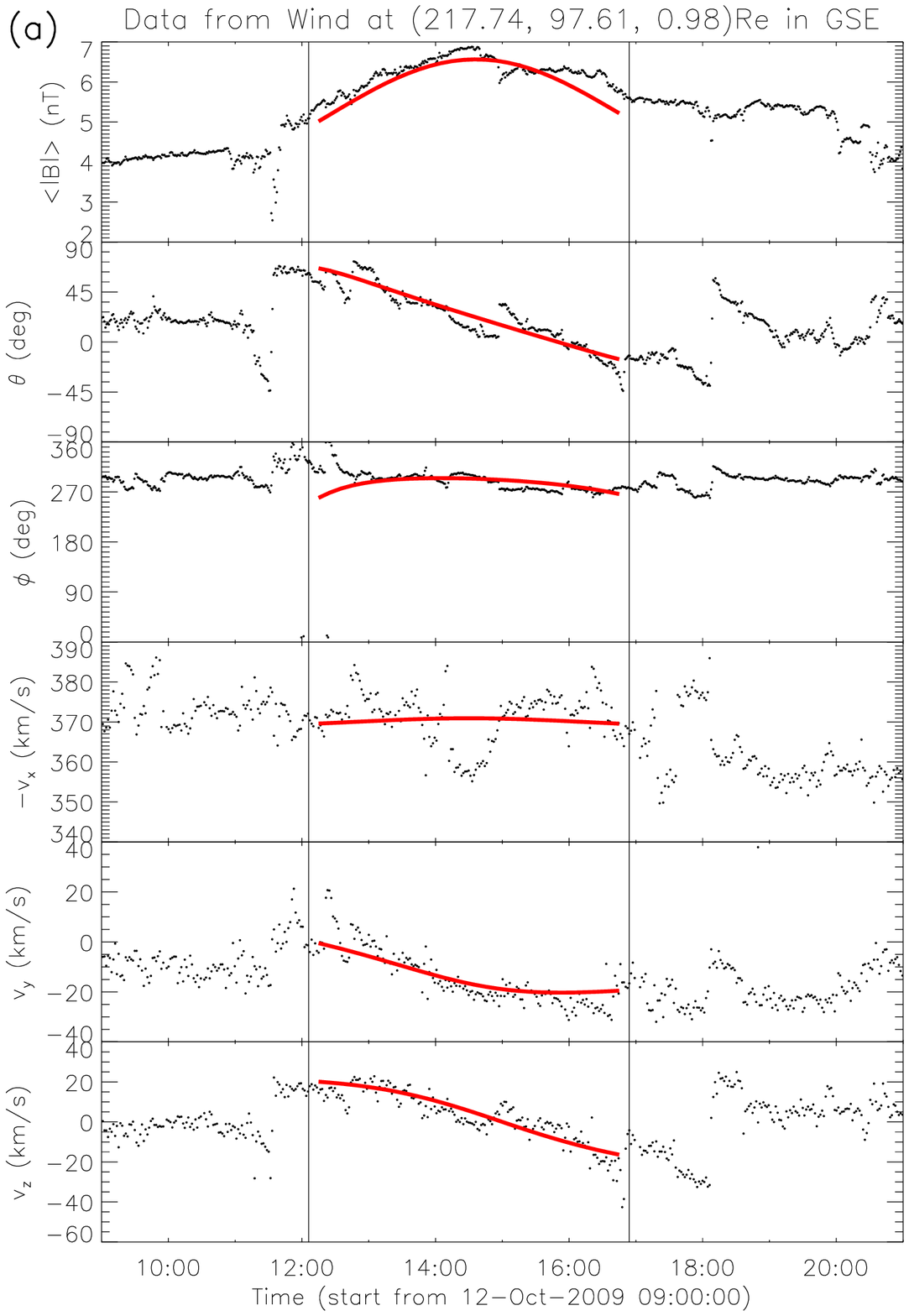}
  \includegraphics[width=0.60\hsize]{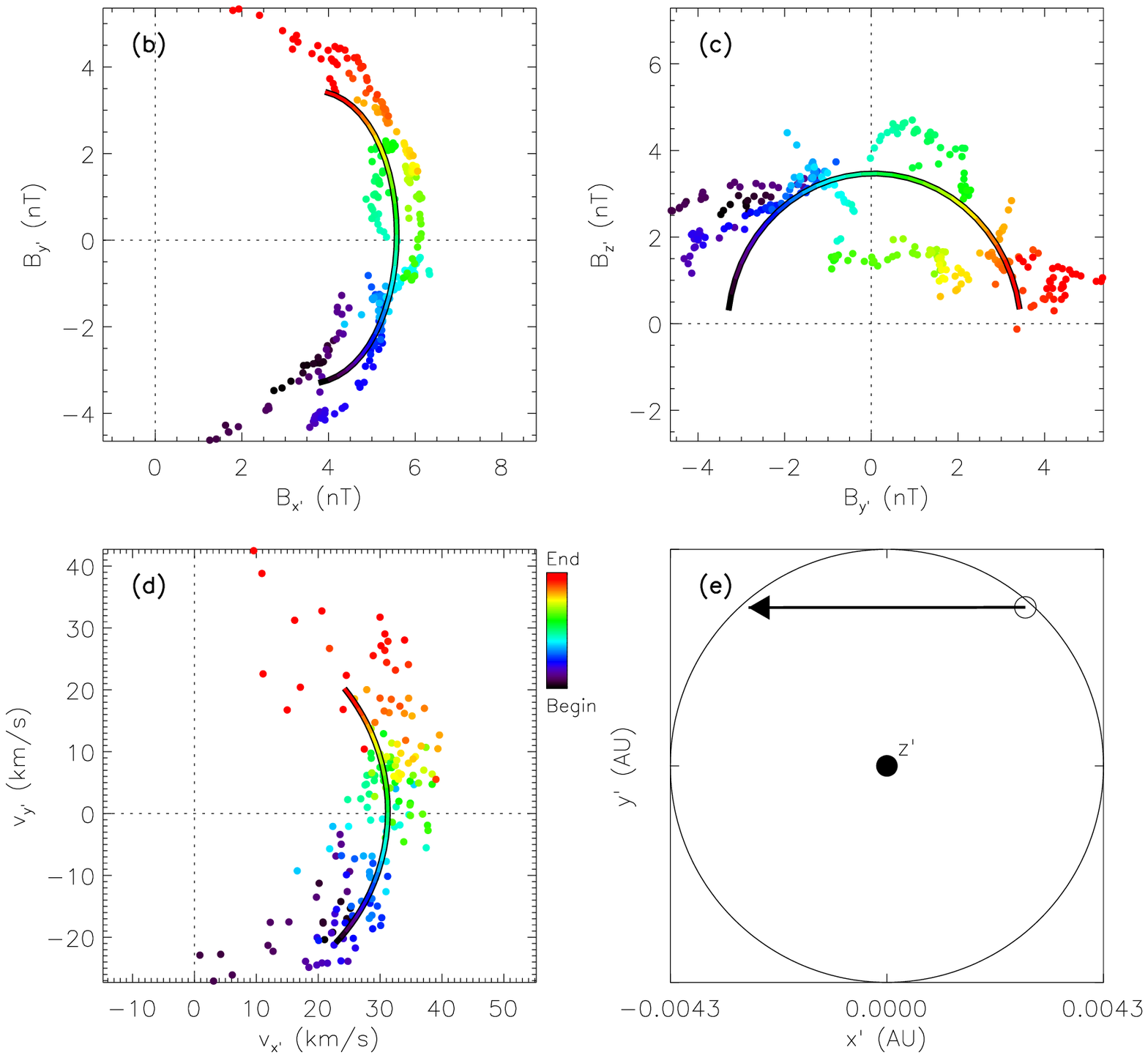}
  \caption{An MC observed on October 12, 2009. (a) The fitting results of the velocity-modified model. The arrangement of the panels is as the same
  as that in Fig.~\ref{fg_casebv}. (b) and (c) The magnetic field components in the $(x',y')$ and $(y',z')$ plane of the MC frame, respectively. The data points are
  color-coded, indicating the time from the beginning (black, corresponding to the first vertical line in the plot (a)) of the MC to the end (red, corresponding to the second vertical line in the plot (b)). The color-coded lines are the fitting results. (d) The velocity in the $(x',y')$ plane. (e) The derived cross-section of the MC (the circle) and the observational path (the arrow). See main text for more details.}\label{fg_vcircase1}
\end{figure*}

\begin{figure*}[tb]
  \centering
  \includegraphics[width=0.395\hsize]{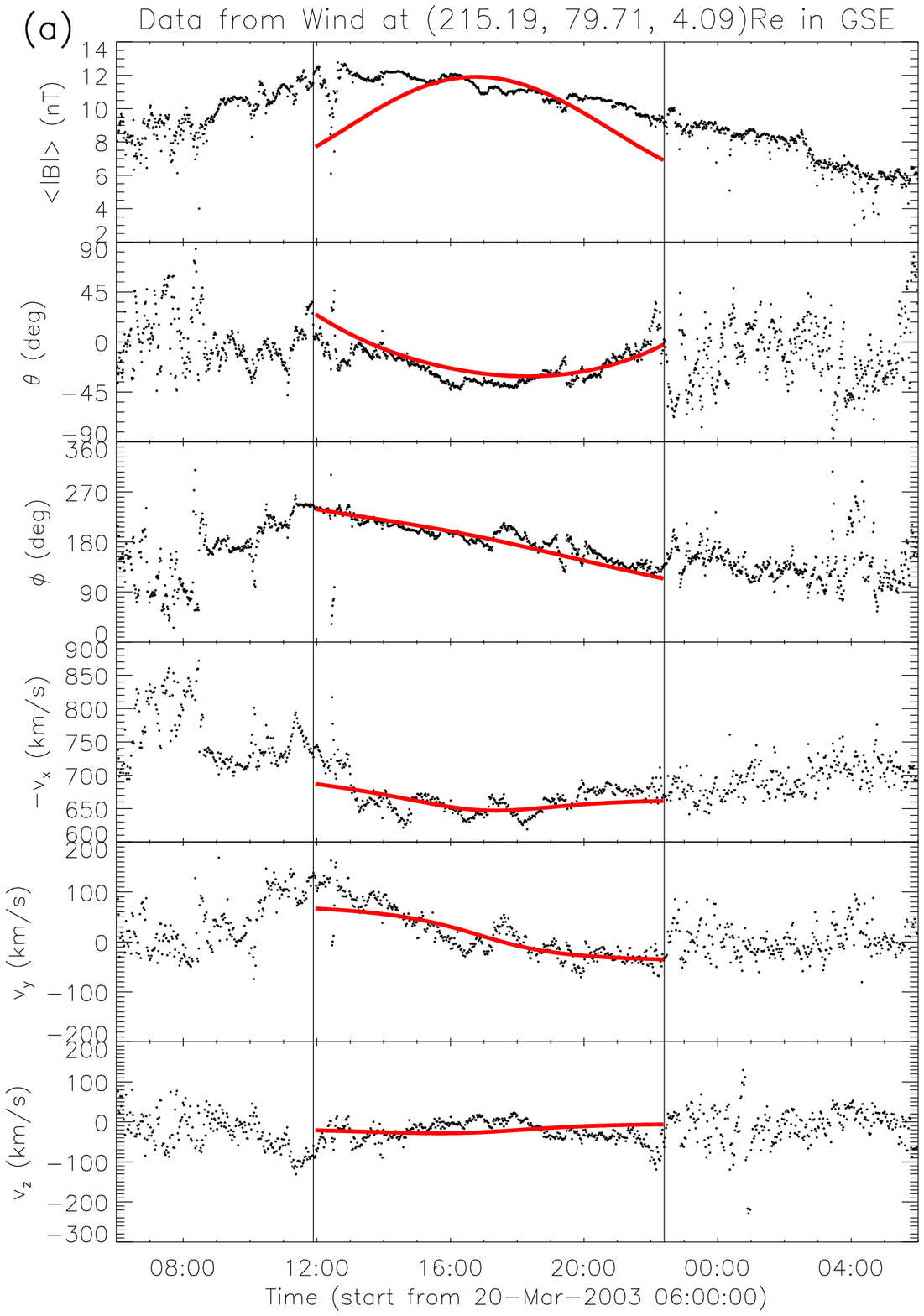}
  \includegraphics[width=0.595\hsize]{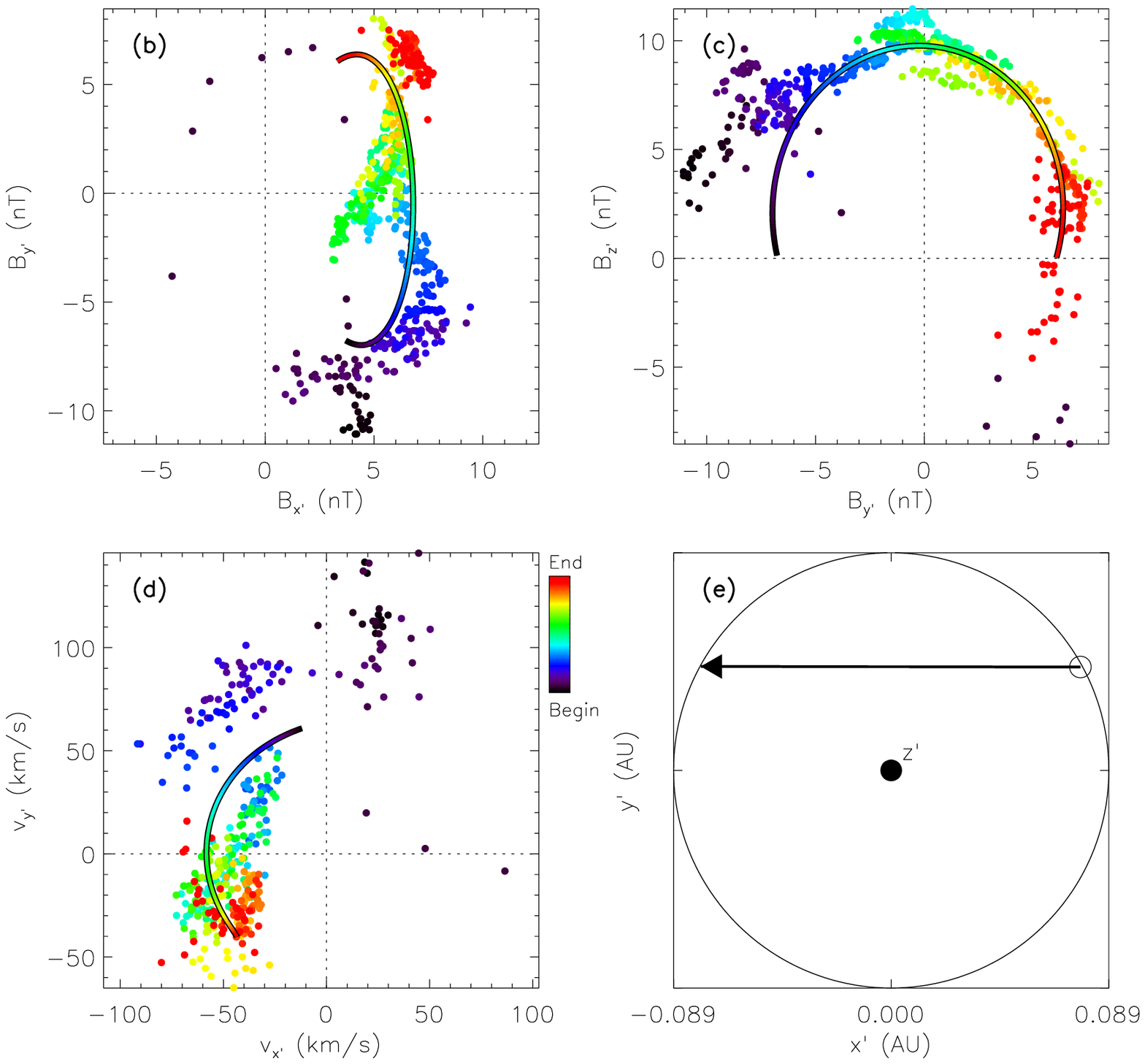}
  \caption{Similar to Fig.~\ref{fg_vcircase1} but for an MC observed on March 20, 2003.}\label{fg_vcircase2}
\end{figure*}

The distribution of poloidal speed is shown in Figure~\ref{fg_vcir}. The median value is about 10 km s$^{-1}$.
By comparing the poloidal speed with other parameters, we cannot find any significant correlation
among them (all the correlation coefficients are no more than 0.40).
Obviously, the poloidal speed, if any, is less significant than the expansion speed on average.
This fact may cause the observational signature of the poloidal motion in an MC unclear.
To most clearly show the poloidal motion, we choose the events without significant expansion and
convert the velocity into the Cartesian frame, $(x',y',z')$, of the MC (see Sec.\ref{sec_model1} for
the definition of the coordinates). A nice example is shown in Figure~\ref{fg_vcircase1}, which was observed
on 2009 October 12. The two vertical lines mark the beginning and the end of the MC, and the fitting results are given by the
red curves in Figure~\ref{fg_vcircase1}a. According to the modeled parameters,
the path of Wind spacecraft in the MC frame is shown in Figure~\ref{fg_vcircase1}e.
Along the path from the beginning to the end of the MC, the rotation of
the magnetic field vector is well presented in both the $(x',y')$ and $(y',z')$ planes
(see Fig.\ref{fg_vcircase1}b and \ref{fg_vcircase1}c). The color-coded dots are observations and the color-coded
lines are fitting curves. More interestingly, the poloidal motion is evident in the MC frame (Fig.\ref{fg_vcircase1}d).
In this event, the expansion speed is almost zero and the poloidal speed is about $-31$ km s$^{-1}$ at time of $t_c$.
Figure~\ref{fg_vcircase2} shows another event on 2003 March 20, in which the modeled poloidal speed is $58$ km s$^{-1}$,
more significant than the modeled expansion speed of about $19$ km s$^{-1}$, and therefore the poloidal
motion can be also recognized in observations.

\begin{figure*}[tbh]
  \centering
  \includegraphics[width=0.49\hsize]{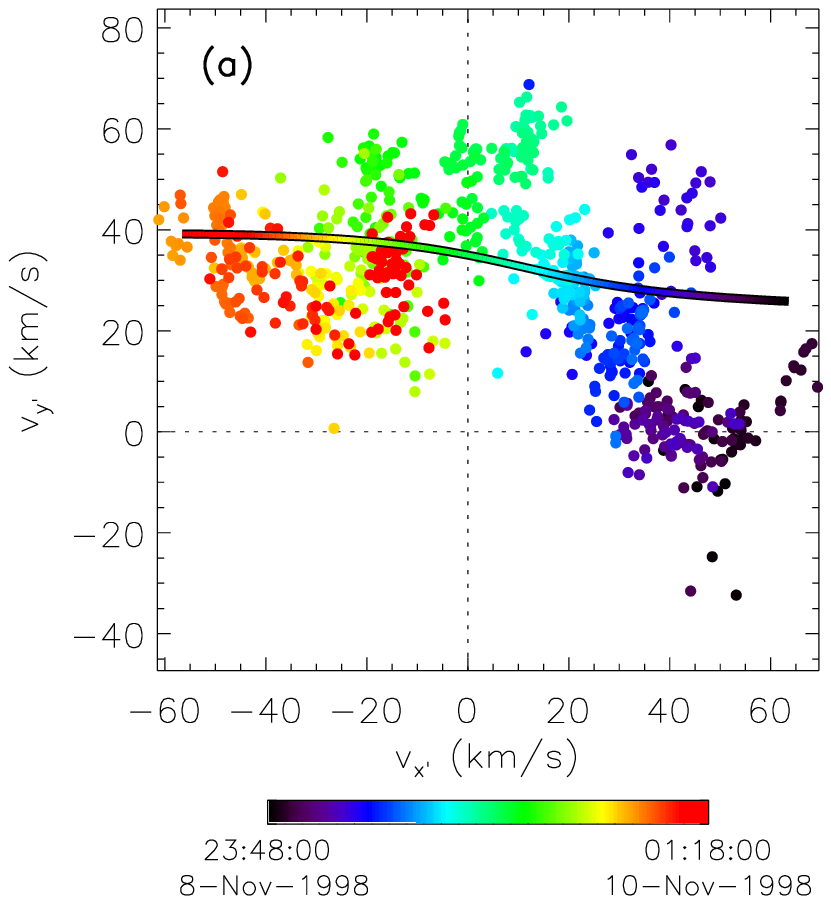}
  \includegraphics[width=0.5\hsize]{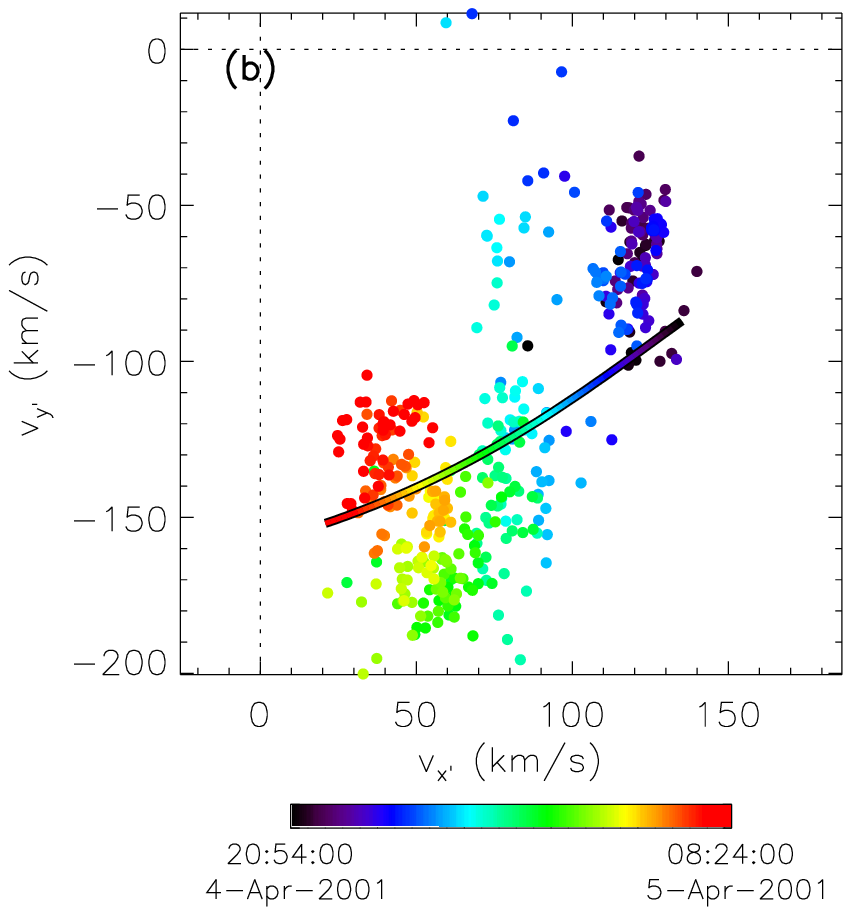}
  \caption{The scatter plots of the velocity components in the $(x',y')$ plane for (a) the MC observed during November 8--10, 1998
  and (b) the MC during April 4--5, 2001. The arrangement is as the same as that in Fig.~\ref{fg_vcircase1}d.}\label{fg_vscatter}
\end{figure*}

Based on the model, we could expect that the data points of velocity will form
an arc which is symmetric about the axis of $v_{y'}=0$ if the
poloidal motion was significant and dominant in an MC. The above two cases just show
the pattern. However, the arc's symmetrical axis will change to $v_{x'}=0$ if
the expansion was significant and dominant. Figure~\ref{fg_vscatter}a roughly shows the case,
which was observed during 1998 November 8 -- 10 and the values of $v_{e}$ and $v_{p}$
are 69 and $-8$ km s$^{-1}$, respectively. If both expansion and poloidal motion are
significant, the `symmetric' axis will rotate, and the velocity data points may
deviate slightly from a symmetric distribution. The 2001 April 4 -- 5 event shown in
Figure~\ref{fg_vscatter}b is an example to show the change.

\begin{figure*}[tbh]
  \centering
  \includegraphics[width=0.24\hsize]{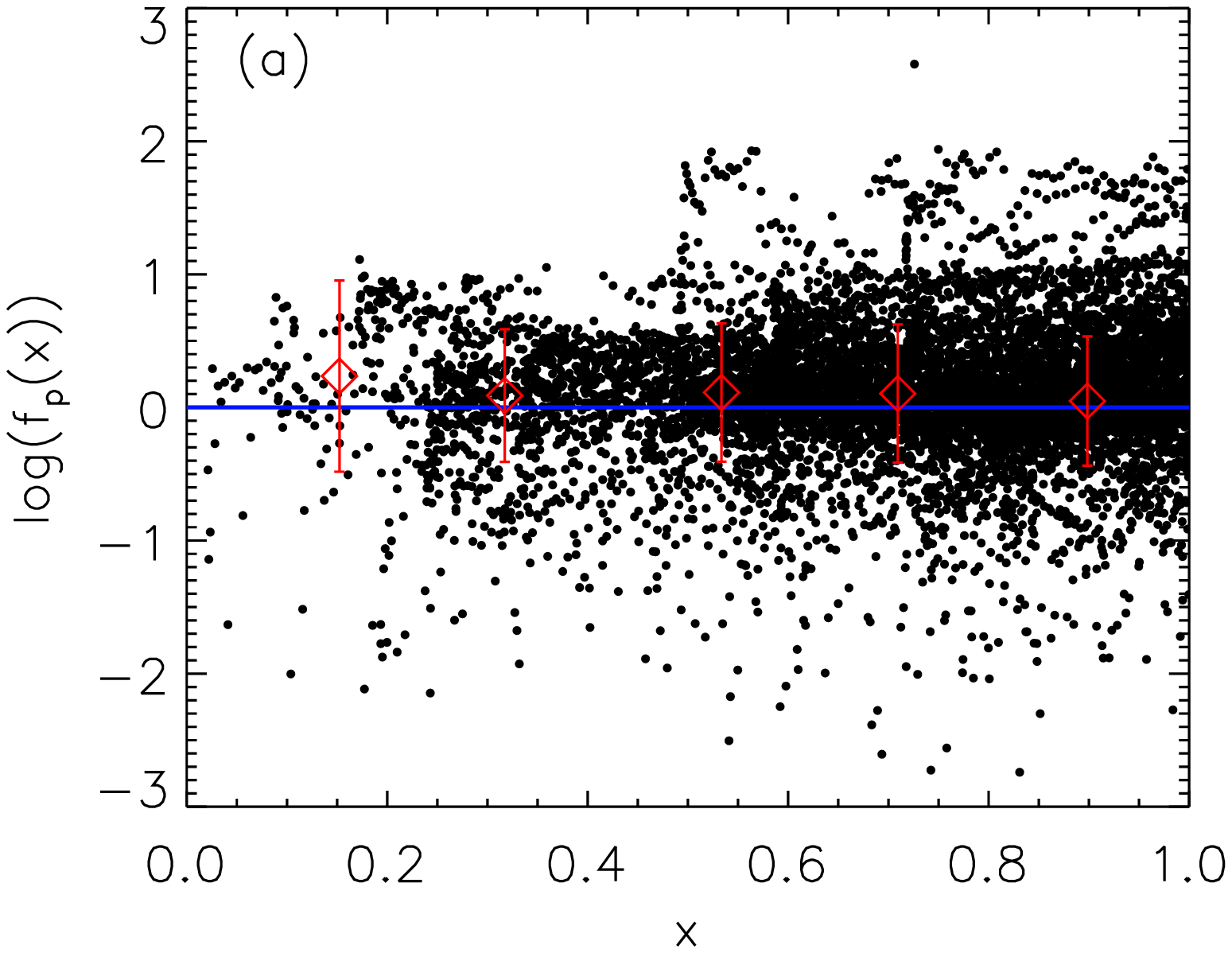}
  \includegraphics[width=0.24\hsize]{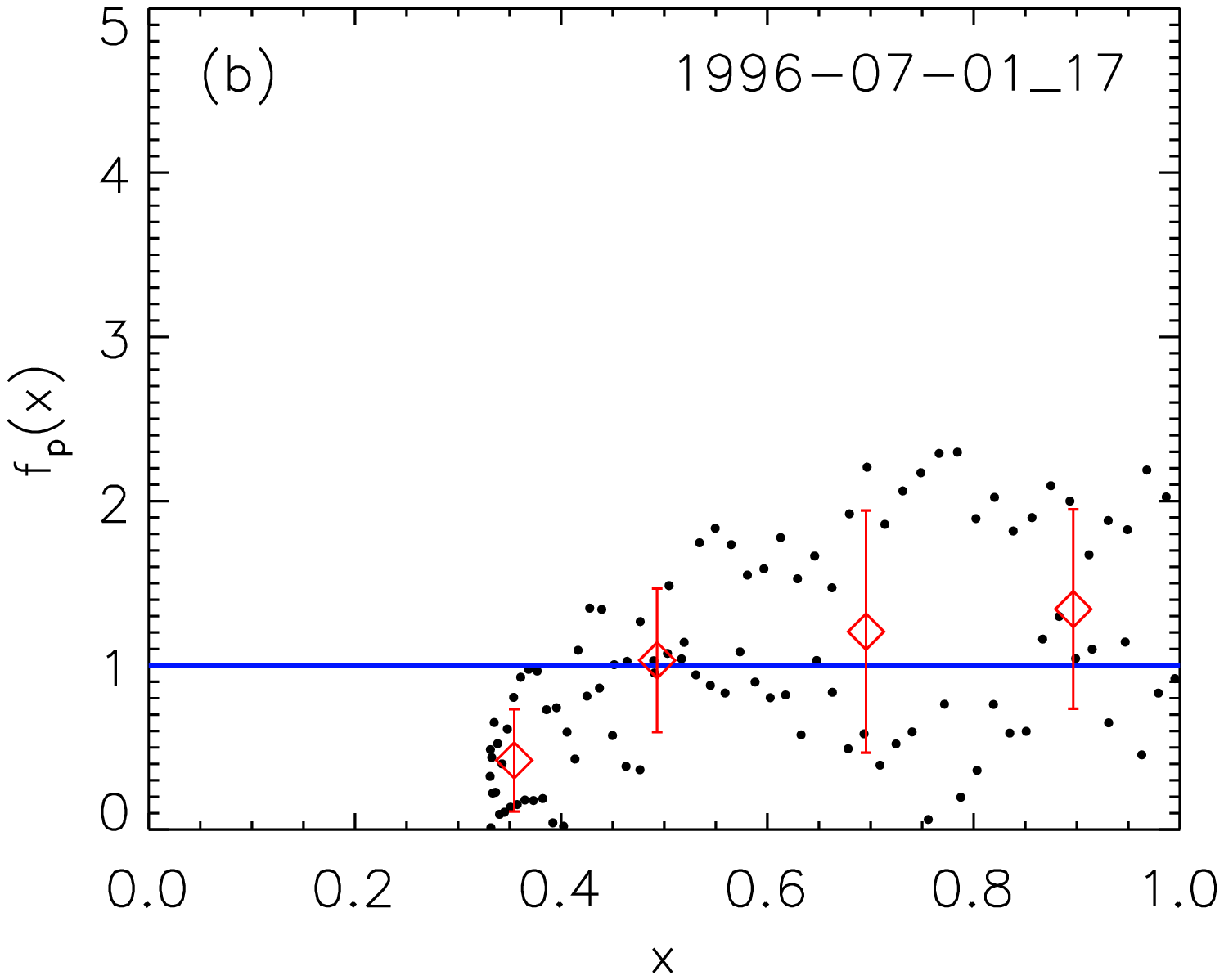}
  \includegraphics[width=0.24\hsize]{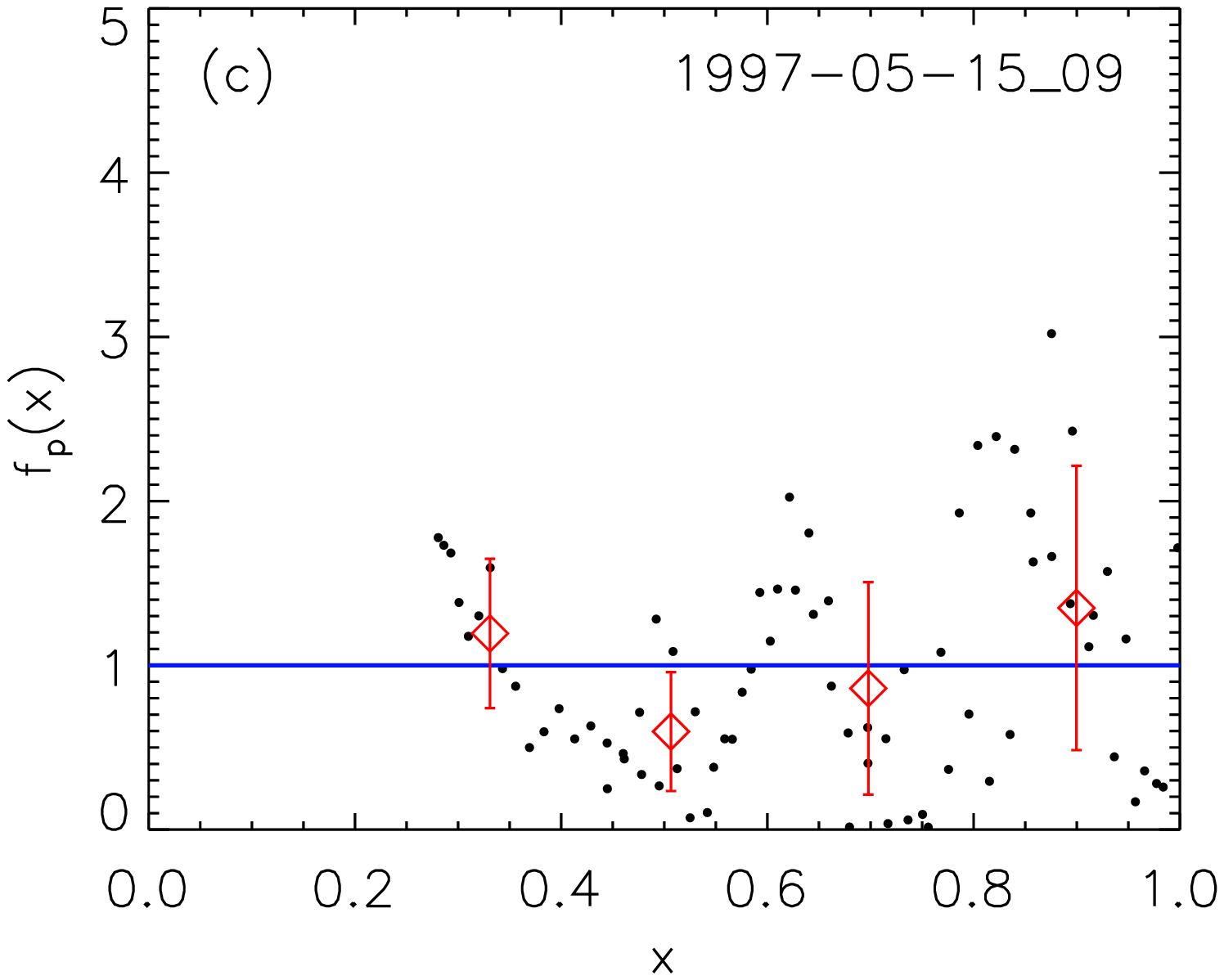}
  \includegraphics[width=0.24\hsize]{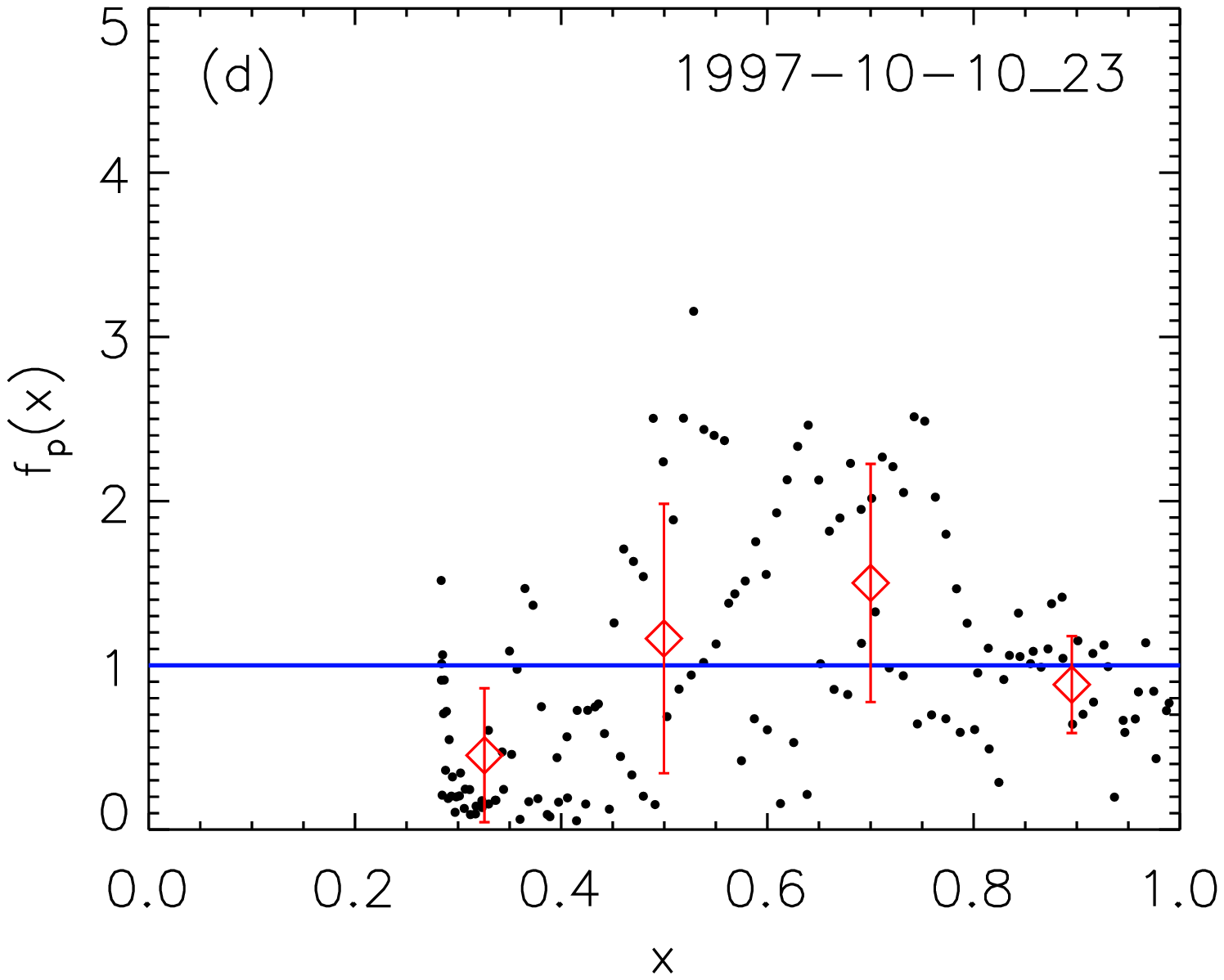}\\
  \includegraphics[width=0.24\hsize]{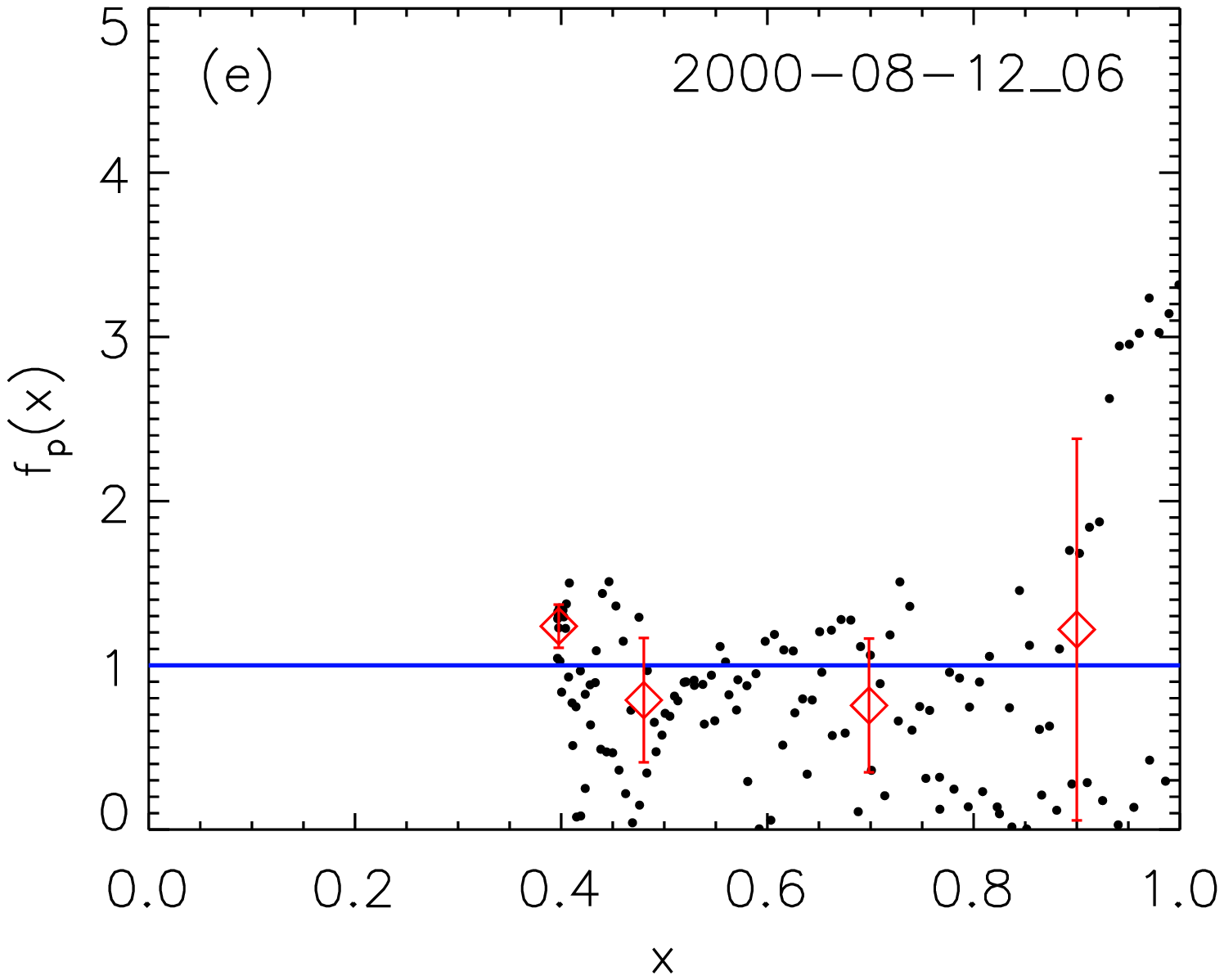}
  \includegraphics[width=0.24\hsize]{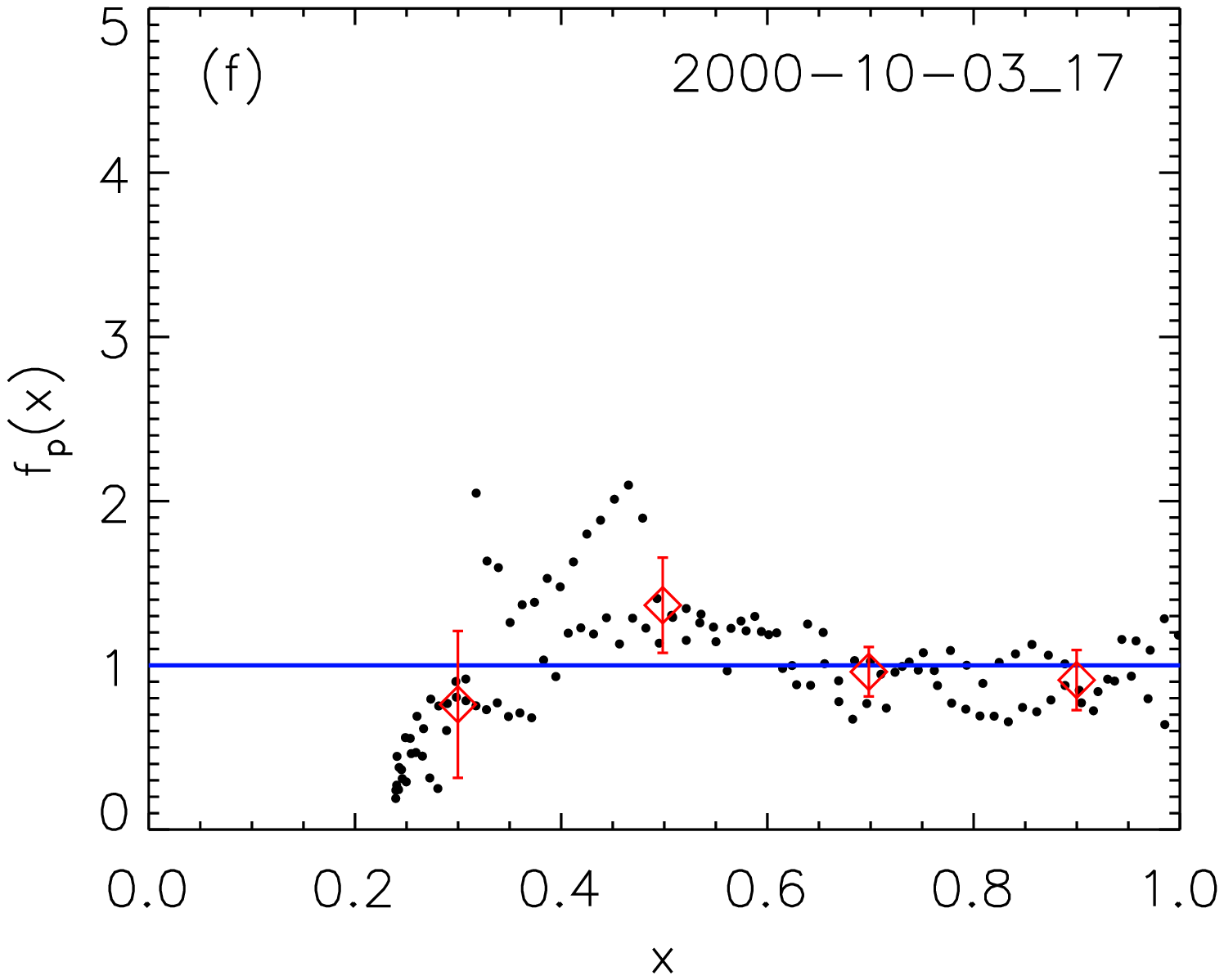}
  \includegraphics[width=0.24\hsize]{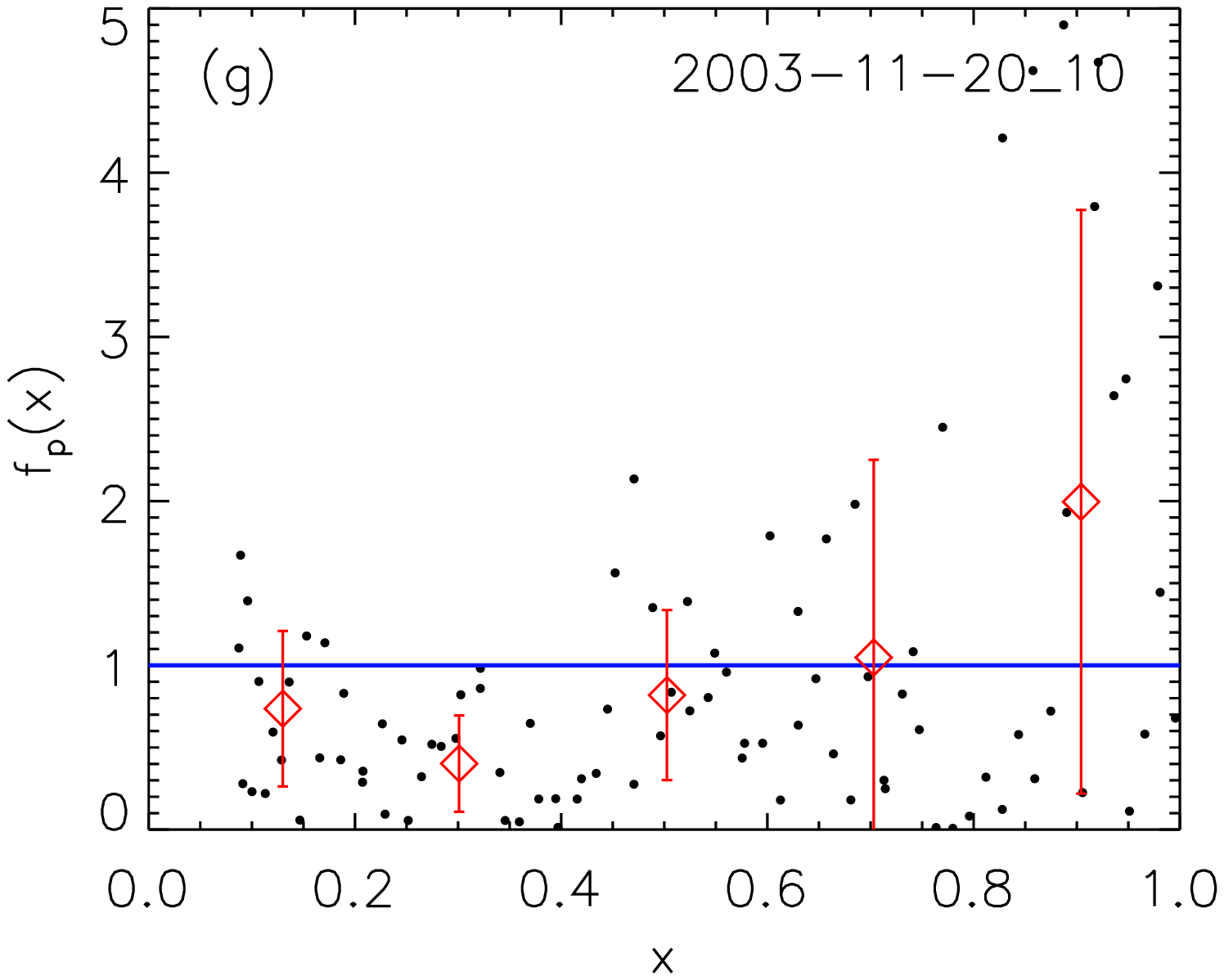}
  \includegraphics[width=0.24\hsize]{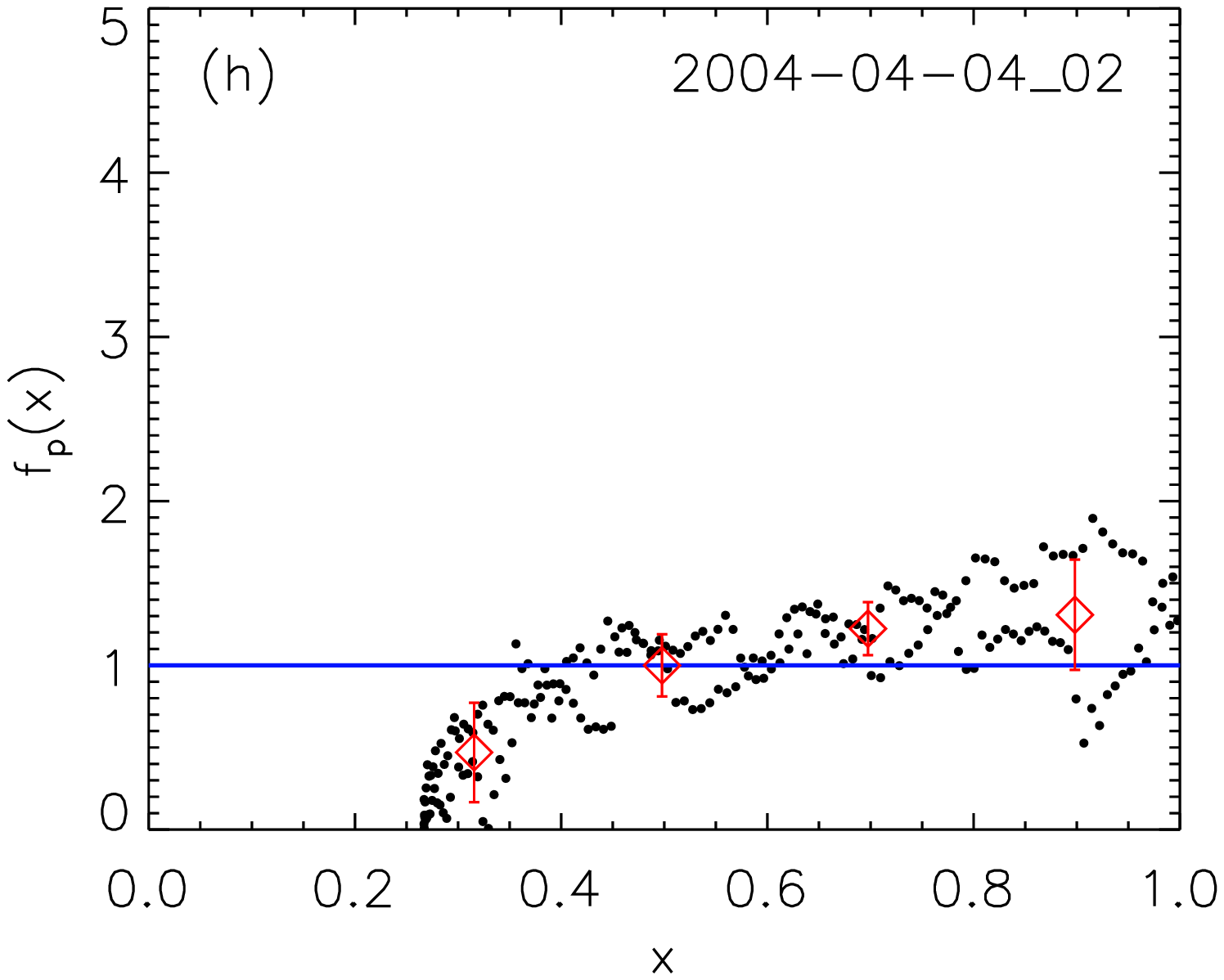}
  \caption{Scatter plot of measurements during the MC events showing the validity of the assumption of $f_p(x)=1$ in Eq.\ref{eq_vphi}. Panel (a) contains all the data points of the 72 MC events, and each of the other panels display the data points in a selected MC. See the main text for more details.}\label{fg_k2}
\end{figure*}

It should be noted that $f_p(x)$ in the equation of poloidal speed (Eq.\ref{eq_vphi})
is assumed to be unity. Whether or not is this assumption reasonable? We check it by investigating
the correlation between $f_p(x)=\frac{v_\varphi R}{k_1}$ and $x$ as shown in 
Figure~\ref{fg_k2}. $v_\varphi$ is derived from velocity measurements in the MC frame, and $R$
are obtained from model results. $k_1$ is an unknown constant, that may change from one event to another.
We determine the value of $k_1$ by making the value of $f_p(x)$ of most data points around unity. 
All the data points of the 72 events are plotted together to show the statistical trend (Fig.\ref{fg_k2}a).
The red diamonds in the figure are the mean values of the data points in the bins with a horizontal
size of $0.2$, and the error bars indicate the standard deviations. The blue line gives
$f_p(x)=1$. It is clear that data points are generally distributed around the
blue line no matter which value $x$ is, though a large scattering is evident, suggesting that
$f_p(x)=1$ is an appropriate assumption from the view of statistics. 

However, the form of $f_p(x)$ is case dependent. Is the assumption of $f_p(x)=1$ generally appropriate 
for an individual MC? It is further checked in Figure~\ref{fg_k2}b--\ref{fg_k2}h, in which the dependence
of $f_p(x)$ on $x$ for 7 selected MCs are presented. These 7 MCs are selected for investigation because 
(1) the derived poloidal speed is significant, larger than 10 km s$^{-1}$, and (2) the closest 
approach is smaller than $0.5R$ so that there is a complete scan of $x$. For different events, the pattern is indeed different.
Figure~\ref{fg_k2}b and \ref{fg_k2}h suggest a general increase of $f_p(x)$ with increasing $x$, 
Figure~\ref{fg_k2}d and \ref{fg_k2}f show that $f_p(x)$ has a unimodal distribution, and 
Figure~\ref{fg_k2}c, \ref{fg_k2}e and \ref{fg_k2}g show a reversed unimodal distribution of $f_p(x)$. 
If considering the errors, we find that $f_p(x)$ is almost independent on $x$ for all the cases except the last one 
(Fig.\ref{fg_k2}h). Even for the last case, $f_p(x)$ is almost invariant for $x>0.4$. 
Thus, based on the above analysis, we may treat the simplest assumption, $f_p(x)=1$, as an acceptable approximation.

\section{conclusions and discussion}\label{sec_dis}
In this paper, we present a velocity-modified cylindrical force-free flux rope model in details.
Both observations of in-situ magnetic field and plasma velocity are taken into account by our model
to derive the geometrical and kinematic parameters of MCs, which have been summarized in Table~\ref{tb_para}.
The validity of our fitting procedure and the effect of velocity on the fitting results are checked
through the Lepping list. It is found that the values of the modeled radius and orientation of an MC
and the closest approach are more likely to be changed significantly if the velocity is considered.
In our sample, the radius changes its value by more than 20\% in 22\% of the cases, orientation changes
by more than 30$^\circ$ in 15\%, and the closest approach changes by more than 20\% in 17\%.
In a few cases, the handedness may also be changed. We then obtain the statistical properties
of MCs, including the magnetic field strength, radius, orientation, magnetic flux, helicity
and initial magnetic energy, which have been summarized in Figure~\ref{fg_mcprop}.
Furthermore, some findings about the plasma motion of MCs are obtained.
\begin{enumerate}
\item The linear propagation velocity may not be along the radial direction. The value of the
non-radial component of the propagation velocity could be more than 10\% of radial speed
in some cases, which constitutes the direct evidence of the deflected propagation and/or rotation
of a CME in interplanetary space.

\item As previous studies have shown (see Sec.\ref{sec_exp}), the expansion speed is 
correlated with the radial propagation speed with a coefficient of about 0.55, and 
most MCs did not undergo a self-similar expansion at 1 AU in radial direction,
i.e., the radius is not evolving proportionally to the heliocentric distance. In our 
statistics, 62\%/17\% of MCs underexpanded/overexpanded with an expansion rate $<0.8/>1.2$, 
and on average the expansion rate is about 0.6.

\item The poloidal motion is not as significant as expanding motion generally, but does exist
in some cases. Its speed is on the order of 10 km s$^{-1}$ at 1 AU.
\end{enumerate}

The last point is of particular interest. Considering the possible presence of 
small pure-axial velocity (see Sec.\ref{sec_prop}), there may exist helical plasma motion 
following the helical magnetic field lines of an MC, which is worth to be investigated 
further. No matter how the plasma elements move inside the MC, the presence of the poloidal
motion means that MCs may carry non-zero angular momentum.
We think that there are at least
three possible causes of the angular momentum. 

\begin{figure*}[tbh]
  \centering
  \includegraphics[width=0.95\hsize]{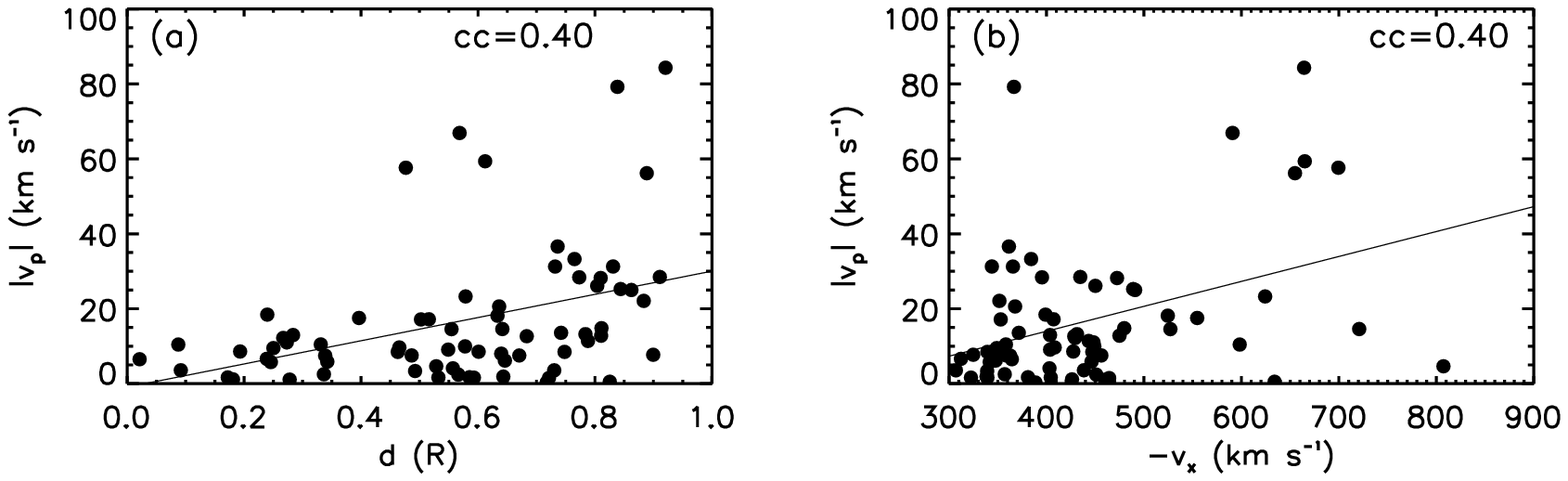}
  \caption{(a) Scatter plot showing the correlation between the poloidal speed and the closest approach and (b) that between the poloidal speed and radial propagation speed. The lines give the linear fitting results.}\label{fg_vcir-dep}
\end{figure*}

The first is that the angular
momentum is generated locally through the interaction with ambient
solar wind. If there is the velocity difference between
the solar wind and an MC, the solar wind plasma would stream around
the MC body. The viscosity or some other processes may cause the
poloidal motion inside the MC. If this is true, we would expect that
the poloidal speed, $v_p$, is stronger at the periphery of the flux rope
and therefore the absolute value of the derived $v_p$ is positively correlated with the closest approach, $d$.
Our statistical study suggests that there is a weak correlation between them as shown in Figure~\ref{fg_vcir-dep}a;
the correlation coefficient is about 0.4.
This result implies that the local solar wind interaction perhaps is probably a factor
for the poloidal motion but not the only one.

The second possible cause is that the angular momentum is generated internally
all the time of the CME from its early phase in the corona to propagation phase in interplanetary
space. It is well known that the magnetic energy of a CME will decrease when
it expands \citep[e.g.,][]{Kumar_Rust_1996, Wang_etal_2009, Nakwacki_etal_2011}.
The decreased magnetic energy may go into thermal energy
and kinetic energy through some mechanisms. It means
that the poloidal motion could be somehow generated inside the MC.
If this is true, the decreased magnetic energy must partially go into
the rotational kinetic energy, and we would expect that the poloidal speed,
$v_{p}$, be more or less dependent on the radial propagation speed, $v_X$,
because generally the faster the propagation speed is the larger is the
rate of the magnetic energy decrease. As we can see in Figure~\ref{fg_vcir-dep}b,
there is also a weak correlation with a coefficient of 0.4. It suggests
that the internal process perhaps is another factor for the poloidal motion.
The numerical simulations of the evolution of magnetic flux rope by
\citet{Wei_etal_1991} had shown that the expansion of a flux rope may
cause the significantly azimuthal velocity, i.e., the poloidal speed
in our study, even if the azimuthal velocity was zero initially.

The third one may be that the angular momentum is generated initially at the eruption of the CME in the corona,
and carried all the way to 1 AU. The kinking/unkinking behavior and rotation of the whole
erupting magnetic structure are often observed in coronagraphs \citep[e.g.,][]{Yurchyshyn_2008,
Yurchyshyn_etal_2009, Vourlidas_etal_2011}. Such processes might introduce
poloidal motion in the body of the CME. However, if the angular momentum is
conserved as that we treated in the model (see Eq.\ref{eq_vphi}), we may
estimate that the poloidal speed will be about 200 times of that at 1 AU, or
about 2000 km s$^{-1}$ near the Sun if the poloidal speed is about
10 km s$^{-1}$ at 1 AU. It is interesting to check the imaging data of CMEs to
search such a fast poloidal motion though no such a phenomenon was reported so far.

No matter which of the above speculation(s) is right, the evidence of poloidal
motion inside MCs does shed a new light on the dynamic evolution of CMEs from its
birth to interplanetary space. Some outstanding questions are open. How is the
angular momentum in the solar wind transfered into an MC if the first speculation
is right? How is the magnetic energy of an MC converted into the rotational kinetic
energy if the second speculation is right? How does the poloidal motion depend on
the heliocentric distance? Whether or how does the poloidal motion modify the
macro/micro-scopic properties of an MC? Except further studies based on the current
observations, the upcoming missions, Solar Orbiter and Solar Probe+ will provide
great opportunities to advance our understanding of these questions.

\begin{acknowledgments}
We acknowledge the use of the data from Wind spacecraft. The model
developed in this work could be run and tested online at \url{http://space.ustc.edu.cn/dreams/mc_fitting/}.
We thank Dr. L. C. Lee
for valuable discussion on the poloidal motion, and we are also grateful to the referees for constructive comments.
This work is supported by the grants from MOST 973
key project (2011CB811403), CAS (Key Research
Program KZZD-EW-01-4), NSFC (41131065 and
41121003), MOEC (20113402110001)
and the fundamental research funds for the central universities.
\end{acknowledgments}

\appendix

\section{Derivation of poloidal speed $v_\varphi$}\label{app_poloidal}

Under the self-similar assumption, we may assume the self-similarity variable $x=r/R(t)$,
and the self-similarity solution $\rho(t,r)=\rho_0(t)f_\rho(x)$
and $v_\varphi(t,r)=v_0(t)f_p(x)$, which mean that the shape of the spatial distribution of $\rho$
and $v_\varphi$ along the radial distance in the flux rope will not change with time. Further we
have the self-similarity solution of $v_r(t,r)=v_e(t)x=\frac{v_e(t)}{R(t)}r$, in which
$v_e(t)=\frac{\partial R}{\partial t}$ is the expansion
speed and $R(t)$ is the radius of the flux rope.

The mass conservation requires
\begin{eqnarray}
&&\int \rho rdrd\varphi dz=2\pi l\int_0^R \rho rdr=2\pi lR^2\int_0^1 \rho xdx\nonumber\\
&&=2\pi lR^2\rho_0(t)\int_0^1 f_\rho(x) xdx=M
\end{eqnarray}
where $M$ is the total mass of the flux rope, and the above integral is a constant, say $c_1$.
Then we have
\begin{eqnarray}
\rho_0(t)=\frac{M}{2\pi c_1}R^{-2}l^{-1}
\end{eqnarray}
The conservation of fluxes implies that $l\propto R$ (Eq.\ref{eq_rl}) or $l=c_2R$, and therefore the above equation becomes
\begin{eqnarray}
\rho_0(t)=\frac{M}{2\pi c_1c_2} R^{-3}
\end{eqnarray}

Similarly, if there is no force in $\hat{\varphi}$ direction, the conservation of angular momentum requires
\begin{eqnarray}
&&\int \rho rv_\varphi rdrd\varphi dz=2\pi lR^3\int_0^1 \rho v_\varphi x^2dx\nonumber\\
&&=2\pi lR^3\rho_0(t)v_0(t)\int_0^1 f_\rho(x)f_p(x) x^2dx=L_A 
\end{eqnarray}
where $L_A$ is the total angular momentum, and the integral in the above equation is also a constant, say $c_3$.
Then we get
\begin{eqnarray}
v_0(t)=\frac{L_A}{2\pi c_3}\rho_0^{-1}R^{-3}l^{-1}=\frac{c_1L_A}{c_3M}R^{-1}
\end{eqnarray}
or
\begin{eqnarray}
v_0(t)=k_1R^{-1}
\end{eqnarray}
where $k_1=\frac{c_1L_A}{c_3M}$ is also a constant, and therefore expression of 
$v_\varphi$ writes $k_1f_p(x)R^{-1}$, as written in Eq.\ref{eq_vphi}.

\section{Additional tables}\label{app_tables}

\clearpage

\begin{sidewaystable}\centering
\vspace{-250pt}
\caption{Parameters of 72 MCs derived by the velocity-modified cylindrical force-free flux rope model}
\tabcolsep 4pt 
\begin{tabular}{c|cc|ccccccccccc|ccccccc}
\hline
 & \multicolumn{2}{c|}{MC Interval} & \multicolumn{18}{c}{Modeled Parameters}\\
\cline{2-3}\cline{4-21}
No. & $t_0$ & $\Delta t$ & $B_0$ & $R$ & $H$ & $\theta$ & $\phi$ & $d$ & $v_X$ & $v_Y$ & $v_Z$ & $v_{e}$ & $v_{p}$ & $\Delta t_c$ & $\Theta$ & $\Phi_z$ & $\Phi_\varphi$ & $H_m$ & $E_{m0}$ & $\chi_n$ \\
(1) &
(2) &
(3) &
(4) &
(5) &
(6) &
(7) &
(8) &
(9) &
(10) &
(11) &
(12) &
(13) &
(14) &
(15) &
(16) &
(17) &
(18) &
(19) &
(20) &
(21) \\
\hline
  1 & 1995/02/08 05:48 & 19.0 & 15 & 0.09 & -1 & -10 & 122 & -0.56 & -403 & 2 & 14 & 30 & 4 & 8.8 & 57 & 0.40 & 3.41$\pm0.76$ & -1.72$\pm0.38$ & 2.52$\pm0.56$ & 0.26 \\
  2 & 1995/04/03 07:48 & 27.0 & 14 & 0.22 & 1 & 5 & 75 & -0.84 & -366 & 0 & 68 & 31 & -79 & 12.8 & 75 & 2.0 & 7.3$\pm1.6$ & 18.4$\pm4.1$ & 11.7$\pm2.6$ & 0.27 \\
  3 & 1995/04/06 07:18 & 10.5 & 11 & 0.06 & -1 & 45 & 119 & -0.74 & -361 & -9 & -9 & -17 & -36 & 5.5 & 69 & 0.14 & 1.75$\pm0.39$ & -0.309$\pm0.069$ & 0.66$\pm0.15$ & 0.36 \\
  4 & 1995/08/22 21:18 & 22.0 & 13 & 0.10 & 1 & -26 & 308 & -0.64 & -352 & 10 & -20 & 20 & 7 & 10.4 & 55 & 0.41 & 3.22$\pm0.72$ & 1.65$\pm0.37$ & 2.25$\pm0.50$ & 0.28 \\
  5 & 1995/10/18 19:48 & 29.5 & 24 & 0.10 & 1 & 0 & 314 & -0.17 & -404 & -5 & 13 & -23 & -1 & 15.9 & 45 & 0.8 & 6.0$\pm1.3$ & 6.0$\pm1.3$ & 7.8$\pm1.7$ & 0.29 \\
  6 & 1996/05/27 15:18 & 40.0 & 10 & 0.16 & -1 & 5 & 119 & -0.02 & -364 & -20 & -5 & 12 & -6 & 19.3 & 60 & 0.79 & 3.93$\pm0.87$ & -3.85$\pm0.86$ & 3.35$\pm0.74$ & 0.33 \\
  7 & 1996/07/01 17:18 & 17.0 & 12 & 0.07 & -1 & 4 & 75 & 0.33 & -358 & -20 & -9 & 14 & 10 & 8.1 & 75 & 0.21 & 2.19$\pm0.49$ & -0.56$\pm0.12$ & 1.04$\pm0.23$ & 0.40 \\
  8 & 1996/08/07 12:18 & 22.5 & 9 & 0.11 & 1 & -39 & 236 & 0.65 & -347 & 0 & 14 & 7 & 6 & 11.1 & 64 & 0.33 & 2.32$\pm0.52$ & 0.97$\pm0.21$ & 1.17$\pm0.26$ & 0.36 \\
  9 & 1996/12/24 02:48 & 32.5 & 14 & 0.16 & 1 & 29 & 60 & -0.60 & -350 & -15 & 26 & 25 & -8 & 15.1 & 64 & 1.0 & 5.2$\pm1.1$ & 6.5$\pm1.4$ & 5.8$\pm1.3$ & 0.29 \\
 10 & 1997/01/10 05:18 & 21.0 & 17 & 0.10 & 1 & -29 & 240 & -0.09 & -438 & 11 & -16 & 42 & 3 & 9.4 & 64 & 0.49 & 3.92$\pm0.87$ & 2.41$\pm0.54$ & 3.33$\pm0.74$ & 0.26 \\
 11 & 1997/04/11 05:36 & 13.5 & 26 & 0.02 & 1 & 14 & 1 & -0.80 & -450 & -17 & -34 & 12 & 26 & 6.2 & 14 & 0.04 & 1.37$\pm0.30$ & 0.064$\pm0.014$ & 0.405$\pm0.090$ & 0.45 \\
 12 & 1997/05/15 09:06 & 16.0 & 21 & 0.09 & -1 & -15 & 104 & 0.27 & -448 & 37 & 0 & -14 & -11 & 8.3 & 75 & 0.46 & 4.23$\pm0.95$ & -2.41$\pm0.54$ & 3.88$\pm0.86$ & 0.37 \\
 13 & 1997/06/09 02:18 & 21.0 & 17 & 0.04 & 1 & -4 & 195 & 0.74 & -371 & 1 & 6 & 3 & 13 & 10.3 & 15 & 0.08 & 1.58$\pm0.35$ & 0.157$\pm0.035$ & 0.54$\pm0.12$ & 0.35 \\
 14 & 1997/09/22 00:48 & 16.5 & 23 & 0.10 & -1 & 51 & 33 & 0.68 & -428 & 45 & 3 & 62 & 12 & 7.3 & 58 & 0.71 & 5.54$\pm1.23$ & -4.9$\pm1.1$ & 6.7$\pm1.5$ & 0.17 \\
 15 & 1997/10/01 16:18 & 30.5 & 15 & 0.03 & -1 & 6 & 178 & -0.79 & -443 & 0 & 3 & -7 & 11 & 16.6 & 7 & 0.04 & 1.07$\pm0.24$ & -0.054$\pm0.012$ & 0.248$\pm0.055$ & 0.28 \\
 16 & 1997/10/10 23:48 & 25.0 & 14 & 0.11 & 1 & -8 & 240 & -0.28 & -403 & 13 & -11 & 37 & -12 & 11.3 & 60 & 0.52 & 3.67$\pm0.81$ & 2.37$\pm0.53$ & 2.92$\pm0.65$ & 0.27 \\
 17 & 1997/11/07 15:48 & 12.5 & 26 & 0.08 & 1 & 27 & 313 & -0.78 & -431 & 0 & 9 & 20 & -13 & 6.0 & 52 & 0.5 & 5.1$\pm1.1$ & 3.48$\pm0.77$ & 5.7$\pm1.3$ & 0.16 \\
 18 & 1997/11/08 04:54 & 10.0 & 20 & 0.05 & 1 & 52 & 5 & -0.64 & -367 & -2 & 3 & 18 & 20 & 4.7 & 52 & 0.13 & 2.21$\pm0.49$ & 0.353$\pm0.078$ & 1.06$\pm0.24$ & 0.27 \\
 19 & 1998/01/07 03:18 & 29.0 & 22 & 0.16 & -1 & 60 & 134 & -0.59 & -380 & -26 & -1 & 31 & -1 & 13.4 & 69 & 1.6 & 8.2$\pm1.8$ & -16.4$\pm3.6$ & 14.5$\pm3.2$ & 0.26 \\
 20 & 1998/02/04 04:30 & 42.0 & 14 & 0.11 & -1 & 4 & 32 & -0.59 & -322 & 10 & 14 & 34 & -1 & 17.8 & 32 & 0.55 & 3.86$\pm0.86$ & -2.67$\pm0.59$ & 3.24$\pm0.72$ & 0.27 \\
 21 & 1998/03/04 14:18 & 40.0 & 14 & 0.17 & -1 & 16 & 60 & 0.53 & -339 & -2 & -12 & 20 & -1 & 18.8 & 61 & 1.2 & 5.5$\pm1.2$ & -8.1$\pm1.8$ & 6.7$\pm1.5$ & 0.28 \\
 22 & 1998/06/02 10:36 & 5.3 & 15 & 0.02 & -1 & 14 & 34 & 0.52 & -407 & -21 & -46 & 15 & 17 & 2.5 & 37 & 0.01 & 0.58$\pm0.13$ & -0.009$\pm0.002$ & 0.072$\pm0.016$ & 0.20 \\
 23 & 1998/06/24 16:48 & 29.0 & 16 & 0.03 & -1 & 10 & 174 & -0.46 & -447 & -8 & 17 & 1 & 8 & 14.3 & 11 & 0.05 & 1.24$\pm0.28$ & -0.080$\pm0.018$ & 0.336$\pm0.075$ & 0.28 \\
 24 & 1998/08/20 10:18 & 33.0 & 15 & 0.11 & 1 & 1 & 299 & -0.24 & -312 & -5 & -17 & 25 & 6 & 14.9 & 60 & 0.56 & 3.99$\pm0.89$ & 2.80$\pm0.62$ & 3.46$\pm0.77$ & 0.36 \\
 25 & 1998/09/25 10:18 & 27.0 & 17 & 0.21 & -1 & 60 & 195 & 0.58 & -624 & 94 & 50 & 90 & -23 & 11.6 & 61 & 2.33 & 8.72$\pm1.94$ & -25.4$\pm5.6$ & 16.5$\pm3.7$ & 0.23 \\
 26 & 1998/11/08 23:48 & 25.5 & 19 & 0.15 & 1 & -60 & 134 & 0.49 & -456 & -33 & 10 & 68 & -7 & 10.9 & 69 & 1.3 & 6.8$\pm1.5$ & 11.0$\pm2.4$ & 10.0$\pm2.2$ & 0.31 \\
 27 & 1999/08/09 10:48 & 29.0 & 12 & 0.04 & -1 & 15 & 176 & -0.49 & -339 & 5 & -18 & -10 & -3 & 15.8 & 15 & 0.07 & 1.28$\pm0.28$ & -0.112$\pm0.025$ & 0.355$\pm0.079$ & 0.30 \\
 28 & 2000/07/01 08:48 & 18.5 & 12 & 0.11 & -1 & 60 & 178 & -0.77 & -384 & 23 & 9 & -14 & 33 & 9.4 & 60 & 0.48 & 3.32$\pm0.74$ & -1.99$\pm0.44$ & 2.39$\pm0.53$ & 0.25 \\
 29 & 2000/07/28 21:06 & 13.0 & 21 & 0.11 & -1 & -11 & 240 & -0.81 & -474 & -14 & 1 & 4 & -12 & 6.5 & 60 & 0.7 & 5.4$\pm1.2$ & -5.0$\pm1.1$ & 6.2$\pm1.4$ & 0.29 \\
 30 & 2000/08/12 06:06 & 23.0 & 29 & 0.14 & -1 & 9 & 60 & 0.40 & -554 & -1 & -35 & 63 & -17 & 10.2 & 60 & 1.8 & 9.9$\pm2.2$ & -22.0$\pm4.9$ & 21.1$\pm4.7$ & 0.42 \\
\hline
\end{tabular}
\label{tb_parbv1}
\end{sidewaystable}
\clearpage

\begin{sidewaystable}\centering
\vspace{-250pt}
\caption{Parameters of 72 MCs derived by the velocity-modified cylindrical force-free flux rope model (continued)}
\tabcolsep 4pt 
\begin{tabular}{c|cc|ccccccccccc|ccccccc}
\hline
 & \multicolumn{2}{c|}{MC Interval} & \multicolumn{18}{c}{Modeled Parameters}\\
\cline{2-3}\cline{4-21}
No. & $t_0$ & $\Delta t$ & $B_0$ & $R$ & $H$ & $\theta$ & $\phi$ & $d$ & $v_X$ & $v_Y$ & $v_Z$ & $v_{e}$ & $v_{p}$ & $\Delta t_c$ & $\Theta$ & $\Phi_z$ & $\Phi_\varphi$ & $H_m$ & $E_{m0}$ & $\chi_n$ \\
(1) &
(2) &
(3) &
(4) &
(5) &
(6) &
(7) &
(8) &
(9) &
(10) &
(11) &
(12) &
(13) &
(14) &
(15) &
(16) &
(17) &
(18) &
(19) &
(20) &
(21) \\
\hline
 31 & 2000/10/03 17:06 & 21.0 & 19 & 0.09 & 1 & 33 & 59 & 0.24 & -398 & 7 & -13 & 16 & -18 & 10.0 & 65 & 0.50 & 4.17$\pm0.93$ & 2.59$\pm0.57$ & 3.78$\pm0.84$ & 0.24 \\
 32 & 2000/10/13 18:24 & 22.5 & 18 & 0.02 & -1 & 6 & 188 & 0.77 & -395 & -22 & 10 & 0 & -28 & 11.3 & 10 & 0.03 & 0.94$\pm0.21$ & -0.030$\pm0.007$ & 0.190$\pm0.042$ & 0.21 \\
 33 & 2000/11/06 23:06 & 19.0 & 28 & 0.08 & -1 & 0 & 150 & -0.55 & -527 & -17 & -32 & -13 & -14 & 9.8 & 29 & 0.5 & 5.2$\pm1.1$ & -3.21$\pm0.71$ & 5.8$\pm1.3$ & 0.21 \\
 34 & 2001/03/19 23:18 & 19.0 & 25 & 0.02 & -1 & -11 & 185 & -0.55 & -403 & -13 & -25 & -9 & 9 & 10.6 & 12 & 0.02 & 0.99$\pm0.22$ & -0.025$\pm0.006$ & 0.213$\pm0.047$ & 0.20 \\
 35 & 2001/04/04 20:54 & 11.5 & 17 & 0.23 & -1 & 18 & 272 & -0.92 & -664 & -39 & -80 & 132 & 84 & 5.3 & 87 & 2.8 & 9.5$\pm2.1$ & -33.0$\pm7.3$ & 19.5$\pm4.3$ & 0.29 \\
 36 & 2001/04/12 07:54 & 10.0 & 26 & 0.07 & 1 & 6 & 196 & 0.89 & -654 & 83 & -63 & 97 & 56 & 4.2 & 17 & 0.41 & 4.47$\pm0.99$ & 2.29$\pm0.51$ & 4.34$\pm0.96$ & 0.31 \\
 37 & 2001/04/22 00:54 & 24.5 & 15 & 0.10 & -1 & -45 & 309 & 0.34 & -357 & 0 & 4 & 33 & -2 & 11.0 & 63 & 0.45 & 3.62$\pm0.80$ & -2.05$\pm0.45$ & 2.84$\pm0.63$ & 0.18 \\
 38 & 2001/04/29 01:54 & 11.0 & 16 & 0.13 & -1 & 19 & 60 & 0.83 & -634 & 15 & -39 & 39 & 0 & 5.3 & 62 & 0.8 & 4.9$\pm1.1$ & -4.9$\pm1.1$ & 5.2$\pm1.1$ & 0.19 \\
 39 & 2001/05/28 11:54 & 22.5 & 13 & 0.07 & -1 & -14 & 25 & 0.57 & -450 & 26 & 42 & -1 & 2 & 11.3 & 29 & 0.21 & 2.27$\pm0.50$ & -0.61$\pm0.14$ & 1.12$\pm0.25$ & 0.22 \\
 40 & 2001/07/10 17:18 & 39.5 & 8 & 0.13 & 1 & 15 & 224 & -0.25 & -348 & -4 & -12 & 2 & -9 & 19.6 & 46 & 0.37 & 2.35$\pm0.52$ & 1.09$\pm0.24$ & 1.19$\pm0.27$ & 0.41 \\
 41 & 2002/03/19 22:54 & 16.5 & 24 & 0.09 & 1 & 11 & 136 & 0.83 & -343 & -3 & 4 & -12 & -31 & 8.5 & 44 & 0.5 & 4.9$\pm1.1$ & 3.33$\pm0.74$ & 5.3$\pm1.2$ & 0.20 \\
 42 & 2002/03/24 03:48 & 43.0 & 16 & 0.18 & 1 & 29 & 314 & -0.19 & -427 & -1 & -7 & -1 & 8 & 21.6 & 52 & 1.5 & 6.7$\pm1.5$ & 12.6$\pm2.8$ & 9.7$\pm2.2$ & 0.39 \\
 43 & 2002/04/18 04:18 & 22.0 & 20 & 0.06 & 1 & -6 & 346 & -0.81 & -479 & -2 & -5 & 7 & -14 & 10.6 & 15 & 0.21 & 2.83$\pm0.63$ & 0.75$\pm0.17$ & 1.74$\pm0.39$ & 0.17 \\
 44 & 2002/05/19 03:54 & 19.5 & 18 & 0.25 & -1 & -15 & 87 & 0.91 & -434 & 2 & -59 & 78 & -28 & 9.0 & 87 & 3.3 & 10.6$\pm2.3$ & -43.4$\pm9.6$ & 24.3$\pm5.4$ & 0.29 \\
 45 & 2002/08/02 07:24 & 13.7 & 20 & 0.10 & -1 & -8 & 311 & 0.81 & -472 & 1 & 14 & 25 & -28 & 6.5 & 49 & 0.6 & 4.7$\pm1.0$ & -3.48$\pm0.77$ & 4.8$\pm1.1$ & 0.14 \\
 46 & 2002/09/03 00:18 & 18.5 & 13 & 0.06 & 1 & 29 & 196 & 0.50 & -352 & -3 & -32 & -21 & 17 & 10.1 & 33 & 0.13 & 1.77$\pm0.39$ & 0.276$\pm0.061$ & 0.68$\pm0.15$ & 0.44 \\
 47 & 2003/03/20 11:54 & 10.5 & 14 & 0.09 & -1 & -60 & 208 & 0.48 & -699 & 10 & 0 & 18 & 57 & 5.1 & 64 & 0.35 & 3.03$\pm0.67$ & -1.32$\pm0.29$ & 1.99$\pm0.44$ & 0.34 \\
 48 & 2003/08/18 11:36 & 16.8 & 24 & 0.10 & 1 & -23 & 191 & 0.86 & -490 & -11 & 50 & 21 & 25 & 8.0 & 26 & 0.7 & 5.7$\pm1.3$ & 5.3$\pm1.2$ & 7.1$\pm1.6$ & 0.30 \\
 49 & 2003/11/20 10:48 & 15.5 & 33 & 0.10 & 1 & -60 & 134 & 0.09 & -598 & -6 & 27 & 106 & 10 & 6.3 & 69 & 1.1 & 8.2$\pm1.8$ & 11.1$\pm2.4$ & 14.7$\pm3.3$ & 0.38 \\
 50 & 2004/04/04 02:48 & 36.0 & 18 & 0.17 & -1 & 60 & 25 & -0.27 & -429 & 15 & -7 & 25 & -12 & 16.8 & 63 & 1.6 & 7.3$\pm1.6$ & -14.4$\pm3.2$ & 11.5$\pm2.6$ & 0.33 \\
 51 & 2004/07/24 12:48 & 24.5 & 27 & 0.19 & 1 & -36 & 45 & -0.57 & -590 & 72 & 9 & 11 & -66 & 11.9 & 55 & 2.8 & 12.1$\pm2.7$ & 42.9$\pm9.5$ & 31.8$\pm7.1$ & 0.26 \\
 52 & 2004/08/29 18:42 & 26.1 & 18 & 0.16 & 1 & -16 & 119 & 0.72 & -388 & -7 & -1 & 16 & 0 & 12.6 & 61 & 1.3 & 6.5$\pm1.4$ & 10.3$\pm2.3$ & 9.1$\pm2.0$ & 0.33 \\
 53 & 2004/11/08 03:24 & 13.2 & 24 & 0.03 & -1 & -2 & 8 & 0.61 & -664 & 62 & -27 & 42 & 59 & 5.3 & 8 & 0.07 & 1.81$\pm0.40$ & -0.167$\pm0.037$ & 0.71$\pm0.16$ & 0.49 \\
 54 & 2004/11/09 20:54 & 6.5 & 46 & 0.06 & -1 & 23 & 311 & 0.53 & -807 & 10 & 15 & 37 & 4 & 3.0 & 53 & 0.49 & 6.55$\pm1.46$ & -4.02$\pm0.89$ & 9.3$\pm2.1$ & 0.31 \\
 55 & 2004/11/10 03:36 & 7.5 & 35 & 0.06 & -1 & -41 & 10 & 0.64 & -720 & -50 & -10 & 66 & -14 & 3.3 & 42 & 0.3 & 4.8$\pm1.1$ & -2.07$\pm0.46$ & 5.0$\pm1.1$ & 0.16 \\
 56 & 2005/05/20 07:18 & 22.0 & 12 & 0.12 & -1 & 60 & 225 & -0.34 & -446 & 0 & -2 & 7 & 5 & 10.8 & 69 & 0.51 & 3.43$\pm0.76$ & -2.20$\pm0.49$ & 2.55$\pm0.57$ & 0.47 \\
 57 & 2005/06/12 15:36 & 15.5 & 26 & 0.06 & -1 & -17 & 4 & 0.84 & -488 & 28 & -28 & 31 & 25 & 7.0 & 18 & 0.31 & 3.87$\pm0.86$ & -1.51$\pm0.33$ & 3.26$\pm0.72$ & 0.22 \\
 58 & 2005/07/17 15:18 & 12.5 & 13 & 0.06 & 1 & -29 & 60 & 0.18 & -426 & -8 & -25 & 32 & 1 & 5.8 & 64 & 0.14 & 1.89$\pm0.42$ & 0.333$\pm0.074$ & 0.77$\pm0.17$ & 0.30 \\
 59 & 2005/12/31 14:48 & 20.0 & 13 & 0.02 & 1 & 0 & 353 & -0.72 & -464 & -2 & 17 & -11 & 1 & 11.5 & 6 & 0.02 & 0.60$\pm0.13$ & 0.011$\pm0.002$ & 0.077$\pm0.017$ & 0.22 \\
 60 & 2006/02/05 19:06 & 18.0 & 11 & 0.07 & 1 & -30 & 119 & 0.25 & -341 & -12 & -5 & 24 & 5 & 8.3 & 64 & 0.16 & 1.81$\pm0.40$ & 0.36$\pm0.08$ & 0.71$\pm0.16$ & 0.38 \\
\hline
\end{tabular}
\label{tb_parbv2}
\end{sidewaystable}
\clearpage

\begin{sidewaystable}\centering
\vspace{-250pt}
\caption{Parameters of 72 MCs derived by the velocity-modified cylindrical force-free flux rope model (continued)}
\tabcolsep 4pt 
\begin{tabular}{c|cc|ccccccccccc|ccccccc}
\hline
 & \multicolumn{2}{c|}{MC Interval} & \multicolumn{18}{c}{Modeled Parameters}\\
\cline{2-3}\cline{4-21}
No. & $t_0$ & $\Delta t$ & $B_0$ & $R$ & $H$ & $\theta$ & $\phi$ & $d$ & $v_X$ & $v_Y$ & $v_Z$ & $v_{e}$ & $v_{p}$ & $\Delta t_c$ & $\Theta$ & $\Phi_z$ & $\Phi_\varphi$ & $H_m$ & $E_{m0}$ & $\chi_n$ \\
(1) &
(2) &
(3) &
(4) &
(5) &
(6) &
(7) &
(8) &
(9) &
(10) &
(11) &
(12) &
(13) &
(14) &
(15) &
(16) &
(17) &
(18) &
(19) &
(20) &
(21) \\
\hline
 61 & 2006/04/13 20:36 & 13.3 & 25 & 0.08 & -1 & -5 & 226 & -0.63 & -524 & 22 & 6 & -2 & -18 & 6.7 & 46 & 0.5 & 4.9$\pm1.1$ & -3.11$\pm0.69$ & 5.2$\pm1.2$ & 0.23 \\
 62 & 2006/08/30 21:06 & 17.8 & 10 & 0.07 & -1 & -5 & 224 & -0.47 & -408 & 0 & 3 & 8 & -9 & 8.7 & 44 & 0.15 & 1.66$\pm0.37$ & -0.302$\pm0.067$ & 0.60$\pm0.13$ & 0.26 \\
 63 & 2007/05/21 22:54 & 14.7 & 14 & 0.02 & -1 & -11 & 350 & -0.58 & -449 & 20 & 26 & -14 & 9 & 8.5 & 14 & 0.01 & 0.58$\pm0.13$ & -0.009$\pm0.002$ & 0.072$\pm0.016$ & 0.39 \\
 64 & 2007/11/19 23:24 & 13.5 & 18 & 0.06 & -1 & -10 & 314 & 0.28 & -460 & -15 & 23 & 18 & -1 & 6.3 & 46 & 0.18 & 2.52$\pm0.56$ & -0.58$\pm0.13$ & 1.38$\pm0.31$ & 0.49 \\
 65 & 2008/12/17 03:06 & 11.3 & 12 & 0.03 & -1 & 0 & 209 & -0.64 & -338 & 0 & 9 & 17 & 1 & 5.2 & 29 & 0.03 & 0.86$\pm0.19$ & -0.034$\pm0.008$ & 0.161$\pm0.036$ & 0.22 \\
 66 & 2009/02/04 00:06 & 10.8 & 14 & 0.04 & 1 & 2 & 146 & 0.67 & -358 & -7 & -12 & 11 & -7 & 5.2 & 33 & 0.06 & 1.22$\pm0.27$ & 0.085$\pm0.019$ & 0.322$\pm0.071$ & 0.33 \\
 67 & 2009/03/12 00:42 & 24.0 & 12 & 0.10 & 1 & 42 & 122 & 0.34 & -362 & -13 & 0 & -26 & 7 & 12.9 & 67 & 0.39 & 2.97$\pm0.66$ & 1.44$\pm0.32$ & 1.92$\pm0.43$ & 0.43 \\
 68 & 2009/07/21 03:54 & 13.2 & 13 & 0.05 & 1 & 2 & 155 & 0.90 & -324 & -1 & -4 & -2 & 7 & 6.7 & 24 & 0.10 & 1.58$\pm0.35$ & 0.198$\pm0.044$ & 0.54$\pm0.12$ & 0.21 \\
 69 & 2009/09/10 10:24 & 6.0 & 8 & 0.02 & 1 & 20 & 150 & 0.73 & -306 & 2 & 0 & 3 & 3 & 3.0 & 35 & 0.008 & 0.356$\pm0.079$ & 0.004$\pm0.001$ & 0.027$\pm0.006$ & 0.22 \\
 70 & 2009/09/30 07:54 & 9.0 & 13 & 0.04 & -1 & 28 & 219 & -0.75 & -339 & 1 & 16 & 1 & 8 & 4.5 & 47 & 0.06 & 1.23$\pm0.27$ & -0.095$\pm0.021$ & 0.326$\pm0.072$ & 0.25 \\
 71 & 2009/10/12 12:06 & 4.8 & 10 & 0.004 & -1 & 4 & 351 & 0.73 & -365 & 10 & -8 & 0 & -31 & 2.5 & 9 & 0.001 & 0.097$\pm0.022$ & -6e-5$\pm$1e-5 & 20e-4$\pm$5e-4 & 0.18 \\
 72 & 2009/11/01 08:48 & 23.0 & 8 & 0.11 & -1 & 10 & 205 & -0.88 & -351 & 10 & -14 & 2 & -22 & 11.4 & 27 & 0.28 & 2.11$\pm0.47$ & -0.73$\pm0.16$ & 0.96$\pm0.21$ & 0.39 \\
\hline
\end{tabular}
\label{tb_parbv3}
Note 1: Column 2 is the begin time of an MC in UT. Column 3 is the duration of the observed MC interval in units of hour. The interpretations of all the other columns could be found in Table~\ref{tb_para} with the difference that Column 15 is $\Delta t_c=t_c-t_0$. The values of $B_0$, $R$ and $v_{p}$ are all obtained at the time of $t_c$. One can refer to Sec.\ref{sec_model} for more details. \\ Note 2: For the modeled parameters, $B_0$ is in units of nT, $R$ in units of AU, $\theta$, $\phi$ and $\Theta$ in units of degree, $d$ in units of $R$, all the speeds are in units of km s$^{-1}$, $\Delta t_c$ in units of hour, $\Phi_z$ and $\Phi_\varphi$ in units of $10^{21}$ Mx, $H_m$ in units of $10^{42}$ Mx$^2$, and $E_{m0}$ in units of $10^{31}$ erg.
\end{sidewaystable}

\clearpage
\bibliographystyle{agufull}

\begin{thebibliography}{76}
\providecommand{\natexlab}[1]{#1}
\expandafter\ifx\csname urlstyle\endcsname\relax
  \providecommand{\doi}[1]{doi:\discretionary{}{}{}#1}\else
  \providecommand{\doi}{doi:\discretionary{}{}{}\begingroup
  \urlstyle{rm}\Url}\fi

\bibitem[{\textit{Al-Haddad et~al.}(2011)\textit{Al-Haddad, Roussev,
  M{\"{o}}stl, Jacobs, Lugaz, Poedts, and Farrugia}}]{Al-Haddad_etal_2011}
Al-Haddad, N., I.~I. Roussev, C.~M{\"{o}}stl, C.~Jacobs, N.~Lugaz, S.~Poedts,
  and C.~J. Farrugia, On the internal structure of the magnetic field in
  magnetic clouds and interplanetary coronal mass ejections: Writhe versus
  twist, \textit{Astrophys. J. Lett.}, \textit{738}, L18(6pp), 2011.

\bibitem[{\textit{Al-Haddad et~al.}(2013)\textit{Al-Haddad, Nieves-Chinchilla,
  Savani, M{\"{o}}stl, Marubashi, Hidalgo, Roussev, Poedts, and
  Farrugia}}]{Al-Haddad_etal_2013}
Al-Haddad, N., T.~Nieves-Chinchilla, N.~P. Savani, C.~M{\"{o}}stl,
  K.~Marubashi, M.~A. Hidalgo, I.~I. Roussev, S.~Poedts, and C.~J. Farrugia,
  Magnetic field configuration models and reconstruction methods for
  interplanetary coronal mass ejections, \textit{Sol. Phys.}, \textit{284},
  129--149, 2013.

\bibitem[{\textit{Berdichevsky et~al.}(2003)\textit{Berdichevsky, Lepping, and
  Farrugia}}]{Berdichevsky_etal_2003}
Berdichevsky, D.~B., R.~P. Lepping, and C.~J. Farrugia, Geometric
  considerations of the evolution of magnetic flux ropes, \textit{Phys. Rev.
  E}, \textit{67}, 036,405, 2003.

\bibitem[{\textit{Berger}(2003)}]{Berger_2003}
Berger, M.~A., Topological quantities in magnetohydrodynamics, in
  \textit{Advances in Nonlinear Dynamics}, edited by A.~Ferriz-Mas and
  M.~N{\'u}{\~n}ez, p. 345, London: Taylor and Francis Group, 2003.

\bibitem[{\textit{Bothmer and Schwenn}(1998)}]{Bothmer_Schwenn_1998}
Bothmer, V., and R.~Schwenn, The structure and origin of magnetic clouds in the
  solar wind, \textit{Ann. Geophys.}, \textit{16}, 1, 1998.

\bibitem[{\textit{Burlaga et~al.}(1981)\textit{Burlaga, Sittler, Mariani, and
  Schwenn}}]{Burlaga_etal_1981}
Burlaga, L., E.~Sittler, F.~Mariani, and R.~Schwenn, Magnetic loop behind an
  interplanetary shock: {Voyager}, {Helios}, and {IMP} 8 observations,
  \textit{J. Geophys. Res.}, \textit{86}, 6673--6684, 1981.

\bibitem[{\textit{Burlaga}(1988)}]{Burlaga_1988}
Burlaga, L.~F., Magnetic clouds and force-free field with constant alpha,
  \textit{J. Geophys. Res.}, \textit{93}, 7217, 1988.

\bibitem[{\textit{Cid et~al.}(2002)\textit{Cid, Hidalgo, Nieves-Chinchilla,
  Sequeiros, and Vi\~{n}as}}]{Cid_etal_2002}
Cid, C., M.~A. Hidalgo, T.~Nieves-Chinchilla, J.~Sequeiros, and A.~F.
  Vi\~{n}as, Plasma and magnetic field inside magnetic clouds: a global study,
  \textit{Sol. Phys.}, \textit{207}, 187--198, 2002.

\bibitem[{\textit{Dasso et~al.}(2005)\textit{Dasso, Mandrini, D\'{e}moulin,
  Luoni, and Gulisano}}]{Dasso_etal_2005}
Dasso, S., C.~H. Mandrini, P.~D\'{e}moulin, M.~L. Luoni, and A.~M. Gulisano,
  Large scale mhd properties of interplanetary magnetic clouds, \textit{Adv. in
  Space Res.}, \textit{35}, 711--724, 2005.

\bibitem[{\textit{Dasso et~al.}(2007)\textit{Dasso, Nakwacki, D\'{e}moulin, and
  Mandrini}}]{Dasso_etal_2007}
Dasso, S., M.~S. Nakwacki, P.~D\'{e}moulin, and C.~H. Mandrini, Progressive
  transformation of a flux rope to an {ICME}. comparative analysis using the
  direct and fitted expansion methods, \textit{Sol. Phys.}, \textit{244},
  115--137, 2007.

\bibitem[{\textit{Dasso et~al.}(2009)\textit{Dasso, Mandrini, Schmieder,
  Cremades, Cid, Cerrato, Saiz, D\'{e}moulin, Zhukov, Rodriguez, Aran,
  Menvielle, and Poedts}}]{Dasso_etal_2009}
Dasso, S., C.~H. Mandrini, B.~Schmieder, H.~Cremades, C.~Cid, Y.~Cerrato,
  E.~Saiz, P.~D\'{e}moulin, A.~N. Zhukov, L.~Rodriguez, A.~Aran, M.~Menvielle,
  and S.~Poedts, Linking two consecutive nonmerging magnetic clouds with their
  solar sources, \textit{J. Geophys. Res.}, \textit{114}, A02,109, 2009.

\bibitem[{\textit{D\'{e}moulin and
  Dasso}(2009{\natexlab{a}})}]{Demoulin_Dasso_2009}
D\'{e}moulin, P., and S.~Dasso, Causes and consequences of magnetic cloud
  expansion, \textit{Astron. \& Astrophys.}, \textit{498}, 551--566,
  2009{\natexlab{a}}.

\bibitem[{\textit{D\'{e}moulin and
  Dasso}(2009{\natexlab{b}})}]{Demoulin_Dasso_2009a}
D\'{e}moulin, P., and S.~Dasso, Magnetic cloud models with bent and oblate
  cross-section boundaries, \textit{Astron. \& Astrophys.}, \textit{507},
  969--980, 2009{\natexlab{b}}.

\bibitem[{\textit{D\'{e}moulin et~al.}(2008)\textit{D\'{e}moulin, Nakwacki,
  Dasso, and Mandrini}}]{Demoulin_etal_2008}
D\'{e}moulin, P., M.~S. Nakwacki, S.~Dasso, and C.~H. Mandrini, Expected in
  situ velocities from a hierarchical model for expanding interplanetary
  coronal mass ejections, \textit{Sol. Phys.}, \textit{250}, 347--374, 2008.

\bibitem[{\textit{D\'{e}moulin et~al.}(2013)\textit{D\'{e}moulin, Dasso, and
  Janvier}}]{Demoulin_etal_2013}
D\'{e}moulin, P., S.~Dasso, and M.~Janvier, Does spacecraft trajectory strongly
  affect detection of magnetic clouds?, \textit{Astron. \& Astrophys.},
  \textit{550}, A3, 2013.

\bibitem[{\textit{Farrugia et~al.}(1993)\textit{Farrugia, Burlaga, Osherovich,
  Richardson, Freeman, Lepping, and Lazarus}}]{Farrugia_etal_1993a}
Farrugia, C.~J., L.~F. Burlaga, V.~A. Osherovich, I.~G. Richardson, M.~P.
  Freeman, R.~P. Lepping, and A.~J. Lazarus, A study of an expanding
  interplanetary magnetic cloud and its interaction with the earth's
  magnetosphere -- the interplanetary aspect, \textit{J. Geophys. Res.},
  \textit{98(A5)}, 7621--7632, 1993.

\bibitem[{\textit{Farrugia et~al.}(1995)\textit{Farrugia, Osherovich, and
  Burlaga}}]{Farrugia_etal_1995}
Farrugia, C.~J., V.~A. Osherovich, and L.~F. Burlaga, Magnetic flux rope versus
  the spheromak as models for interplanetary magnetic clouds, \textit{J.
  Geophys. Res.}, \textit{100}, 12,293, 1995.

\bibitem[{\textit{Goldstein}(1983)}]{Goldstein_1983}
Goldstein, H., On the field configuration in magnetic clouds, in \textit{Sol.
  Wind Five}, p. 731, NASA Conf. Publ. 2280, Washington D. C., 1983.

\bibitem[{\textit{Gulisano et~al.}(2010)\textit{Gulisano, D\'{e}moulin, Dasso,
  Ruiz, and Marsch}}]{Gulisano_etal_2010}
Gulisano, A.~M., P.~D\'{e}moulin, S.~Dasso, M.~E. Ruiz, and E.~Marsch, Global
  and local expansion of magnetic clouds in the inner heliosphere,
  \textit{Astron. \& Astrophys.}, \textit{509}, A39, 2010.

\bibitem[{\textit{Gulisano et~al.}(2012)\textit{Gulisano, D\'{e}moulin, Dasso,
  and Rodriguez}}]{Gulisano_etal_2012}
Gulisano, A.~M., P.~D\'{e}moulin, S.~Dasso, and L.~Rodriguez, Expansion of
  magnetic clouds in the outer heliosphere, \textit{Astron. \& Astrophys.},
  \textit{543}, A107, 2012.

\bibitem[{\textit{Hidalgo}(2003)}]{Hidalgo_2003}
Hidalgo, M.~A., A study of the expansion and distortion of the cross section of
  magnetic clouds in the interplanetary medium, \textit{J. Geophys. Res.},
  \textit{108}, 1320, 2003.

\bibitem[{\textit{Hidalgo and
  Nieves-Chinchilla}(2012)}]{Hidalgo_Nieves-Chinchilla_2012}
Hidalgo, M.~A., and T.~Nieves-Chinchilla, A global magnetic topology model for
  magnetic clouds. i., \textit{Astrophys. J.}, \textit{748}, 109(7pp), 2012.

\bibitem[{\textit{Hidalgo et~al.}(2002{\natexlab{a}})\textit{Hidalgo, Cid,
  Vinas, and Sequeiros}}]{Hidalgo_etal_2002}
Hidalgo, M.~A., C.~Cid, A.~F. Vinas, and J.~Sequeiros, A non-force-free
  approach to the topology of magnetic clouds in the solar wind, \textit{J.
  Geophys. Res.}, \textit{107}, 1002, 2002{\natexlab{a}}.

\bibitem[{\textit{Hidalgo et~al.}(2002{\natexlab{b}})\textit{Hidalgo,
  Nieves-Chinchilla, and Cid}}]{Hidalgo_etal_2002a}
Hidalgo, M.~A., T.~Nieves-Chinchilla, and C.~Cid, Elliptical cross-section
  model for the magnetic topology of magnetic clouds, \textit{Geophys. Res.
  Lett.}, \textit{29}, 1637, 2002{\natexlab{b}}.

\bibitem[{\textit{Hu and Sonnerup}(2001)}]{Hu_Sonnerup_2001}
Hu, Q., and B.~U.~{\"O}. Sonnerup, Reconstruction of magnetic flux ropes in the
  solar wind, \textit{Geophys. Res. Lett.}, \textit{28}, 467--470, 2001.

\bibitem[{\textit{Hu and Sonnerup}(2002)}]{Hu_Sonnerup_2002}
Hu, Q., and B.~U.~O. Sonnerup, Reconstruction of magnetic clouds in the solar
  wind: Orientations and configurations, \textit{J. Geophys. Res.},
  \textit{107}, 1142, 2002.

\bibitem[{\textit{Isavnin et~al.}(2013)\textit{Isavnin, Vourlidas, and
  Kilpua}}]{Isavnin_etal_2013}
Isavnin, A., A.~Vourlidas, and E.~K.~J. Kilpua, Three-dimensional evolution of
  erupted flux ropes from the {Sun} (2--20 r$\odot$) to 1 {AU}, \textit{Sol.
  Phys.}, \textit{284}, 203--215, 2013.

\bibitem[{\textit{Isavnin et~al.}(2014)\textit{Isavnin, Vourlidas, and
  Kilpua}}]{Isavnin_etal_2014}
Isavnin, A., A.~Vourlidas, and E.~K.~J. Kilpua, Three-dimensional evolution of
  flux-rope {CMEs} and its relation to the local orientation of the
  heliospheric current sheet, \textit{Sol. Phys.}, \textit{289}, 2141--2156,
  2014.

\bibitem[{\textit{Janvier et~al.}(2013)\textit{Janvier, D\'{e}moulin, and
  Dasso}}]{Janvier_etal_2013}
Janvier, M., P.~D\'{e}moulin, and S.~Dasso, Global axis shape of magnetic
  clouds deduced from the distribution of their local axis orientation,
  \textit{Astron. \& Astrophys.}, \textit{556}, A50, 2013.

\bibitem[{\textit{Jian et~al.}(2006)\textit{Jian, Russell, Luhmann, and
  Skoug}}]{Jian_etal_2006}
Jian, L., C.~T. Russell, J.~G. Luhmann, and R.~M. Skoug, Properties of
  interplanetary coronal mass ejections at one au during 1995 -- 2004,
  \textit{Sol. Phys.}, \textit{239}, 393--436, 2006.

\bibitem[{\textit{Kilpua et~al.}(2009)\textit{Kilpua, Pomoell, Vourlidas,
  Vainio, Luhmann, Li, Schroeder, Galvin, and Simunac}}]{kilpua_etal_2009}
Kilpua, E. K.~J., J.~Pomoell, A.~Vourlidas, R.~Vainio, J.~Luhmann, Y.~Li,
  P.~Schroeder, A.~B. Galvin, and K.~Simunac, {STEREO} observations of
  interplanetary coronal mass ejections and prominence deflection during solar
  minimum period, \textit{Ann. Geophys.}, \textit{27}, 4491--4503, 2009.

\bibitem[{\textit{Klein and Burlaga}(1982)}]{Klein_Burlaga_1982}
Klein, L.~W., and L.~F. Burlaga, Interplanetary magnetic clouds at 1 {AU},
  \textit{J. Geophys. Res.}, \textit{87}, 613--624, 1982.

\bibitem[{\textit{Kumar and Rust}(1996)}]{Kumar_Rust_1996}
Kumar, A., and D.~M. Rust, Interplanetary magnetic clouds, helicity
  conservation, and current-core flux-ropes, \textit{J. Geophys. Res.},
  \textit{101}, 15,667--15,684, 1996.

\bibitem[{\textit{Larson et~al.}(1997)\textit{Larson, Lin, McTiernan, McFadden,
  Ergun, McCarthy, R\`{e}me, Sanderson, Kaiser, Lepping, and
  Mazur}}]{Larson_etal_1997}
Larson, D.~E., R.~P. Lin, J.~M. McTiernan, J.~P. McFadden, R.~E. Ergun,
  M.~McCarthy, H.~R\`{e}me, T.~R. Sanderson, M.~Kaiser, R.~P. Lepping, and
  J.~Mazur, Tracing the topology of the october 18-20, 1995, magnetic cloud
  with $\sim0.1-10^2$ kev electrons, \textit{Geophys. Res. Lett.},
  \textit{24(15)}, 1911--1914, 1997.

\bibitem[{\textit{Lepping et~al.}(1990)\textit{Lepping, Jones, and
  Burlaga}}]{Lepping_etal_1990}
Lepping, R.~P., J.~A. Jones, and L.~F. Burlaga, Magnetic field structure of
  interplanetary magnetic clouds at 1 {AU}, \textit{J. Geophys. Res.},
  \textit{95}, 11,957--11,965, 1990.

\bibitem[{\textit{Lepping et~al.}(2006)\textit{Lepping, Berdichevsky, Wu,
  Szabo, Narock, Mariani, Lazarus, and Quivers}}]{Lepping_etal_2006}
Lepping, R.~P., D.~B. Berdichevsky, C.-C. Wu, A.~Szabo, T.~Narock, F.~Mariani,
  A.~J. Lazarus, and A.~J. Quivers, A summary of {WIND} magnetic clouds for
  years 1995--2003: model-fitted parameters, associated errors and
  classifications, \textit{Ann. Geophys.}, \textit{24}, 215--245, 2006.

\bibitem[{\textit{Lugaz}(2010)}]{Lugaz_2010}
Lugaz, N., Accuracy and limitations of fitting and stereoscopic methods to
  determine the direction of coronal mass ejections from heliospheric imagers
  observations, \textit{Sol. Phys.}, \textit{267}, 411--429, 2010.

\bibitem[{\textit{Lugaz et~al.}(2013)\textit{Lugaz, Farrugia, {Manchester, IV},
  and Schwadron}}]{Lugaz_etal_2013}
Lugaz, N., C.~J. Farrugia, W.~B. {Manchester, IV}, and N.~Schwadron, The
  interaction of two coronal mass ejections: Influence of relative orientation,
  \textit{Astrophys. J.}, \textit{778}, 20(14pp), 2013.

\bibitem[{\textit{Lundquist}(1950)}]{Lundquist_1950}
Lundquist, S., Magnetohydrostatic fields, \textit{Ark. Fys.}, \textit{2}, 361,
  1950.

\bibitem[{\textit{Markwardt}(2009)}]{Markwardt_2009}
Markwardt, C.~B., Non-linear least squares fitting in {IDL} with {MPFIT}, in
  \textit{Astronomical Data Analysis Software and Systems XVIII}, \textit{ASP
  Conference Series}, vol. 411, edited by D.~Bohlender, P.~Dowler, and
  D.~Durand, pp. 251--254, 2009.

\bibitem[{\textit{Marubashi}(1986)}]{Marubashi_1986}
Marubashi, K., Structure of the interplanetary magnetic clouds and their solar
  origins, \textit{Adv. in Space Res.}, \textit{6}, 335--338, 1986.

\bibitem[{\textit{Marubashi}(1997)}]{Marubashi_1997}
Marubashi, K., Interplanetary magnetic flux ropes and solar filaments, in
  \textit{Coronal Mass Ejections}, edited by N.~Crooker, J.~A. Joselyn, and
  J.~Feynman, Geophys. Monogr. Ser. 99, pp. 147--156, AGU, 1997.

\bibitem[{\textit{Marubashi and Lepping}(2007)}]{Marubashi_Lepping_2007}
Marubashi, K., and R.~P. Lepping, Long-duration magnetic clouds: a comparison
  of analyses using torus- and cylinder-shaped flux rope models, \textit{Ann.
  Geophys.}, \textit{25}, 2453--2477, 2007.

\bibitem[{\textit{Marubashi et~al.}(2012)\textit{Marubashi, Cho, Kim, Park, and
  Park}}]{Marubashi_etal_2012}
Marubashi, K., K.-S. Cho, Y.-H. Kim, Y.-D. Park, and S.-H. Park, Geometry of
  the 20 {November} 2003 magnetic cloud, \textit{J. Geophys. Res.},
  \textit{117}, A01,101, 2012.

\bibitem[{\textit{More}(1978)}]{More_1978}
More, J., The {Levenberg-Marquardt} algorithm: Implementation and theory, in
  \textit{Numerical Analysis}, vol. 630, edited by G.~A. Watson, p. 105, 1978.

\bibitem[{\textit{Mulligan and Russell}(2001)}]{Mulligan_Russell_2001}
Mulligan, T., and C.~T. Russell, Multispacecraft modeling of the flux rope
  structure of interplanetary coronal mass ejections: Cylindrically symmetric
  versus nonsymmetric topologies, \textit{J. Geophys. Res.}, \textit{106(A6)},
  10,581--10,596, 2001.

\bibitem[{\textit{Nakwacki et~al.}(2008)\textit{Nakwacki, Dasso, Mandrini, and
  D{\'e}moulin}}]{Nakwacki_etal_2008}
Nakwacki, M.~S., S.~Dasso, C.~H. Mandrini, and P.~D{\'e}moulin, Analysis of
  large scale {MHD} quantities in expanding magnetic clouds, \textit{J. Atmos.
  Solar-Terres. Phys.}, \textit{70}, 1318--1326, 2008.

\bibitem[{\textit{Nakwacki et~al.}(2011)\textit{Nakwacki, Dasso, D{\'e}moulin,
  Mandrini, and Gulisano}}]{Nakwacki_etal_2011}
Nakwacki, M.~S., S.~Dasso, P.~D{\'e}moulin, C.~H. Mandrini, and A.~M. Gulisano,
  Dynamical evolution of a magnetic cloud from the sun to 5.4 {AU},
  \textit{Astron. \& Astrophys.}, \textit{535}, A52, 2011.

\bibitem[{\textit{Nieves-Chinchilla et~al.}(2012)\textit{Nieves-Chinchilla,
  Colaninno, Vourlidas, Szabo, Lepping, Boardsen, Anderson, and
  Korth}}]{Nieves-Chinchilla_etal_2012}
Nieves-Chinchilla, T., R.~Colaninno, A.~Vourlidas, A.~Szabo, R.~P. Lepping,
  S.~A. Boardsen, B.~J. Anderson, and H.~Korth, Remote and in situ observations
  of an unusual earth-directed coronal mass ejection from multiple viewpoints,
  \textit{J. Geophys. Res.}, \textit{117}, A06,106, 2012.

\bibitem[{\textit{Owens et~al.}(2006)\textit{Owens, Merkin, and
  Riley}}]{Owens_etal_2006}
Owens, M.~J., V.~G. Merkin, and P.~Riley, A kinematically distorted flux rope
  model for magnetic clouds, \textit{J. Geophys. Res.}, \textit{111}, A03,104,
  2006.

\bibitem[{\textit{Qiu et~al.}(2007)\textit{Qiu, Hu, Howard, and
  Yurchyshyn}}]{Qiu_etal_2007}
Qiu, J., Q.~Hu, T.~A. Howard, and V.~B. Yurchyshyn, On the magnetic flux budget
  in low-corona magnetic reconnection and interplanetary coronal mass
  ejections, \textit{Astrophys. J.}, \textit{659}, 758--772, 2007.

\bibitem[{\textit{Riley and Crooker}(2004)}]{Riley_Crooker_2004}
Riley, P., and N.~U. Crooker, Kinematic treatment of coronal mass ejection
  evolution in the solar wind, \textit{Astrophys. J.}, \textit{600},
  1035--1042, 2004.

\bibitem[{\textit{Riley et~al.}(2003)\textit{Riley, Linker, Mikic, Odstrcil,
  Zurbuchen, and Lario}}]{Riley_etal_2003}
Riley, P., J.~A. Linker, Z.~Mikic, D.~Odstrcil, T.~H. Zurbuchen, and R.~P.
  Lario, D. amd~Lepping, Using an {MHD} simulation to interpret the global
  context of a coronal mass ejection observed by two spacecraft, \textit{J.
  Geophys. Res.}, \textit{108}, 1272, 2003.

\bibitem[{\textit{Riley et~al.}(2004)\textit{Riley, Linker, Lionello,
  Miki\'{c}, Odstrcil, Hidalgo, Cid, Hu, Lepping, Lynch, and
  Rees}}]{Riley_etal_2004}
Riley, P., J.~A. Linker, R.~Lionello, Z.~Miki\'{c}, D.~Odstrcil, M.~A. Hidalgo,
  C.~Cid, Q.~Hu, R.~P. Lepping, B.~J. Lynch, and A.~Rees, Fitting flux ropes to
  a global {MHD} solution: a comparison of techniques, \textit{J. Atmos.
  Solar-Terres. Phys.}, \textit{66}, 1321--1331, 2004.

\bibitem[{\textit{Rodriguez et~al.}(2011)\textit{Rodriguez, Mierla, Zhukov,
  West, and Kilpua}}]{Rodriguez_etal_2011}
Rodriguez, L., M.~Mierla, A.~Zhukov, M.~West, and E.~Kilpua, Linking
  remote-sensing and in situ observations of coronal mass ejections using
  {STEREO}, \textit{Sol. Phys.}, \textit{270}, 561--573, 2011.

\bibitem[{\textit{Romashets and Vandas}(2003)}]{Romashets_Vandas_2003}
Romashets, E.~P., and M.~Vandas, Force-free field inside a toroidal magnetic
  cloud, \textit{Geophys. Res. Lett.}, \textit{30}, 2065, 2003.

\bibitem[{\textit{Rust et~al.}(2005)\textit{Rust, Anderson, Andrews, Acuna,
  Russell, Schuck, and Mulligan}}]{Rust_etal_2005}
Rust, K., B.~J. Anderson, M.~D. Andrews, M.~H. Acuna, C.~T. Russell, P.~W.
  Schuck, and T.~Mulligan, Comparison of interplanetary disturbances at the
  {NEAR} spacecraft with coronal mass ejections at the {SUN},
  \textit{Astrophys. J.}, \textit{621}, 524--536, 2005.

\bibitem[{\textit{Savani et~al.}(2009)\textit{Savani, Rouillard, Davies, Owens,
  Forsyth, Davis, and Harrison}}]{Savani_etal_2009}
Savani, N.~P., A.~P. Rouillard, J.~A. Davies, M.~J. Owens, R.~J. Forsyth, C.~J.
  Davis, and R.~A. Harrison, The radial width of a coronal mass ejection
  between 0.1 and 0.4 {AU} estimated from the heliospheric imager on {STEREO},
  \textit{Ann. Geophys.}, \textit{27}, 4349--4358, 2009.

\bibitem[{\textit{Schwenn et~al.}(2005)\textit{Schwenn, {Dal Lago}, Huttunen,
  and Gonzalez}}]{Schwenn_etal_2005}
Schwenn, R., A.~{Dal Lago}, E.~Huttunen, and W.~D. Gonzalez, The association of
  coronal mass ejections with their effects near the {Earth}, \textit{Ann.
  Geophys.}, \textit{23(3)}, 1033--1059, 2005.

\bibitem[{\textit{Shimazu and Vandas}(2002)}]{Shimazu_Vandas_2002}
Shimazu, H., and M.~Vandas, A self-similar solution of expanding cylindrical
  flux ropes for any polytropic index value, \textit{Earth Planets Space},
  \textit{54}, 783--790, 2002.

\bibitem[{\textit{Vandas and Romashets}(2003)}]{Vandas_Romashets_2003}
Vandas, M., and E.~P. Romashets, A force-free field with constant alpha in an
  oblate cylinder: A generalization of the lundquist solution, \textit{Astron.
  \& Astrophys.}, \textit{398}, 801--807, 2003.

\bibitem[{\textit{Vourlidas et~al.}(2010)\textit{Vourlidas, Howard, Esfandiari, Patsourakos,
Yashiro, and Michalek}}]{Vourlidas_etal_2010}
Vourlidas, A., R.~A. Howard, E.~Esfandiari, S.~Patsourakos, S.~Yashiro, and G.~Michalek, 
Comprehensive analysis of coronal mass ejection mass and energy properties over a full solar cycle, 
\textit{Astrophys. J.}, \textit{722}, 1522--1538, 2010.

\bibitem[{\textit{Vourlidas et~al.}(2011)\textit{Vourlidas, Colaninno,
  Nieves-Chinchilla, and Stenborg}}]{Vourlidas_etal_2011}
Vourlidas, A., R.~Colaninno, T.~Nieves-Chinchilla, and G.~Stenborg, The first
  observation of a rapidly rotating coronal mass ejection in the middle corona,
  \textit{Astrophys. J. Lett.}, \textit{733}, L23(6pp), 2011.

\bibitem[{\textit{Wang et~al.}(2005)\textit{Wang, Du, and
  Richardson}}]{Wang_etal_2005a}
Wang, C., D.~Du, and J.~D. Richardson, Characteristics of the interplanetary
  coronal mass ejections in the heliosphere between 0.3 and 5.4 {AU},
  \textit{J. Geophys. Res.}, \textit{110}, {A10,107}, 2005.

\bibitem[{\textit{Wang et~al.}(2002)\textit{Wang, Ye, Wang, Zhou, and
  Wang}}]{Wang_etal_2002a}
Wang, Y., P.~Z. Ye, S.~Wang, G.~P. Zhou, and J.~X. Wang, A statistical study on
  the geoeffectiveness of earth-directed coronal mass ejections from {March}
  1997 to {December} 2000, \textit{J. Geophys. Res.}, \textit{107(A11)}, 1340,
  doi:10.1029/2002JA009,244, 2002.

\bibitem[{\textit{Wang et~al.}(2004)\textit{Wang, Shen, Ye, and
  Wang}}]{Wang_etal_2004b}
Wang, Y., C.~Shen, P.~Ye, and S.~Wang, Deflection of coronal mass ejection in
  the interplanetary medium, \textit{Sol. Phys.}, \textit{222}, 329--343, 2004.

\bibitem[{\textit{Wang et~al.}(2006{\natexlab{a}})\textit{Wang, Xue, Shen, Ye,
  Wang, and Zhang}}]{Wang_etal_2006a}
Wang, Y., X.~Xue, C.~Shen, P.~Ye, S.~Wang, and J.~Zhang, Impact of the major
  coronal mass ejections on geo-space during {September} 7 -- 13, 2005,
  \textit{Astrophys. J.}, \textit{646}, 625--633, 2006{\natexlab{a}}.

\bibitem[{\textit{Wang et~al.}(2006{\natexlab{b}})\textit{Wang, Zhou, Ye, Wang,
  and Wang}}]{Wang_etal_2006c}
Wang, Y., G.~Zhou, P.~Ye, S.~Wang, and J.~Wang, A study on the orientation of
  interplanetary magnetic clouds and solar filaments, \textit{Astrophys. J.},
  \textit{651}, 1245--1255, 2006{\natexlab{b}}.

\bibitem[{\textit{Wang et~al.}(2009)\textit{Wang, Zhang, and
  Shen}}]{Wang_etal_2009}
Wang, Y., J.~Zhang, and C.~Shen, An analytical model probing the internal state
  of coronal mass ejections based on observations of their expansions and
  propagations, \textit{J. Geophys. Res.}, \textit{114}, A10,104, 2009.

\bibitem[{\textit{Wang et~al.}(2014)\textit{Wang, Wang, Shen, Shen, and
  Lugaz}}]{Wang_etal_2014}
Wang, Y., B.~Wang, C.~Shen, F.~Shen, and N.~Lugaz, Deflected propagation of a
  coronal mass ejection from the corona to interplanetary space, \textit{J.
  Geophys. Res.}, \textit{accepted}, 2014.

\bibitem[{\textit{Wei et~al.}(1991)\textit{Wei, Lee, Wang, and
  Akasofu}}]{Wei_etal_1991}
Wei, C.~Q., L.~C. Lee, S.~Wang, and S.-I. Akasofu, Evolution of magnetic flux
  ropes associated with flux transfer events and interplanetary magnetic
  clouds, \textit{J. Geophys. Res.}, \textit{96}, 1619--1632, 1991.

\bibitem[{\textit{Xiong et~al.}(2006)\textit{Xiong, Zheng, Wang, and
  Wang}}]{Xiong_etal_2006b}
Xiong, M., H.~Zheng, Y.~Wang, and S.~Wang, Magnetohydrodynamic simulation of
  the interaction between interplanetary strong shock and magnetic cloud and
  its consequent geoeffectiveness: 2. oblique collision, \textit{J. Geophys.
  Res.}, \textit{111}, A11,102, 2006.

\bibitem[{\textit{Xiong et~al.}(2007)\textit{Xiong, Zheng, Wu, Wang, and
  Wang}}]{Xiong_etal_2007}
Xiong, M., H.~Zheng, S.~T. Wu, Y.~Wang, and S.~Wang, Magnetohydrodynamic
  simulation of the interaction between two interplanetary magnetic clouds and
  its consequent geoeffectiveness, \textit{J. Geophys. Res.}, \textit{112},
  A11,103, 2007.

\bibitem[{\textit{Yurchyshyn}(2008)}]{Yurchyshyn_2008}
Yurchyshyn, V., Relationship between {EIT} posteruption arcades, coronal mass
  ejections, the coronal neutral line, and magnetic clouds, \textit{Astrophys.
  J. Lett.}, \textit{675}, L49--L52, 2008.

\bibitem[{\textit{Yurchyshyn et~al.}(2009)\textit{Yurchyshyn, Abramenko, and
  Tripathi}}]{Yurchyshyn_etal_2009}
Yurchyshyn, V., V.~Abramenko, and D.~Tripathi, Rotation of white-light coronal
  mass ejection structures as inferred from {LASCO} coronagraph,
  \textit{Astrophys. J.}, \textit{705}, 426--435, 2009.

\end{thebibliography}

\end{document}